\documentclass[10pt]{llncs}\usepackage[]{graphicx}\usepackage[]{color}
\makeatletter
\def\maxwidth{ %
  \ifdim\Gin@nat@width>\linewidth
    \linewidth
  \else
    \Gin@nat@width
  \fi
}
\makeatother

\definecolor{fgcolor}{rgb}{0.345, 0.345, 0.345}

\usepackage{framed}
\makeatletter
 {\par\unskip\endMakeFramed%
 \at@end@of@kframe}
\makeatother

\definecolor{shadecolor}{rgb}{.97, .97, .97}
\definecolor{messagecolor}{rgb}{0, 0, 0}
\definecolor{warningcolor}{rgb}{1, 0, 1}
\definecolor{errorcolor}{rgb}{1, 0, 0}

\usepackage{alltt} %

\usepackage{epsfig,endnotes,url}

\usepackage{xspace}

\usepackage[obeyDraft,colorinlistoftodos]{todonotes}

\usepackage[english]{babel}
\usepackage[latin1]{inputenc}
\usepackage{amsmath, amssymb, latexsym} %

\usepackage{graphicx}
\usepackage{multirow}
\usepackage{booktabs,longtable}
\usepackage{rotating}
\usepackage{afterpage}
\usepackage{wasysym}

\usepackage{paralist}

\usepackage{balance}

\usepackage{versions}
\excludeversion{DocumentVersionConference}
\includeversion{DocumentVersionTR}
\includeversion{DocumentVersionLNCS}

\usepackage[caption=false]{subfig}

\newsubfloat{figure}

\newsubfloat{table}

\newtheorem{researchquestion}{RQ}

\newcommand{\vari}[1]{\ensuremath{\mathit{#1}\xspace}}
\newcommand{\const}[1]{\ensuremath{\mathsf{#1}\xspace}}

\newcommand{\CASCAde}{ERC Starting Grant CASCAde (GA n\textsuperscript{o}716980)}

\newcommand{\EXCL}{\ensuremath{\varnothing}\xspace}

\IfFileExists{upquote.sty}{\usepackage{upquote}}{}
\begin{document}

\newcommand{\contingencyVenueYear}{

\begin{table}[ht]
\centering
\caption{Sample composition by \textsf{venue} and \textsf{year}.} 
\label{tab:contingencyVenueYear}
\begingroup\small
\begin{tabular}{rrrrrrrrrrrrr}
  \hline
 & \begin{sideways} 2006 \end{sideways} & \begin{sideways} 2007 \end{sideways} & \begin{sideways} 2008 \end{sideways} & \begin{sideways} 2009 \end{sideways} & \begin{sideways} 2010 \end{sideways} & \begin{sideways} 2011 \end{sideways} & \begin{sideways} 2012 \end{sideways} & \begin{sideways} 2013 \end{sideways} & \begin{sideways} 2014 \end{sideways} & \begin{sideways} 2015 \end{sideways} & \begin{sideways} 2016 \end{sideways} & \begin{sideways} Sum \end{sideways} \\ 
  \hline
SOUPS & 6 & 3 & 4 & 6 & 8 & 4 & 10 & 8 & 13 & 9 & 6 & 77 \\ 
  USEC & 0 & 0 & 0 & 0 & 0 & 0 & 0 & 0 & 4 & 0 & 0 & 4 \\ 
  CCS & 0 & 0 & 0 & 0 & 0 & 0 & 0 & 0 & 4 & 1 & 3 & 8 \\ 
  USENIX & 0 & 0 & 0 & 1 & 0 & 0 & 4 & 1 & 1 & 0 & 0 & 7 \\ 
  PETS & 1 & 0 & 0 & 0 & 0 & 1 & 1 & 0 & 0 & 1 & 2 & 6 \\ 
  TISSEC & 0 & 0 & 0 & 0 & 0 & 1 & 0 & 0 & 0 & 0 & 2 & 3 \\ 
  LASER & 0 & 0 & 0 & 0 & 0 & 0 & 1 & 0 & 0 & 0 & 1 & 2 \\ 
  S\&P & 0 & 0 & 0 & 0 & 0 & 0 & 0 & 0 & 1 & 1 & 0 & 2 \\ 
  TDSC & 0 & 1 & 0 & 0 & 0 & 0 & 1 & 0 & 1 & 0 & 1 & 4 \\ 
  WEIS & 0 & 0 & 0 & 0 & 0 & 0 & 0 & 0 & 0 & 1 & 0 & 1 \\ 
  Sum & 7 & 4 & 4 & 7 & 8 & 6 & 17 & 9 & 24 & 13 & 15 & 114 \\ 
   \hline
\end{tabular}
\endgroup
\end{table}
}
\newcommand{\contingencySCOutcomeYear}{
\begin{table}[ht]
\centering
\caption{Contingency table of aggregated paper \textsf{statcheck} outcomes by \textsf{year}, FET $p = .458$.} 
\label{tab:contingencySCOutcomeYear}
\begingroup\small
\begin{tabular}{rrrrrrrrrrrr}
  \hline
 & \begin{sideways} 2006 \end{sideways} & \begin{sideways} 2007 \end{sideways} & \begin{sideways} 2008 \end{sideways} & \begin{sideways} 2009 \end{sideways} & \begin{sideways} 2010 \end{sideways} & \begin{sideways} 2011 \end{sideways} & \begin{sideways} 2012 \end{sideways} & \begin{sideways} 2013 \end{sideways} & \begin{sideways} 2014 \end{sideways} & \begin{sideways} 2015 \end{sideways} & \begin{sideways} 2016 \end{sideways} \\ 
  \hline
CorrectNHST &   2 &   1 &   1 &   2 &   2 &   4 &   5 &   1 &   5 &   3 &   1 \\ 
  Inconsistency &   2 &   0 &   0 &   1 &   1 &   0 &   0 &   0 &   3 &   3 &   2 \\ 
  DecisionError &   0 &   1 &   0 &   1 &   1 &   0 &   1 &   1 &   0 &   0 &   1 \\ 
  Incomplete &   3 &   2 &   3 &   3 &   4 &   2 &  11 &   7 &  16 &   7 &  11 \\ 
   \hline
\end{tabular}
\endgroup
\end{table}
}
\newcommand{\contingencySCOutcomeVenue}{
\begin{table}[ht]
\centering
\caption{Contingency table of aggregated paper \textsf{statcheck} outcomes by \textsf{venue}, FET $p = .964$.} 
\label{tab:contingencySCOutcomeVenue}
\begingroup\small
\begin{tabular}{rrrrrrrrrrr}
  \hline
 & \begin{sideways} SOUPS \end{sideways} & \begin{sideways} USEC \end{sideways} & \begin{sideways} CCS \end{sideways} & \begin{sideways} USENIX \end{sideways} & \begin{sideways} PETS \end{sideways} & \begin{sideways} TISSEC \end{sideways} & \begin{sideways} LASER \end{sideways} & \begin{sideways} S\&P \end{sideways} & \begin{sideways} TDSC \end{sideways} & \begin{sideways} WEIS \end{sideways} \\ 
  \hline
CorrectNHST &  19 &   0 &   1 &   2 &   3 &   1 &   0 &   0 &   1 &   0 \\ 
  Inconsistency &  10 &   1 &   1 &   0 &   0 &   0 &   0 &   0 &   0 &   0 \\ 
  DecisionError &   5 &   0 &   0 &   0 &   0 &   0 &   1 &   0 &   0 &   0 \\ 
  Incomplete &  43 &   3 &   6 &   5 &   3 &   2 &   1 &   2 &   3 &   1 \\ 
   \hline
\end{tabular}
\endgroup
\end{table}
}
\newcommand{\contingencyTestsSCOutcomeYear}{
\begin{table}[ht]
\centering
\caption{Contingency table of individual test \textsf{statcheck} outcomes by \textsf{year}, FET $p < .001$.} 
\label{tab:contingencyTestsSCOutcomeYear}
\begingroup\small
\begin{tabular}{rrrrrrrrrrrr}
  \hline
 & \begin{sideways} 2006 \end{sideways} & \begin{sideways} 2007 \end{sideways} & \begin{sideways} 2008 \end{sideways} & \begin{sideways} 2009 \end{sideways} & \begin{sideways} 2010 \end{sideways} & \begin{sideways} 2011 \end{sideways} & \begin{sideways} 2012 \end{sideways} & \begin{sideways} 2013 \end{sideways} & \begin{sideways} 2014 \end{sideways} & \begin{sideways} 2015 \end{sideways} & \begin{sideways} 2016 \end{sideways} \\ 
  \hline
CorrectNHST &  13 &  24 &  14 &  18 &  26 &  13 &  22 &   9 &  37 &  28 &  14 \\ 
  Inconsistency &   2 &   1 &   0 &   1 &   2 &   0 &   2 &   4 &   4 &   3 &   5 \\ 
  DecisionError &   0 &   5 &   0 &   1 &   1 &   0 &   1 &   1 &   0 &   0 &   1 \\ 
  Incomplete &  53 &  57 &  28 & 105 &  96 &  59 & 347 & 123 & 270 & 170 & 215 \\ 
   \hline
\end{tabular}
\endgroup
\end{table}
}
\newcommand{\contingencyTestsSCOutcomeVenue}{
\begin{table}[ht]
\centering
\caption{Contingency table of individual test \textsf{statcheck} outcomes by \textsf{venue}, FET $p = .033$.} 
\label{tab:contingencyTestsSCOutcomeVenue}
\begingroup\small
\begin{tabular}{rrrrrrrrrrr}
  \hline
 & \begin{sideways} SOUPS \end{sideways} & \begin{sideways} USEC \end{sideways} & \begin{sideways} CCS \end{sideways} & \begin{sideways} USENIX \end{sideways} & \begin{sideways} PETS \end{sideways} & \begin{sideways} TISSEC \end{sideways} & \begin{sideways} LASER \end{sideways} & \begin{sideways} S\&P \end{sideways} & \begin{sideways} TDSC \end{sideways} & \begin{sideways} WEIS \end{sideways} \\ 
  \hline
CorrectNHST & 170 &   1 &   9 &   4 &  11 &   6 &   5 &   0 &  12 &   0 \\ 
  Inconsistency &  19 &   1 &   3 &   0 &   0 &   0 &   1 &   0 &   0 &   0 \\ 
  DecisionError &   9 &   0 &   0 &   0 &   0 &   0 &   1 &   0 &   0 &   0 \\ 
  Incomplete & 1028 &  33 & 122 & 100 &  72 &  71 &  19 &  11 &  60 &   7 \\ 
   \hline
\end{tabular}
\endgroup
\end{table}
}

\newcommand{\histogramPDiff}{
\begin{figure}[tb]
\centering\vspace{-2cm}\begin{minipage}{0.6\textwidth}

\includegraphics[width=\maxwidth]{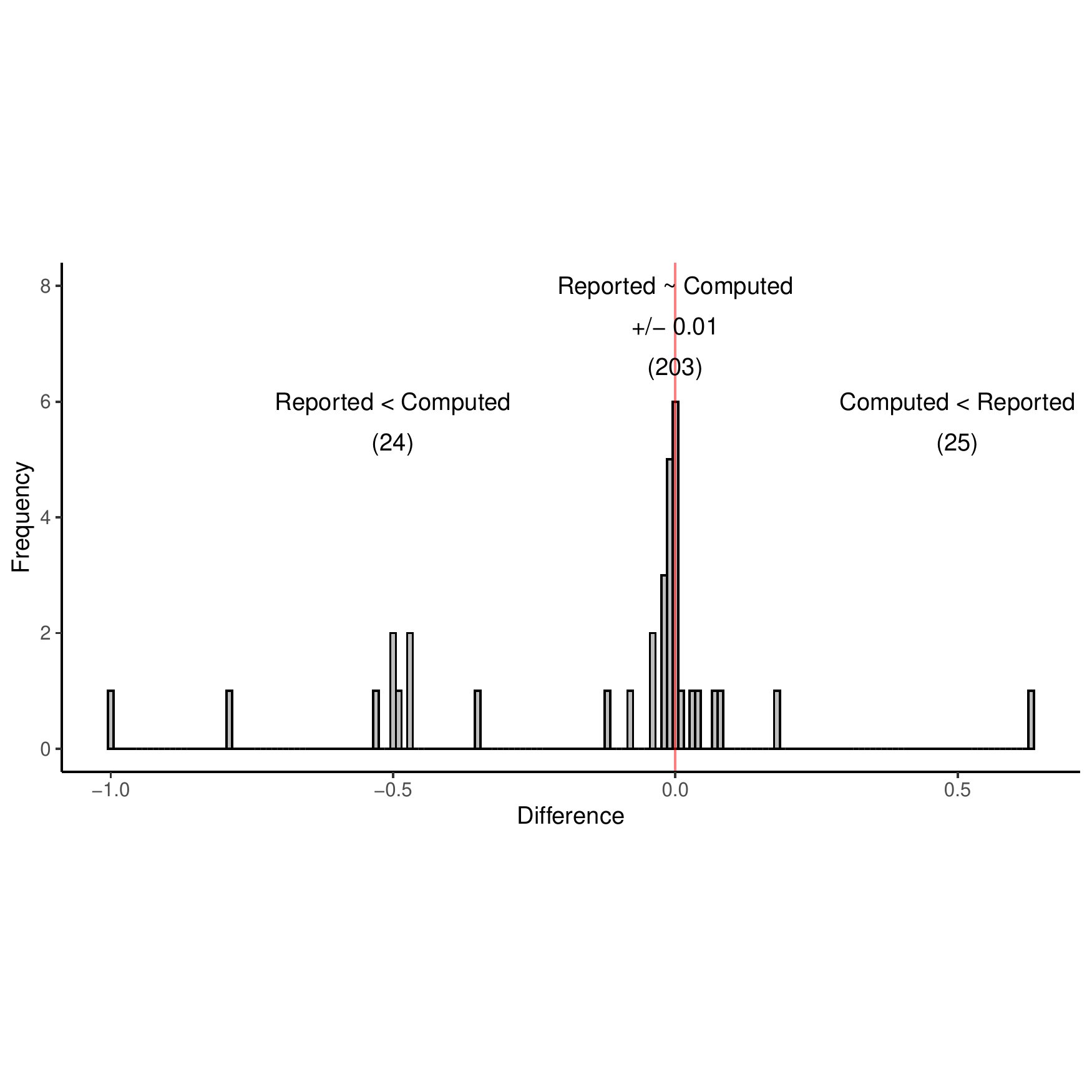} 
\end{minipage}
\vspace{-2cm}\caption{Histogram of difference reported $p$-values minus \textsf{statcheck}-computed $p$-values.}
\label{fig:histogramPValueDifference}
\end{figure}
}

\newcommand{\proportionsplotCombinedSLRArea}{
\begin{figure}[tb]
\centering
\subfloat[Venues]{
\begin{minipage}{0.49\textwidth}

\includegraphics[width=\maxwidth]{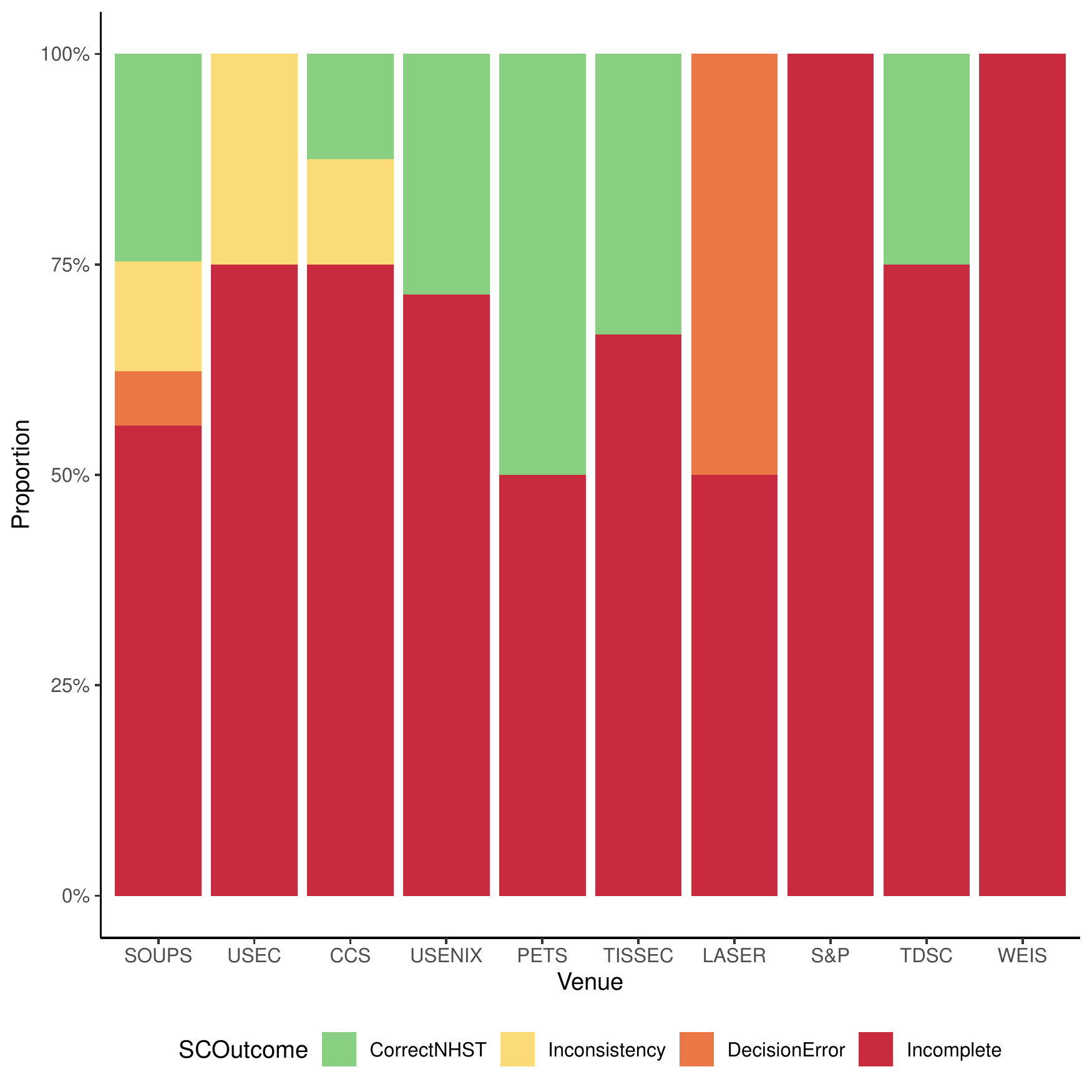} 
\end{minipage}}~
\subfloat[Years]{
\begin{minipage}{0.49\textwidth}

\includegraphics[width=\maxwidth]{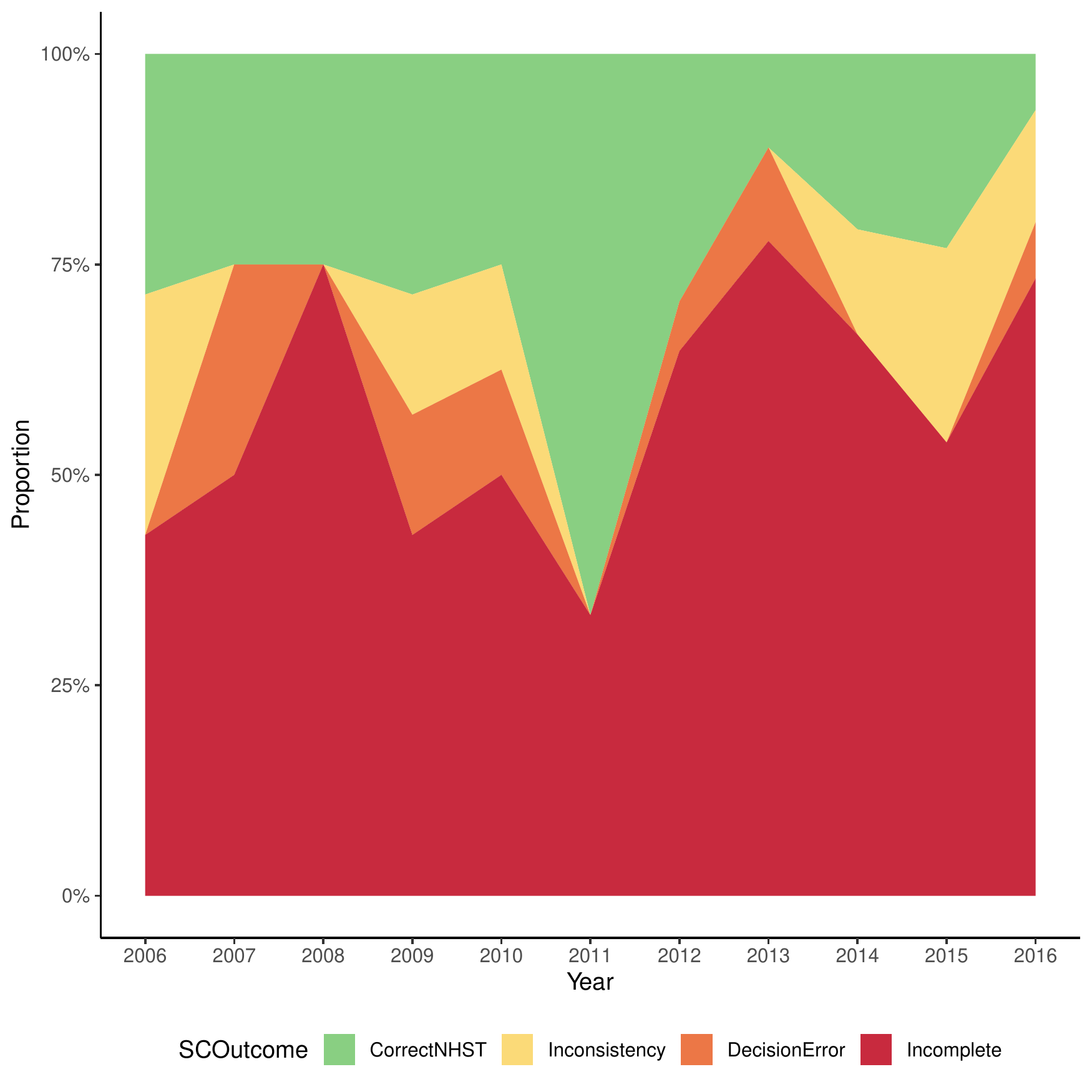} 
\end{minipage}}
\caption{Proportions of per-paper aggregated \textsf{statcheck} outcomes by \textsf{venue} and \textsf{year}. The results by \textsf{year} are shown as area plot to highlight development over time.}
\label{fig:proportionsplotCombinedSLR}
\end{figure}
}

\newcommand{\proportionsplotPropertiesYear}{
\begin{figure}[tb]
\subfloat[MTurk use]{
\label{fig:plotMTurkUse}\begin{minipage}{0.3\linewidth}
\includegraphics[width=\maxwidth]{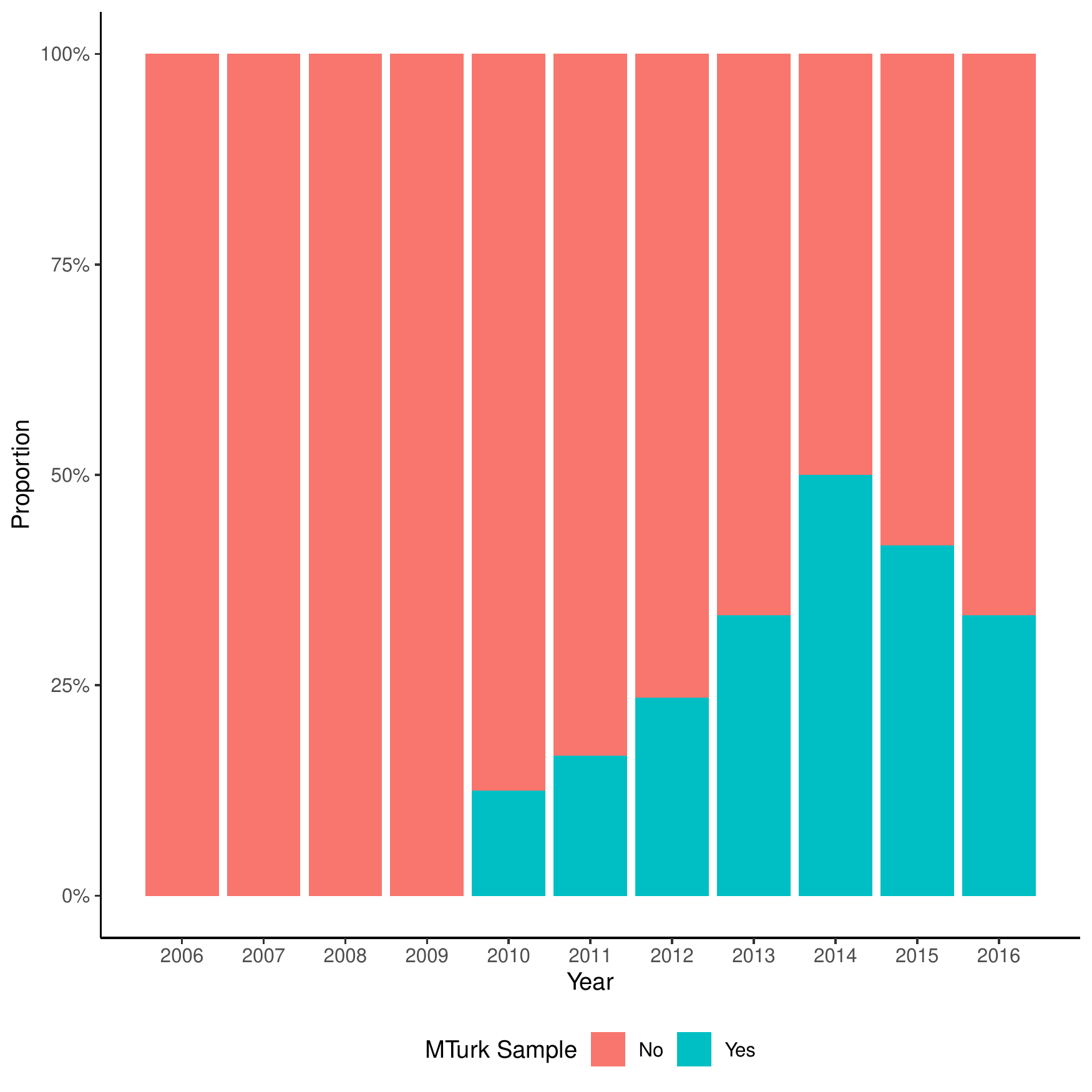} 
\label{fig:proportionsplotMTurkYear}
\end{minipage}
}
\subfloat[MCC]{
\label{fig:plotMCCUse}\begin{minipage}{0.3\linewidth}

\includegraphics[width=\maxwidth]{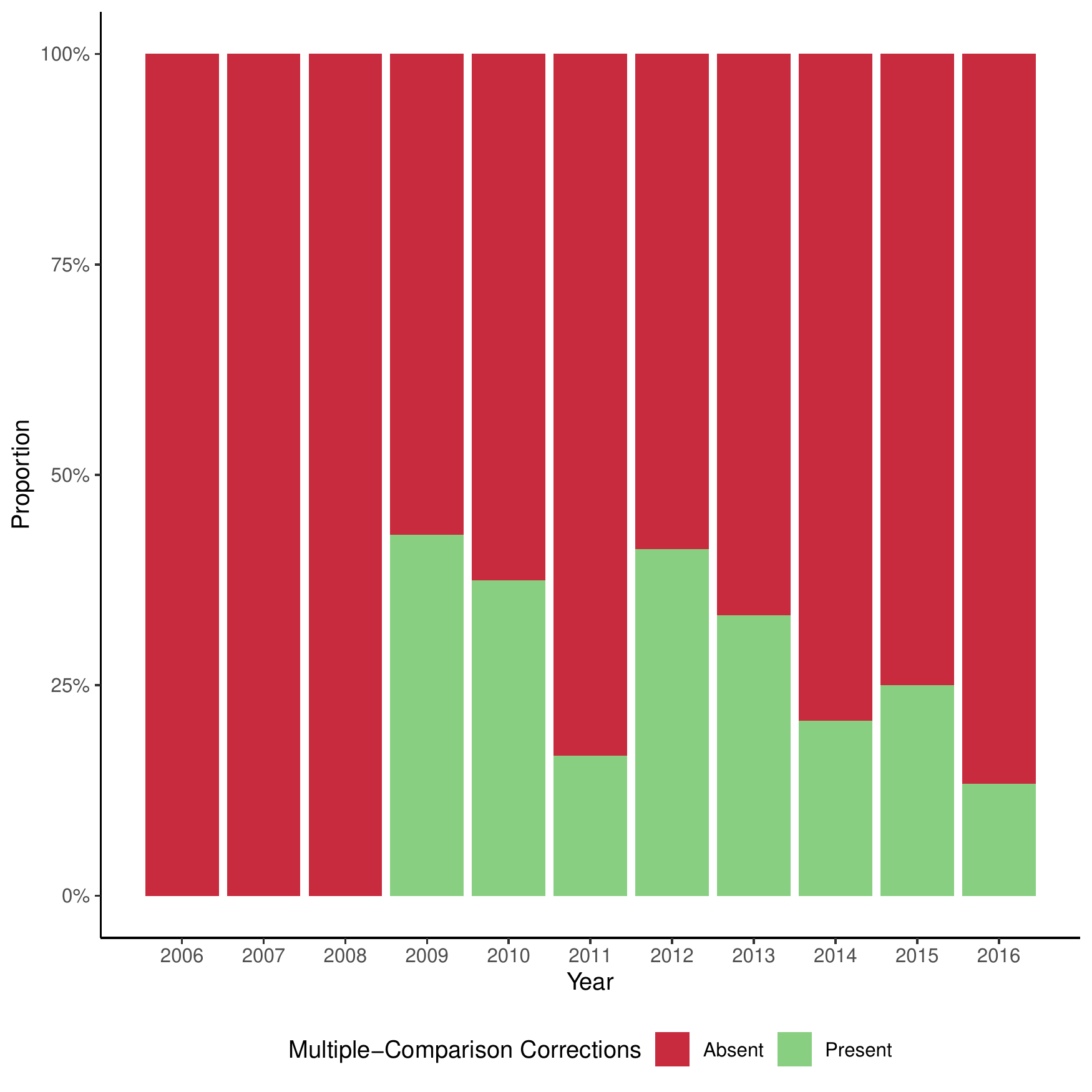} 
\label{fig:proportionsplotMCYear}
\end{minipage}
}
\subfloat[Effect size reporting]{
\label{fig:ESUse}\begin{minipage}{0.3\linewidth}
\includegraphics[width=\maxwidth]{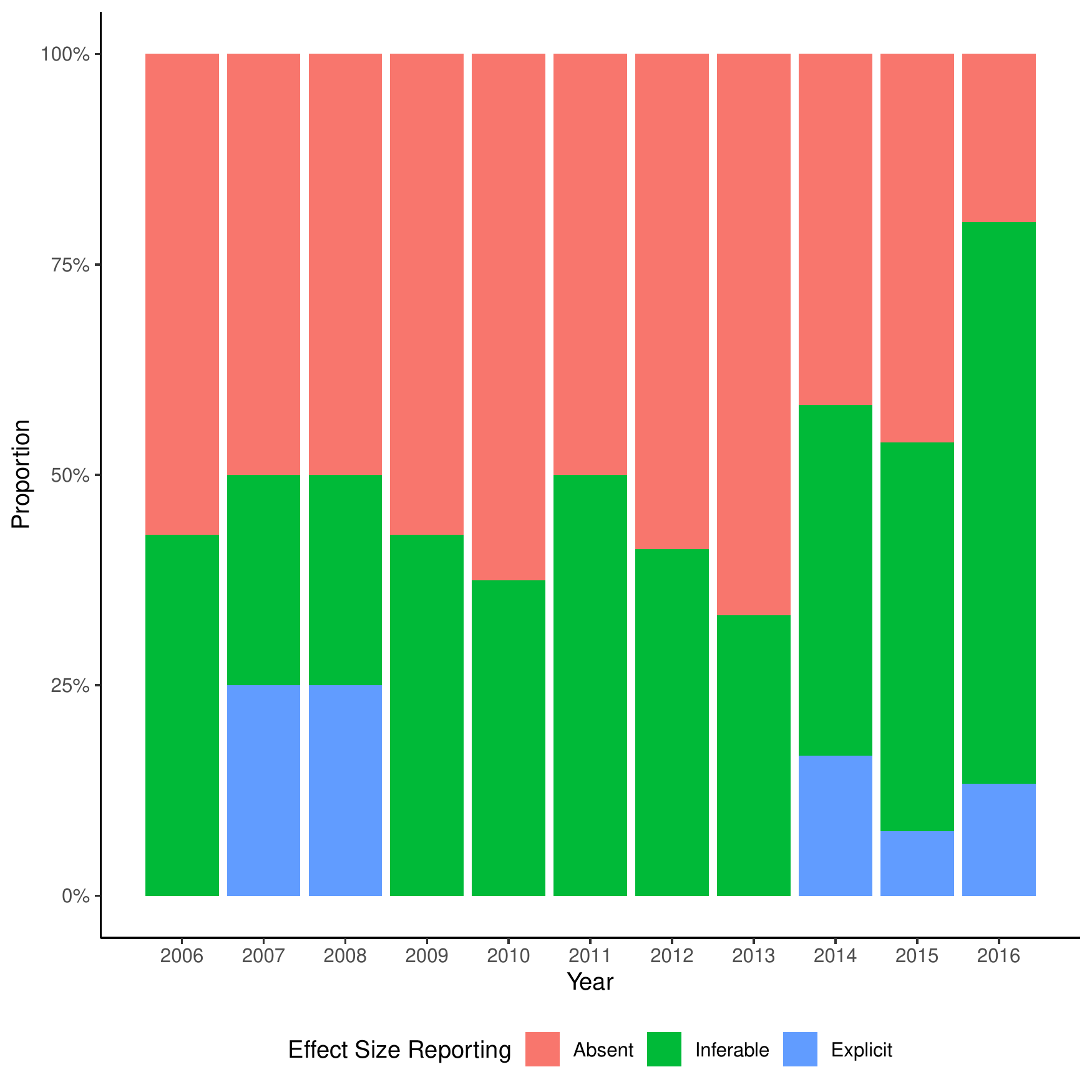} 
\label{fig:proportionsplotESYear}
\end{minipage}
}
\caption{Properties of SLR papers by \textsf{year}. MCC = Multiple-Comparison Corrections}
\label{fig:proportionsplotPropertiesYear}
\end{figure}
}

\newcommand{\proportionsplotPropertiesYearwoMTurk}{
\begin{figure}[tb]
\subfloat[Multiple-comparison corrections (MCC)]{
\label{fig:plotMCCUse}\begin{minipage}{0.49\linewidth}

\includegraphics[width=\maxwidth]{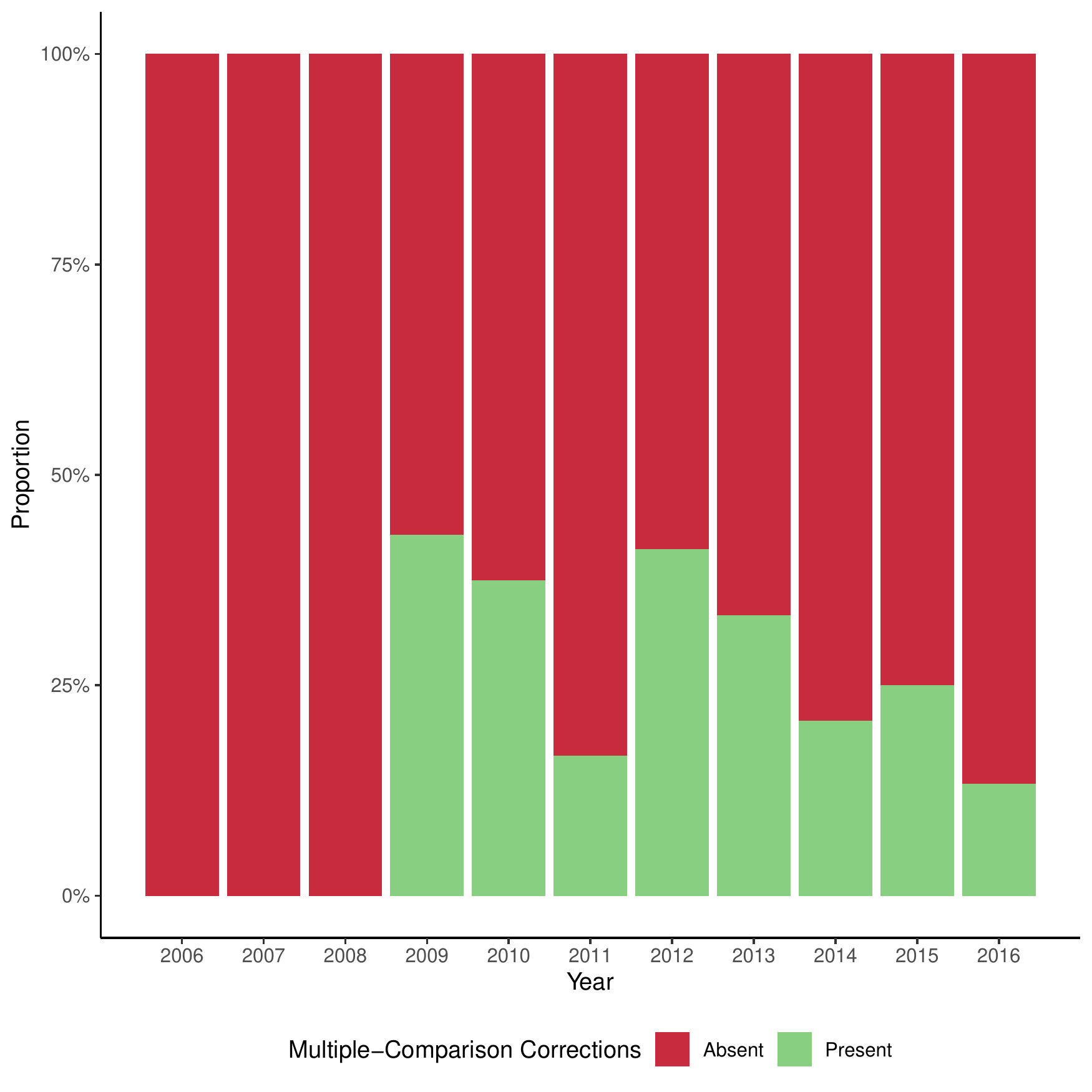} 
\label{fig:proportionsplotMCYear}
\end{minipage}
}
\subfloat[Effect size reporting]{
\label{fig:ESUse}\begin{minipage}{0.49\linewidth}
\includegraphics[width=\maxwidth]{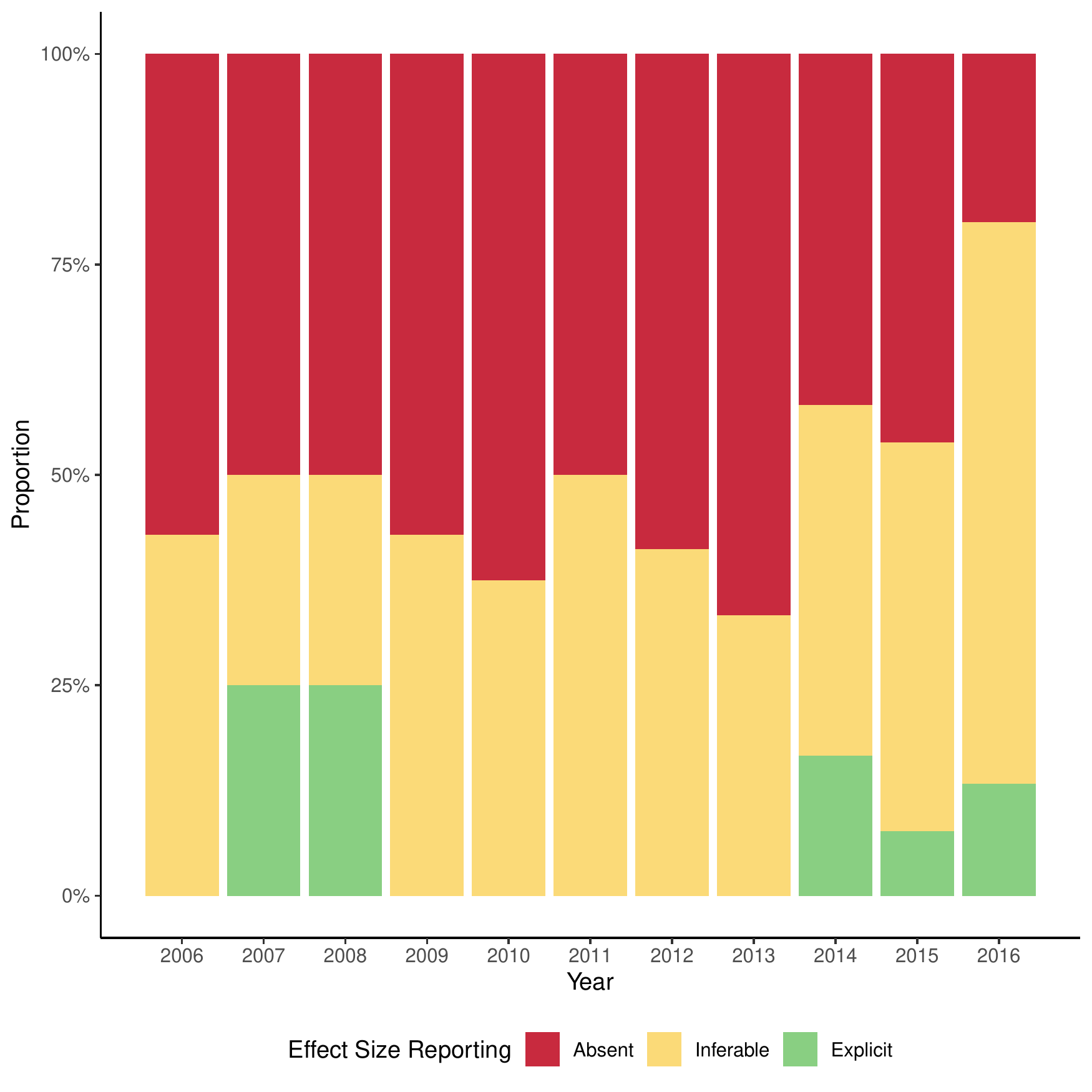} 
\label{fig:proportionsplotESYear}
\end{minipage}
}
\caption{Properties of SLR papers by \textsf{year}.}
\label{fig:proportionsplotPropertiesYear}
\end{figure}
}

\newcommand{\proportionsplotPropertiesYearVert}{
\begin{figure}[p]
\subfloat[MTurk use]{
\label{fig:plotMTurkUse}\begin{minipage}{0.5\linewidth}
\includegraphics[width=\maxwidth]{figure/proportions_plot_properties_year_slr_vertical-1} 
\label{fig:proportionsplotMTurkYear}
\end{minipage}
}

\subfloat[MCC]{
\label{fig:plotMCCUse}\begin{minipage}{0.5\linewidth}

\includegraphics[width=\maxwidth]{figure/proportions_plot_properties_year_slr_vertical-2} 
\label{fig:proportionsplotMCYear}
\end{minipage}
}

\subfloat[Effect size reporting]{
\label{fig:ESUse}\begin{minipage}{0.5\linewidth}
\includegraphics[width=\maxwidth]{figure/proportions_plot_properties_year_slr_vertical-3} 
\label{fig:proportionsplotESYear}
\end{minipage}
}
\caption{Properties of SLR papers by \textsf{year}. MCC = Multiple-Comparison Corrections}
\label{fig:proportionsplotPropertiesYear}
\end{figure}
}

\newcommand{\errorDoubleHistogram}{
\begin{figure}[tb]
\centering\vspace{-1.75cm}
\subfloat[Frequency of inconsistencies]{
\begin{minipage}{0.49\textwidth}

\includegraphics[width=\maxwidth]{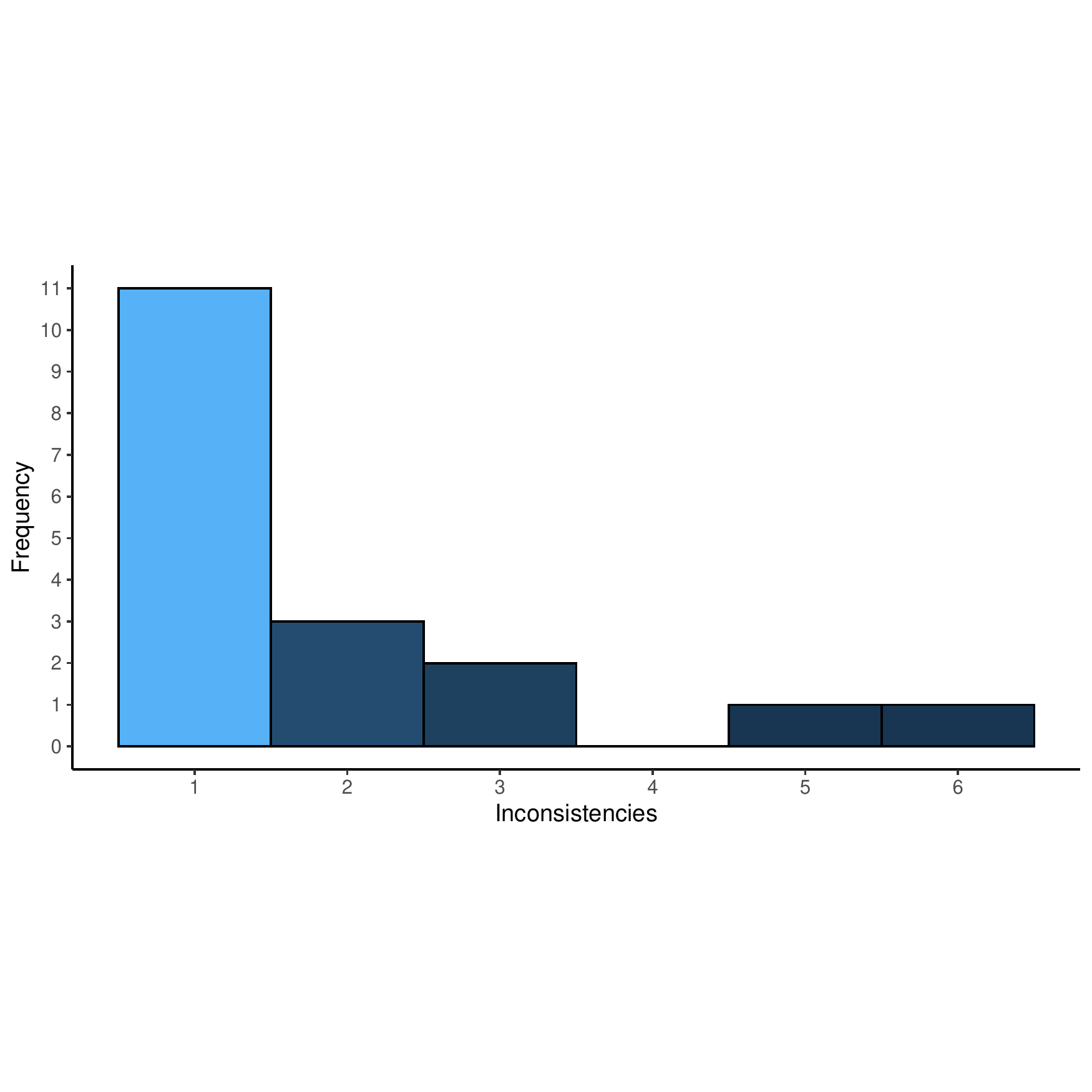} 
\vspace{-1.75cm}
\end{minipage}
}~
\subfloat[Frequency of decision errors]{
\begin{minipage}{0.49\textwidth}

\includegraphics[width=\maxwidth]{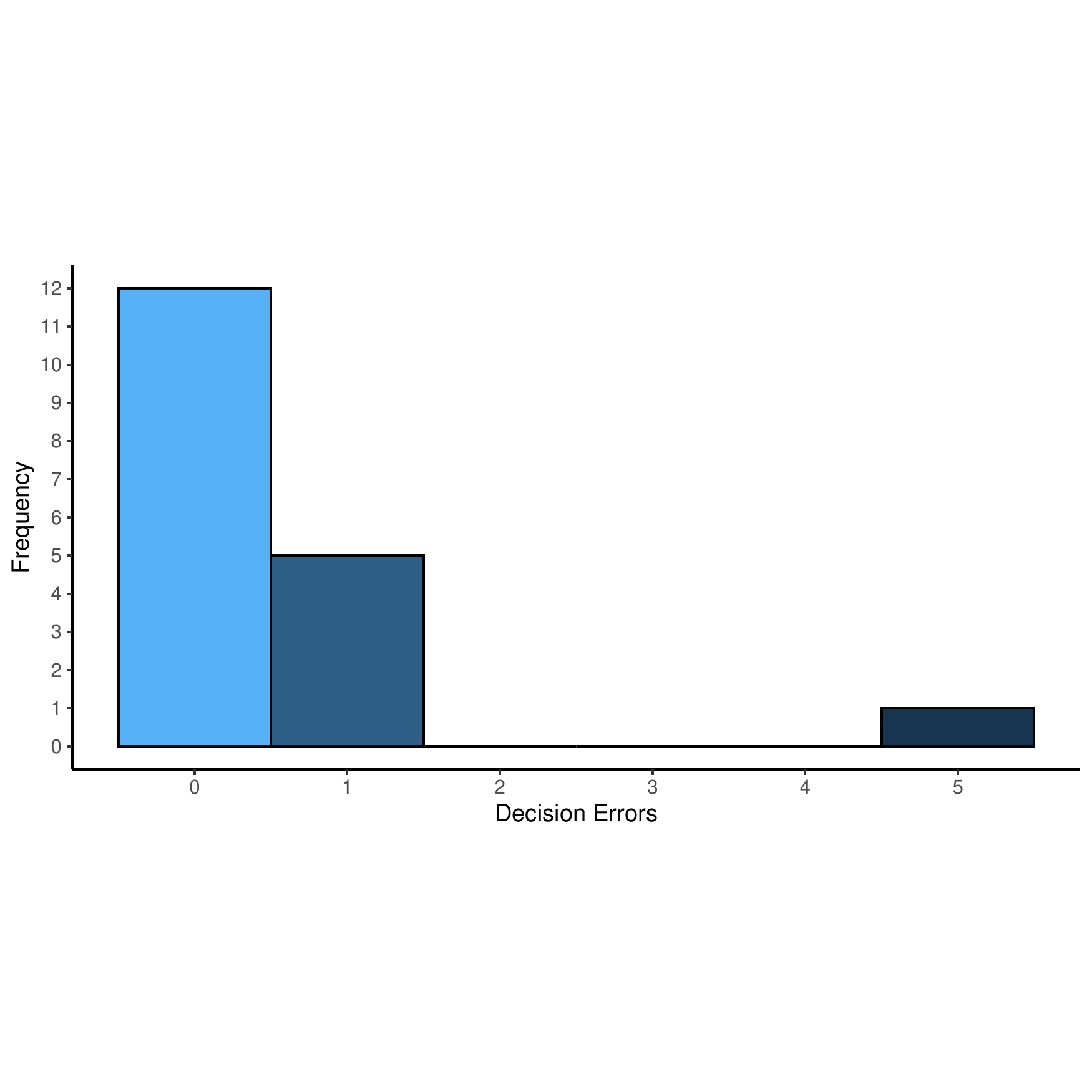} 
\vspace{-1.75cm}
\end{minipage}
}
\caption{Number of errors per paper.}
\label{fig:errorHistogramPlot}
\end{figure}
}

\newcommand{\waffleplotCombined}{
\begin{figure}[tb]
\centering\subfloat[This Study (SLR)]{
\label{fig:waffle.SLR}\label{fig:waffleplotSLR}\centering\begin{minipage}{0.4\textwidth}
\includegraphics[width=\maxwidth]{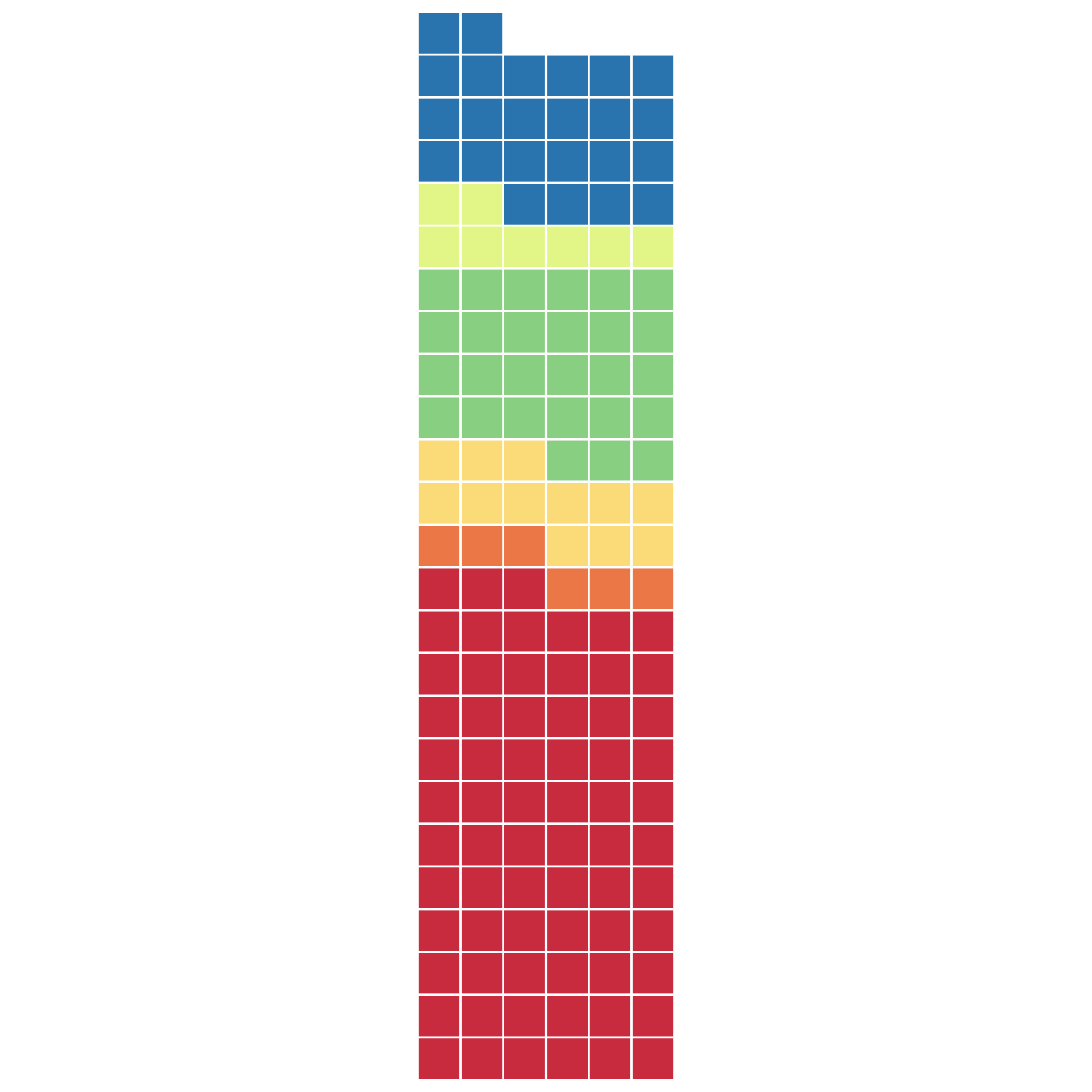} 
\end{minipage}}
~\subfloat[JMP]{
\centering\begin{minipage}{0.4\textwidth}
\includegraphics[width=\maxwidth]{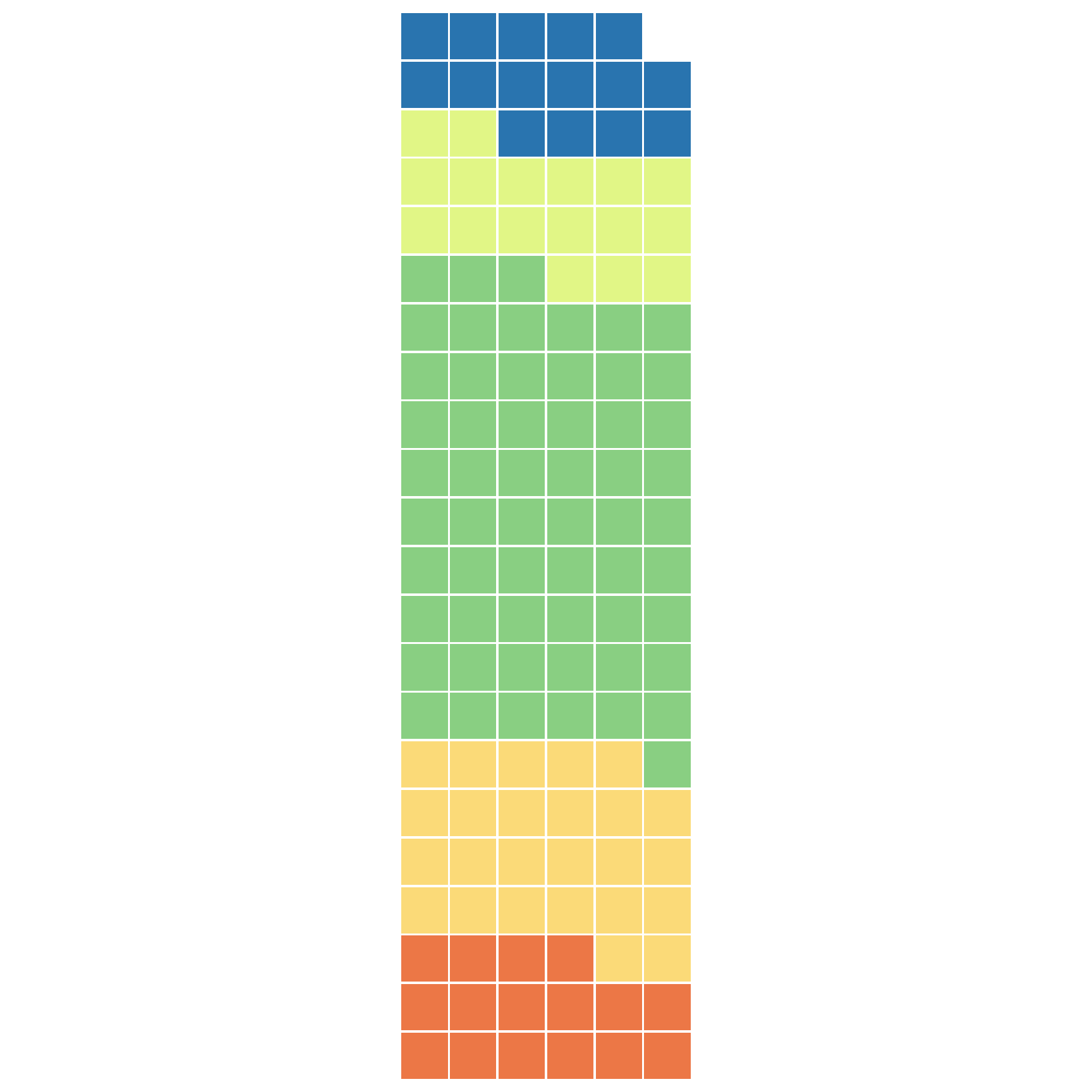} 
\end{minipage}
}
~\subfloat{
\centering\begin{minipage}{0.2\textwidth}\includegraphics[keepaspectratio,width=0.7\textwidth]{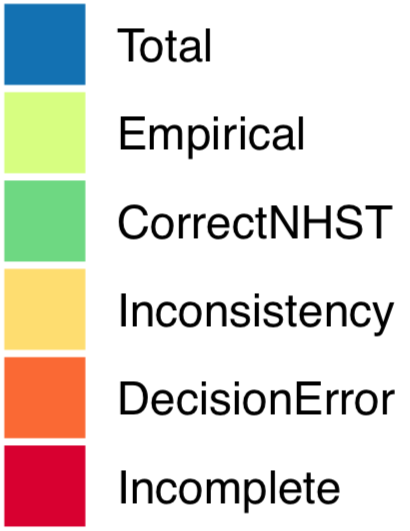}
\end{minipage}
}
\caption{Hierarchical Waffle plots comparing user studies (SLR) in cyber security and the Journal of Media Psychology (JMP) (One square represents one paper).}
\label{fig:waffleplotCombined}
\end{figure}
}

\newcommand{\contingencyTableJMP}{
\begin{table}[ht]
\centering
\caption{Contingency table of comparison between cyber security user studies and studies in the Journal of Psychology (JMP)} 
\label{tab:contingencyJMP}
\begingroup\footnotesize
\begin{tabular}{rrr}
  \hline
 & SLR & JMP \\ 
  \hline
CorrectNHST & 27 & 58 \\ 
  Inconsistency & 12 & 25 \\ 
  DecisionError & 6 & 16 \\ 
  Incomplete & 69 & 0 \\ 
   \hline
\end{tabular}
\endgroup
\end{table}
}

\newcommand{\mosaicplotJMP}{
\begin{figure}[tb]

\includegraphics[width=\maxwidth]{figure/mosaic_slr_jmp-1} 
\caption{Mosaic plot of differences between cyber security user studies and JMP studies.}
\label{fig:mosaicplotJMP}
\end{figure}
}

\newcommand{\combinedplotVY}{
\begin{figure*}[tb]
\begin{subfigure}[c]{0.45\textwidth}

\includegraphics[width=\maxwidth]{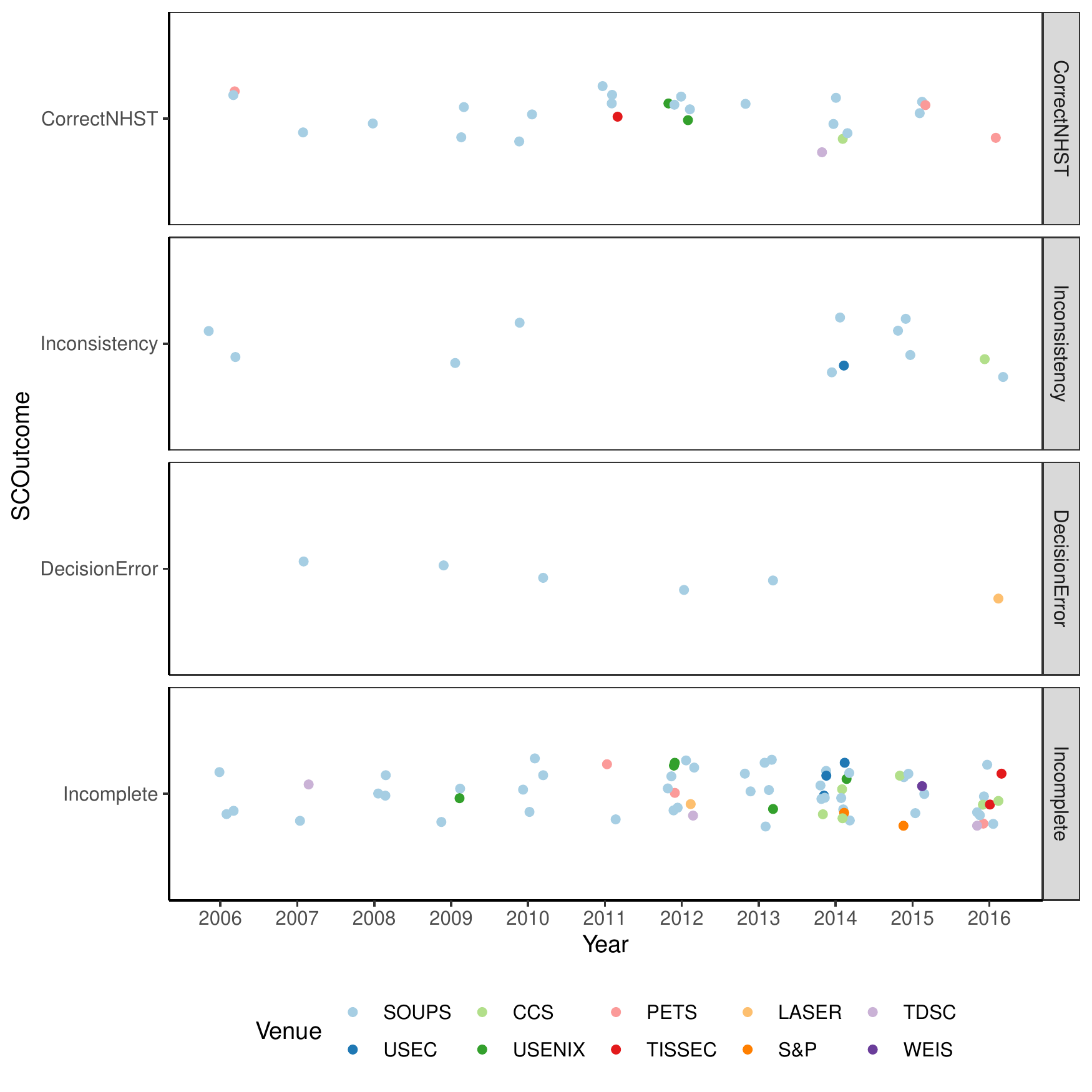} 
\caption{Scatter plot}
\label{fig:scatterplotVY}
\end{subfigure}~
\begin{subfigure}[c]{0.45\textwidth}

\includegraphics[width=\maxwidth]{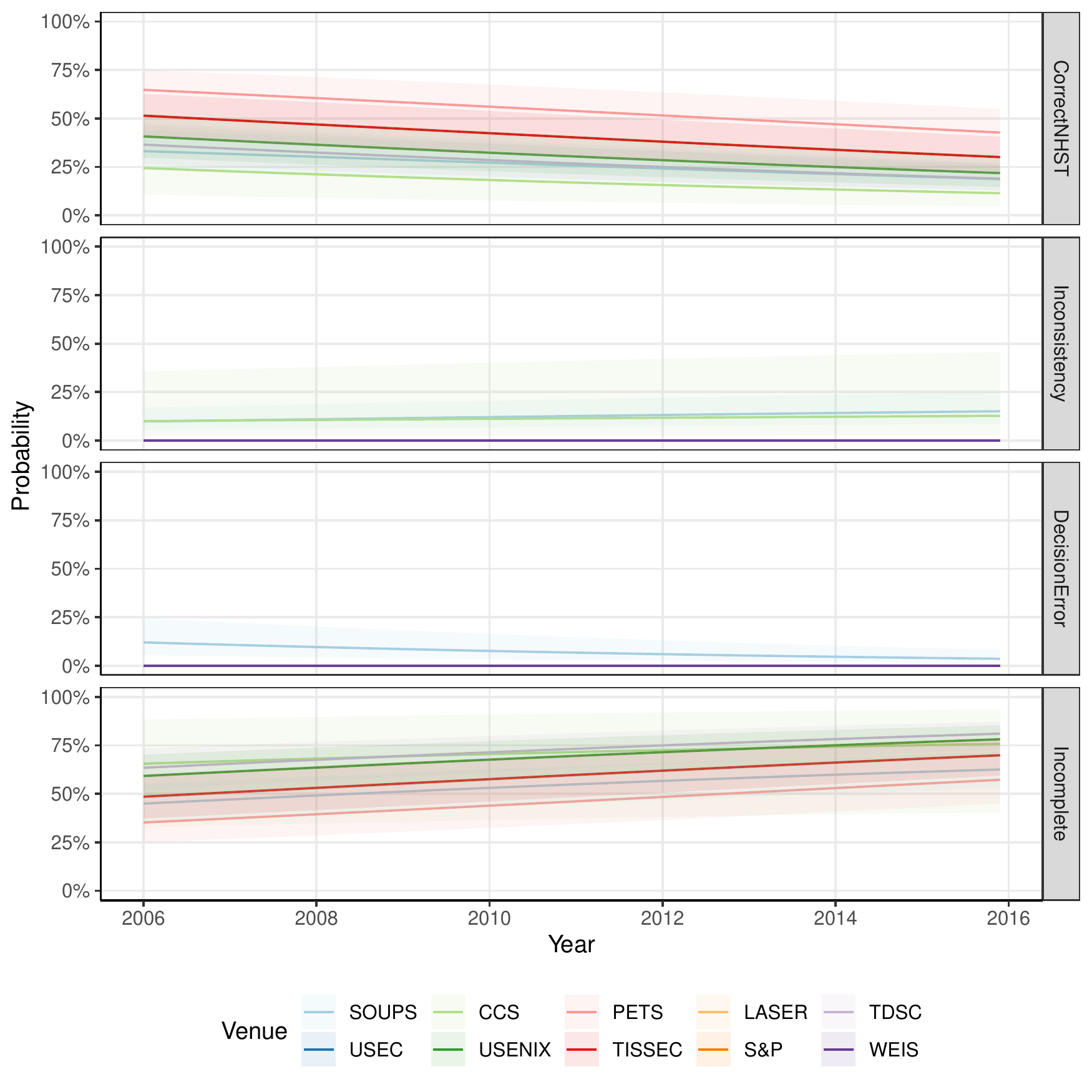} 
\caption{MLR Probabilities with 95\% confidence bands}
\label{fig:mlrplotVY}
\end{subfigure}
\caption{Per-paper aggregated outcomes of \textsf{statcheck} by \textsf{venue} and \textsf{year}. \emph{Note:} The overall multinomial logistic regression (MLR) is not statistically significant, LR Test, $\chi^2(30) = 23.089, p = .812$.}
\label{fig:combinedplotVY}
\end{figure*}
}

\newcommand{\combinedplotYear}{
\begin{figure*}[tb]
\begin{subfigure}[c]{0.45\textwidth}

\includegraphics[width=\maxwidth]{figure/combined_plot_year-1} 
\caption{Scatter plot}
\label{fig:scatterplotYear}
\end{subfigure}~
\begin{subfigure}[c]{0.45\textwidth}

\includegraphics[width=\maxwidth]{figure/combined_plot_year-2} 
\caption{MLR Probabilities with 95\% confidence bands}
\label{fig:mlrplotYear}
\end{subfigure}
\caption{Per-paper aggregated outcomes of \textsf{statcheck} by \textsf{year}. \emph{Note:} The overall multinomial logistic regression (MLR) is not statistically significant, LR Test, $\chi^2(3) = 3.331, p = .343$.}
\label{fig:combinedplotYear}
\end{figure*}
}

\newcommand{\combinedplotTestsVYso}{
\begin{figure*}[tb]
\subfloat[Scatter plot]{
\begin{minipage}{0.49\textwidth}
\vspace{-0.7cm}

\includegraphics[width=\maxwidth]{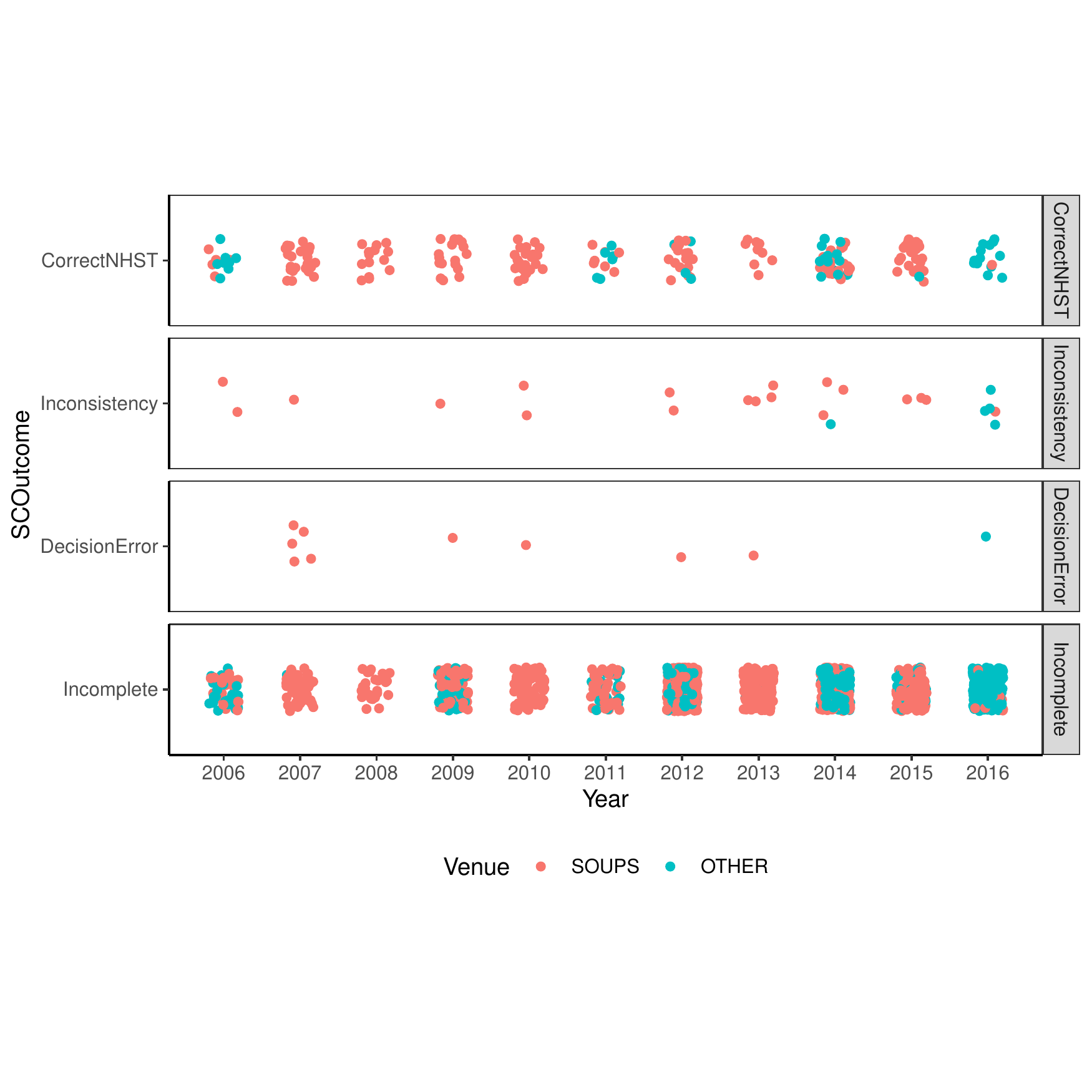} 
\vspace{-1cm}
\label{fig:scatterplotVYso}
\end{minipage}
}~
\subfloat[MLR probabilities]{
\begin{minipage}{0.49\textwidth}
\vspace{-0.7cm}

\includegraphics[width=\maxwidth]{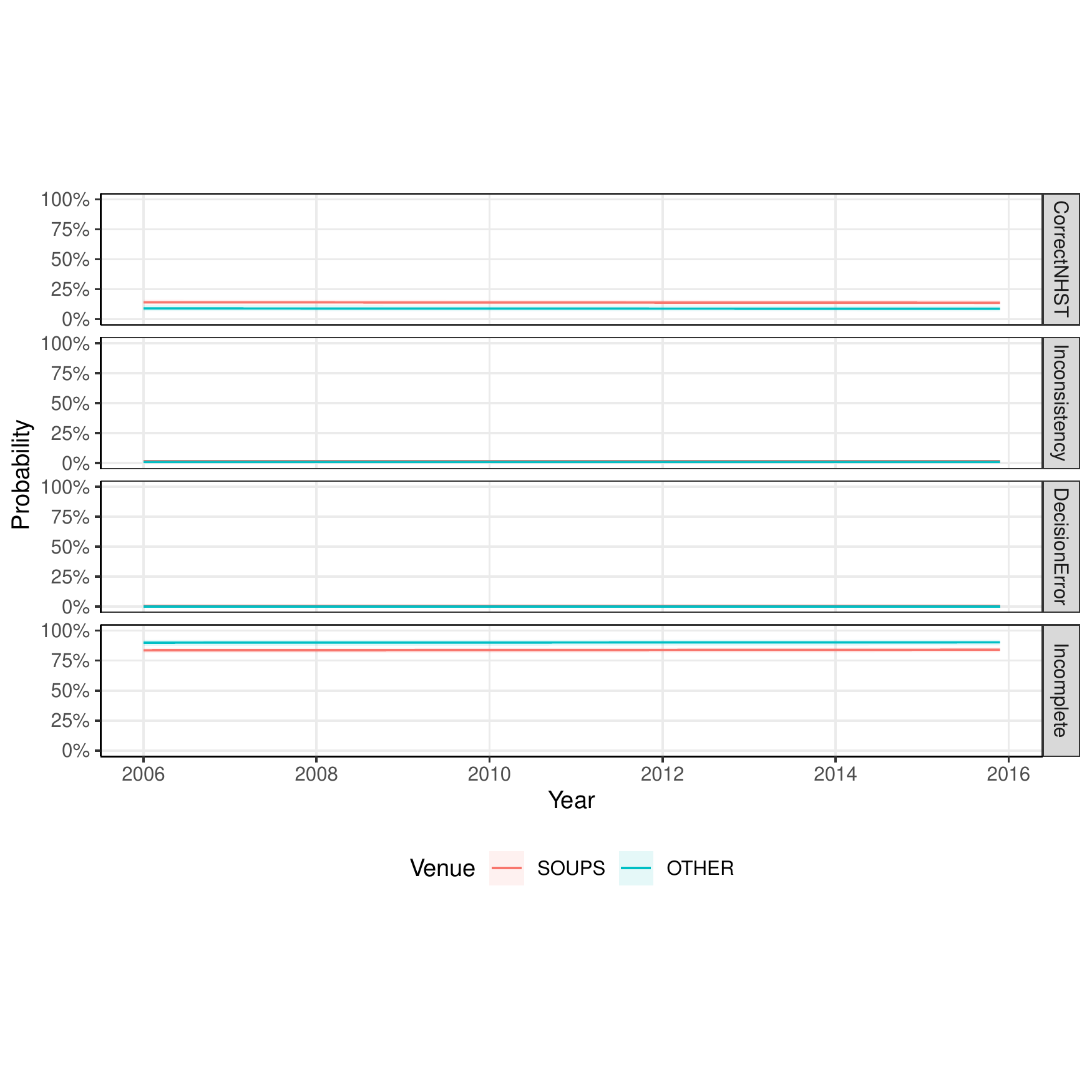} 
\vspace{-1cm}
\label{fig:mlrplotVYso}
\end{minipage}
}
\caption{Per-test\textsf{statcheck} outcomes by \textsf{venue} and \textsf{year}. \emph{Note:} The multinomial logistic regression (MLR) is statistically significant, LR Test, $\chi^2(6) = 15.417, p = .017$.}
\label{fig:combinedplotTestsVYso}
\end{figure*}
}

\newcommand{\coefTabMLRanasoCorrectNHST}{
\begin{table}[tbp]
\centering
\caption{MLR coefficients for CorrectNHST.} 
\label{tab:coefTabMLRanasoCorrectNHST}
\begingroup\footnotesize
\begin{tabular}{rrrlrrrr}
  \toprule
 & b & SE & $z$-value & $p$-Value & OR & LL & UL \\ 
  \midrule
(Intercept) & 4.805 & 0.000 & 31592.632 & $<.001$*** & 122.086 & 122.049 & 122.122 \\ 
  Year & -0.003 & 0.000 & -79.882 & $<.001$*** & 0.997 & 0.997 & 0.997 \\ 
  VenueOTHER & -0.526 & 0.172 & -3.056 & \phantom{<}$.001$** & 0.591 & 0.422 & 0.828 \\ 
   \bottomrule
\end{tabular}
\endgroup
\end{table}
}
\newcommand{\coefTabMLRanasoInconsistency}{
\begin{table}[tbp]
\centering
\caption{MLR coefficients for Inconsistency.} 
\label{tab:coefTabMLRanasoInconsistency}
\begingroup\footnotesize
\begin{tabular}{rrrlrrrr}
  \toprule
 & b & SE & $z$-value & $p$-Value & OR & LL & UL \\ 
  \midrule
(Intercept) & -0.349 & 0.000 & -176649.306 & $<.001$*** & 0.705 & 0.705 & 0.705 \\ 
  Year & -0.002 & 0.000 & -17.656 & $<.001$*** & 0.998 & 0.998 & 0.998 \\ 
  VenueOTHER & -0.633 & 0.002 & -284.961 & $<.001$*** & 0.531 & 0.529 & 0.534 \\ 
   \bottomrule
\end{tabular}
\endgroup
\end{table}
}
\newcommand{\coefTabMLRanasoDecisionError}{
\begin{table}[tbp]
\centering
\caption{MLR coefficients for DecisionError.} 
\label{tab:coefTabMLRanasoDecisionError}
\begingroup\footnotesize
\begin{tabular}{rrrlrrrr}
  \toprule
 & b & SE & $z$-value & $p$-Value & OR & LL & UL \\ 
  \midrule
(Intercept) & 0.633 & 0.000 & 1477755.342 & $<.001$*** & 1.883 & 1.883 & 1.883 \\ 
  Year & -0.003 & 0.000 & -16.893 & $<.001$*** & 0.997 & 0.997 & 0.998 \\ 
  VenueOTHER & -1.464 & 0.000 & -3144.058 & $<.001$*** & 0.231 & 0.231 & 0.231 \\ 
   \bottomrule
\end{tabular}
\endgroup
\end{table}
}

\newcommand{\combinedplotTestsVY}{
\begin{figure*}[tb]
\subfloat[Scatter plot]{
\begin{minipage}{0.49\textwidth}
\vspace{-0.7cm}

\includegraphics[width=\maxwidth]{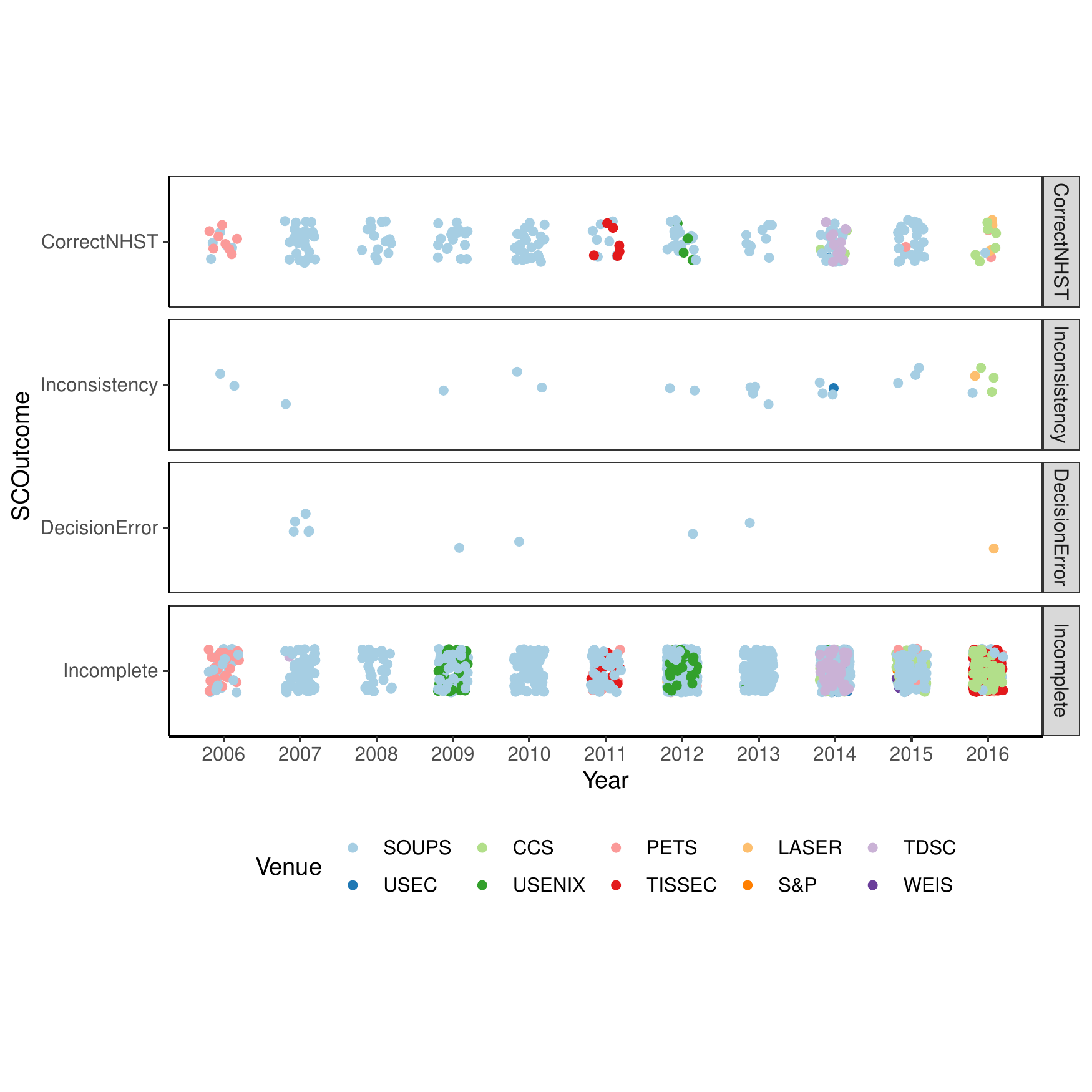} 
\vspace{-0.7cm}
\label{fig:scatterplotVY}
\end{minipage}
}~
\subfloat[MLR Probabilities]{
\begin{minipage}{0.49\textwidth}
\vspace{-0.7cm}

\includegraphics[width=\maxwidth]{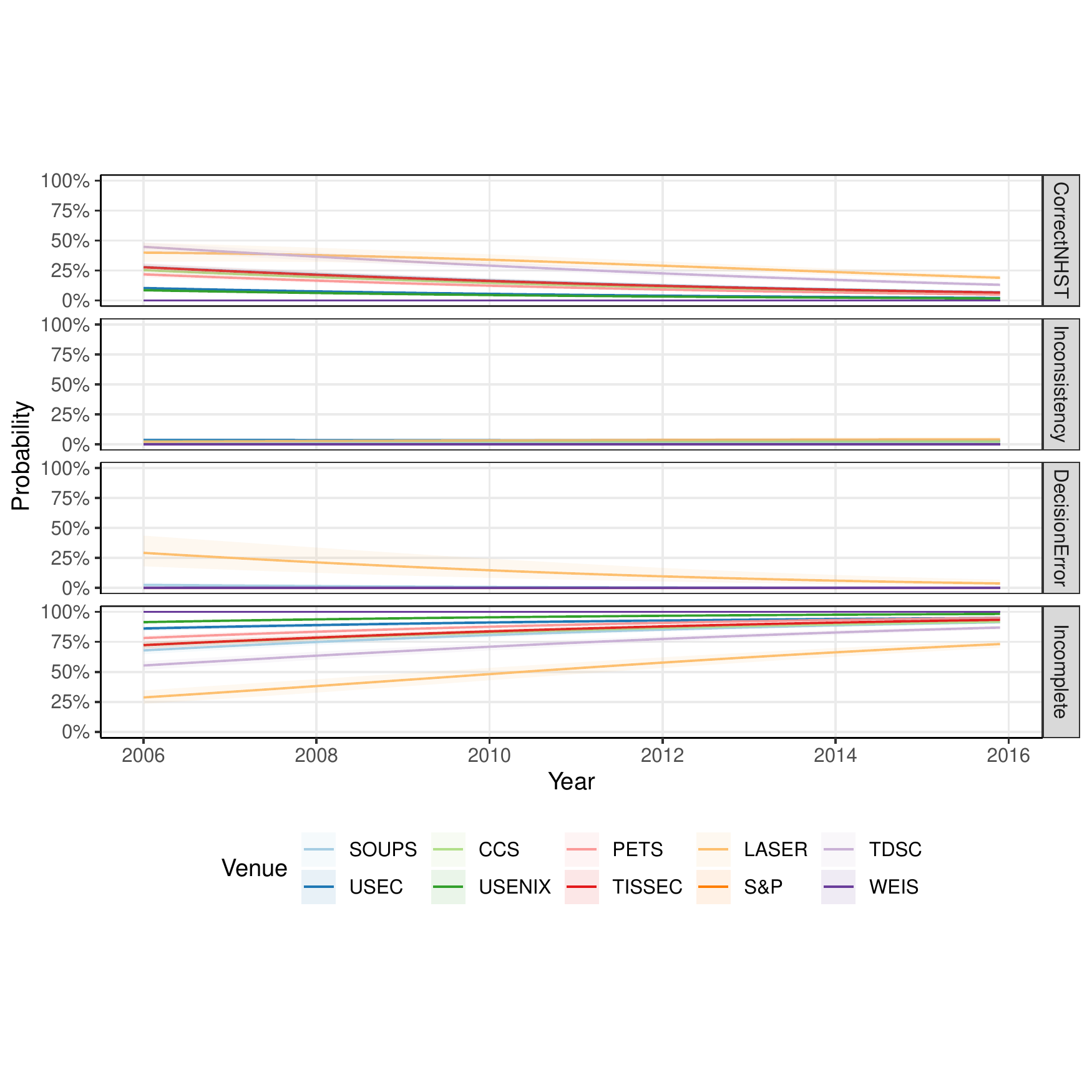} 
\vspace{-0.7cm}
\label{fig:mlrplotVY}
\end{minipage}
}
\caption{Per-test\textsf{statcheck} outcomes by \textsf{venue} and \textsf{year}. \emph{Note:} The multinomial logistic regression (MLR) is statistically significant, LR Test, $\chi^2(30) = 90.713, p < .001$.}
\label{fig:combinedplotTestsVY}
\end{figure*}
}

\newcommand{\combinedplotTestsVYwoUnp}{
\begin{figure*}[tb]
\begin{subfigure}[c]{0.45\textwidth}

\includegraphics[width=\maxwidth]{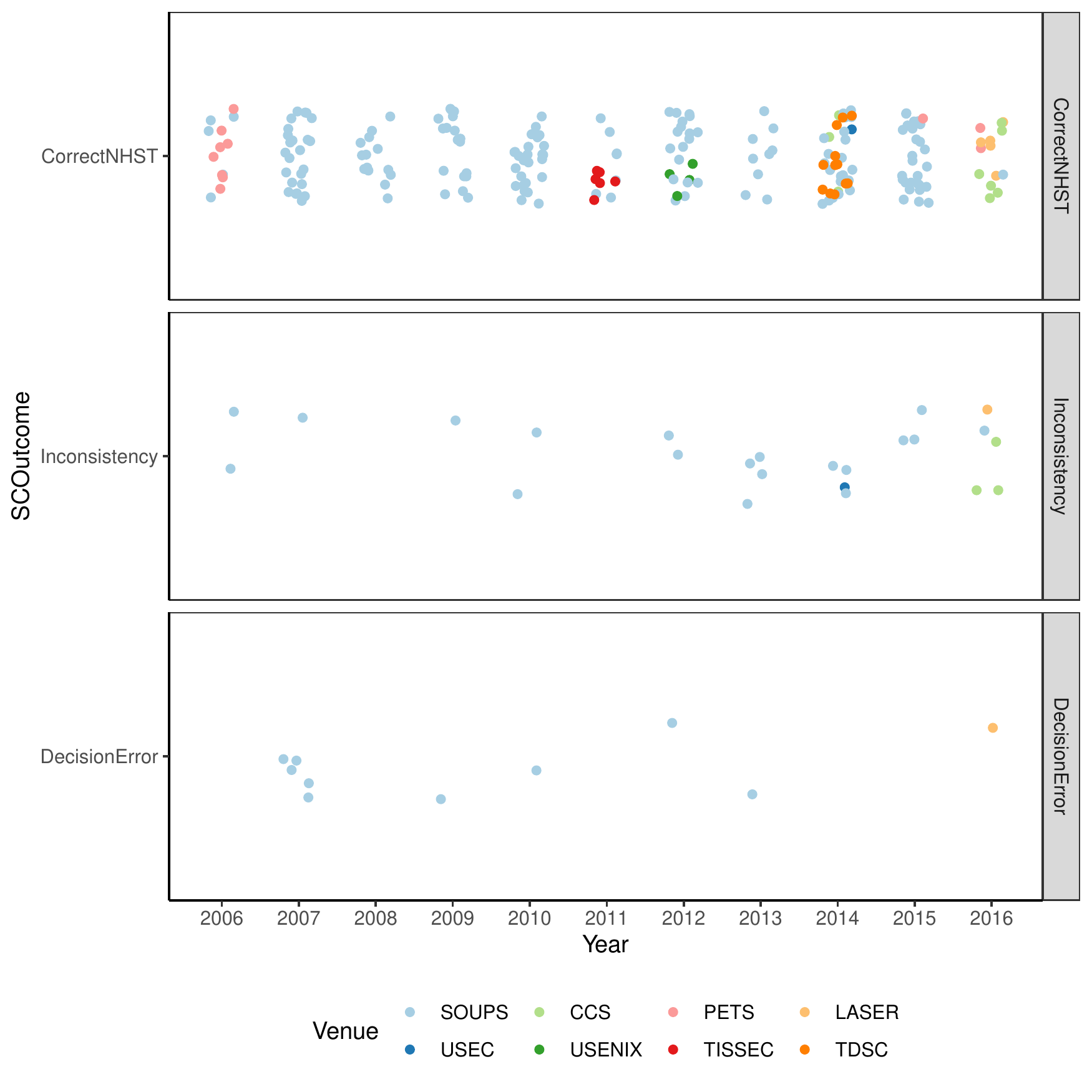} 
\caption{Scatter plot}
\label{fig:scatterplotVY}
\end{subfigure}~
\begin{subfigure}[c]{0.45\textwidth}

\includegraphics[width=\maxwidth]{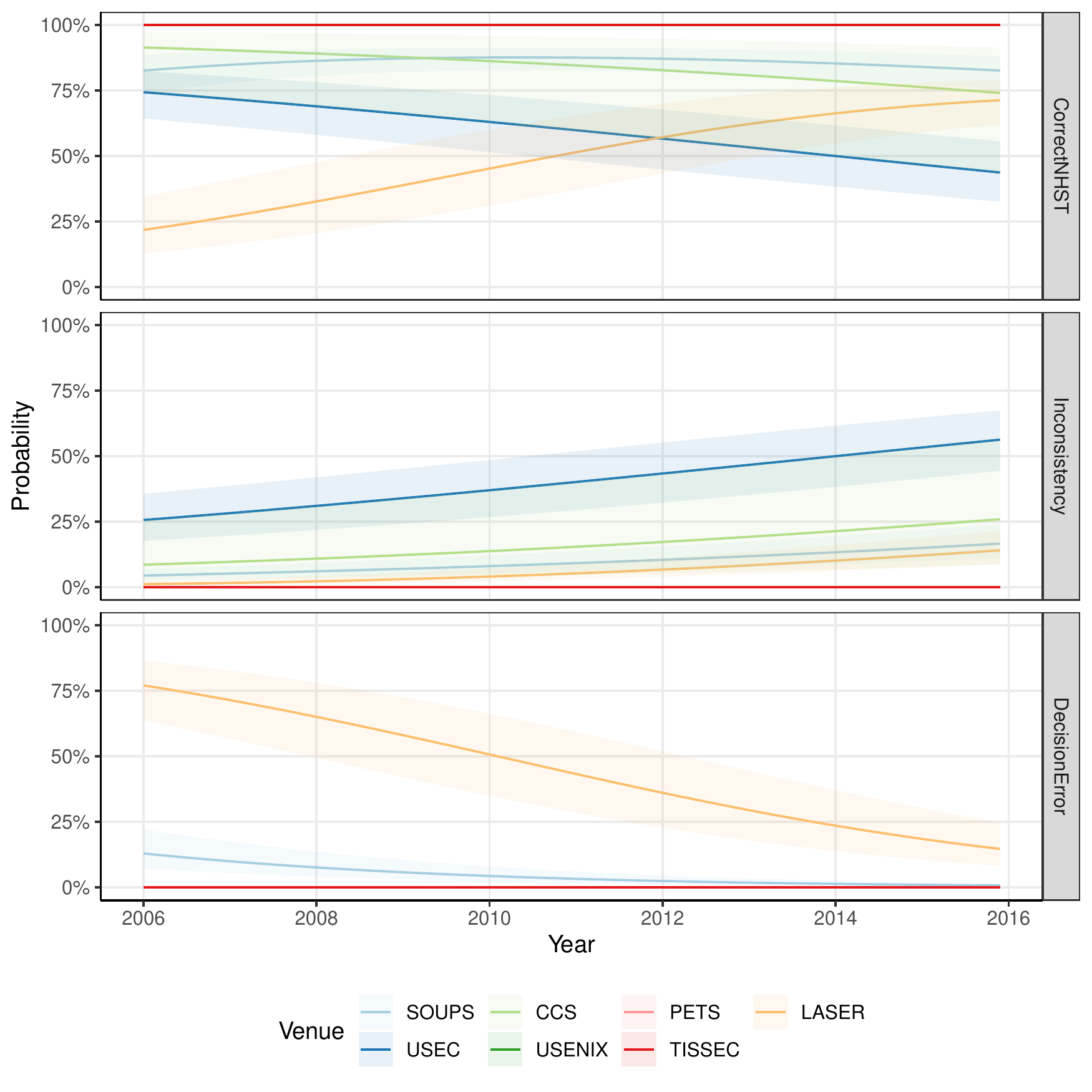} 
\caption{MLR Probabilities with 95\% confidence bands}
\label{fig:mlrplotVY}
\end{subfigure}
\caption{Per-test\textsf{statcheck} outcomes by \textsf{venue} and \textsf{year}. \emph{Note:} The overall multinomial logistic regression (MLR) is not statistically significant, LR Test, $\chi^2(16) = 24.491, p = .079$.}
\label{fig:combinedplotTestsVYwoUnp}
\end{figure*}
}

\newcommand{\combinedplotTestsVYall}{
\begin{figure}[tb]
\renewcommand{\thesubfigure}{a}\subfloat[All tests analyzed, incl. with incomplete test statistics, MLR significant, $\chi^2(30) = 90.713, p < .001$]{
\begin{minipage}{\linewidth}
\vspace{-0.7cm}
\subfloat{
\begin{minipage}{0.49\linewidth}

\includegraphics[width=\maxwidth]{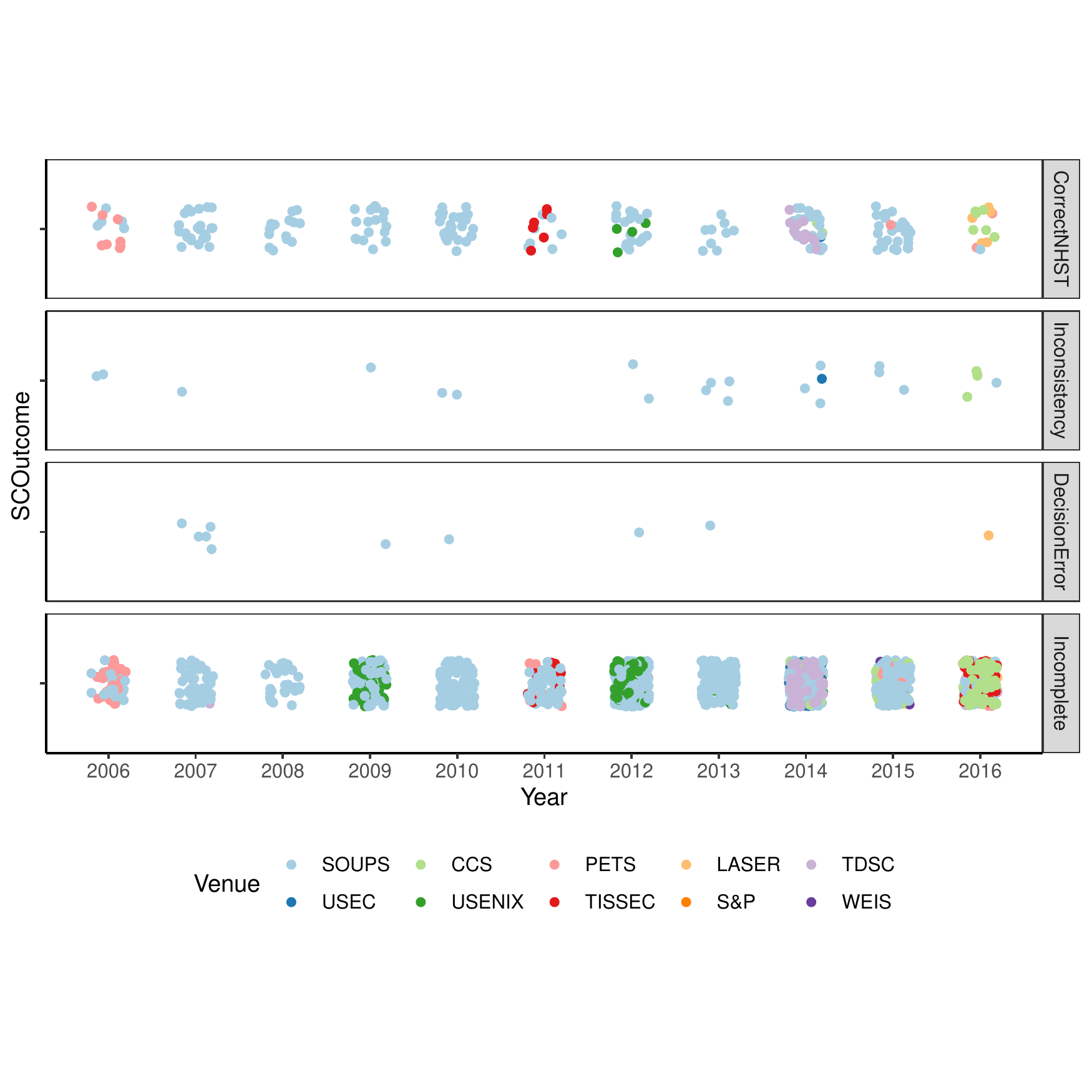} 
\end{minipage}
}~
\subfloat{
\begin{minipage}{0.49\linewidth}
\includegraphics[width=\maxwidth]{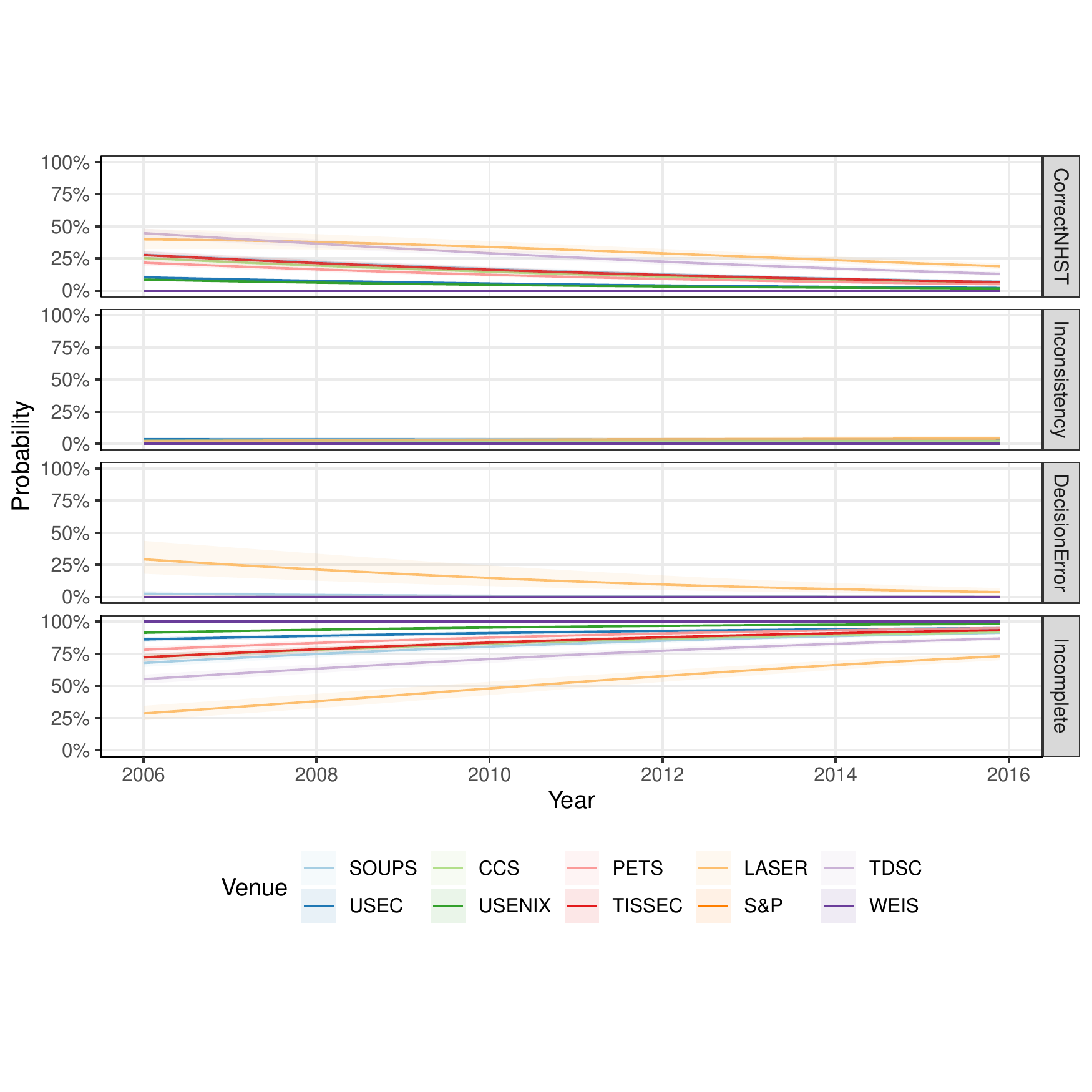} 
\end{minipage}
}
\label{fig:mlrplotVYtests}
\end{minipage}
}

\renewcommand{\thesubfigure}{b}\subfloat[Tests analyzed, excluding Incomplete test statistics, MLR not significant, $\chi^2(16) = 24.491, p = .079$]{
\begin{minipage}{\linewidth}
\vspace{-0.7cm}
\subfloat{
\begin{minipage}{0.49\linewidth}

\includegraphics[width=\maxwidth]{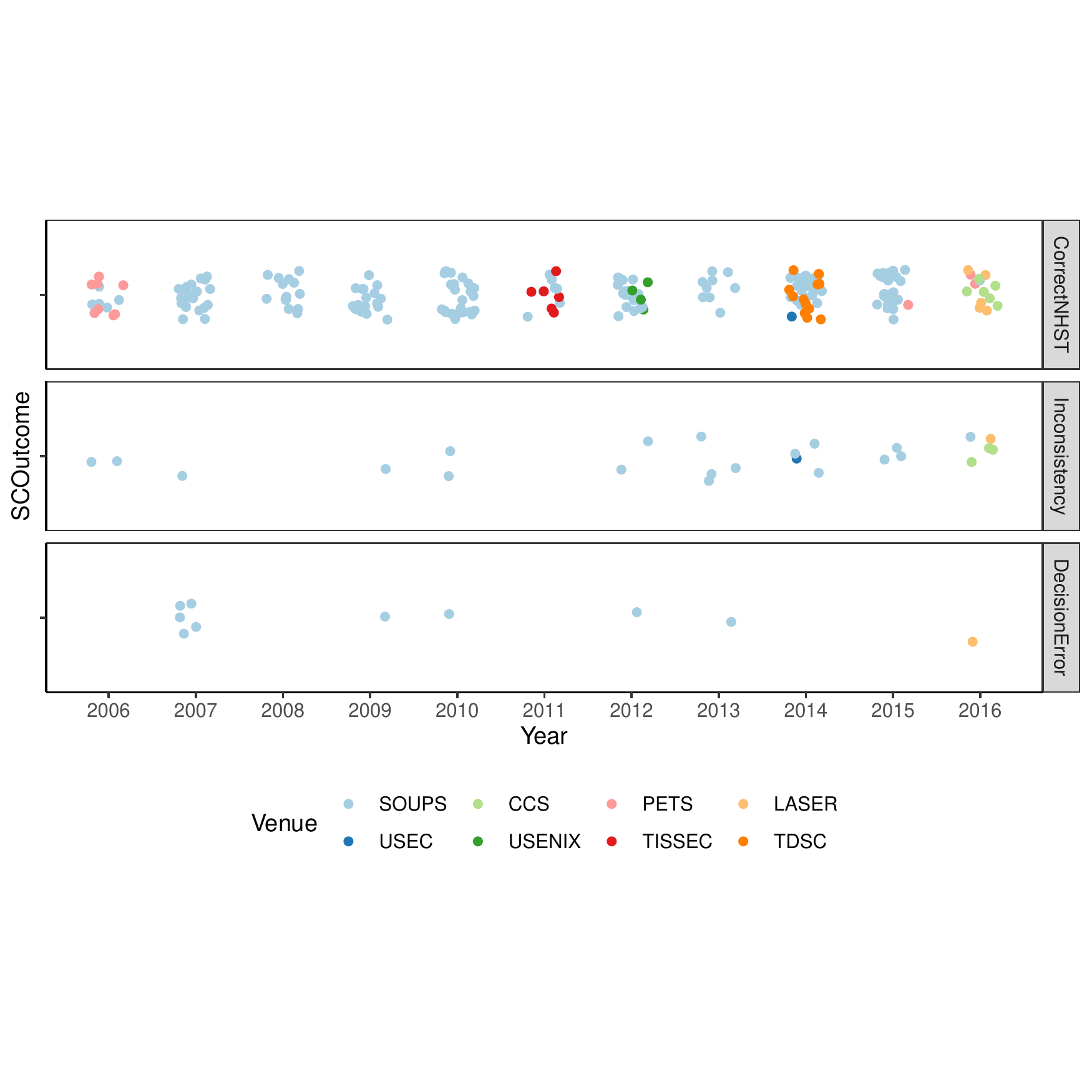} 
\end{minipage}
}~
\subfloat{
\begin{minipage}{0.49\linewidth}
\includegraphics[width=\maxwidth]{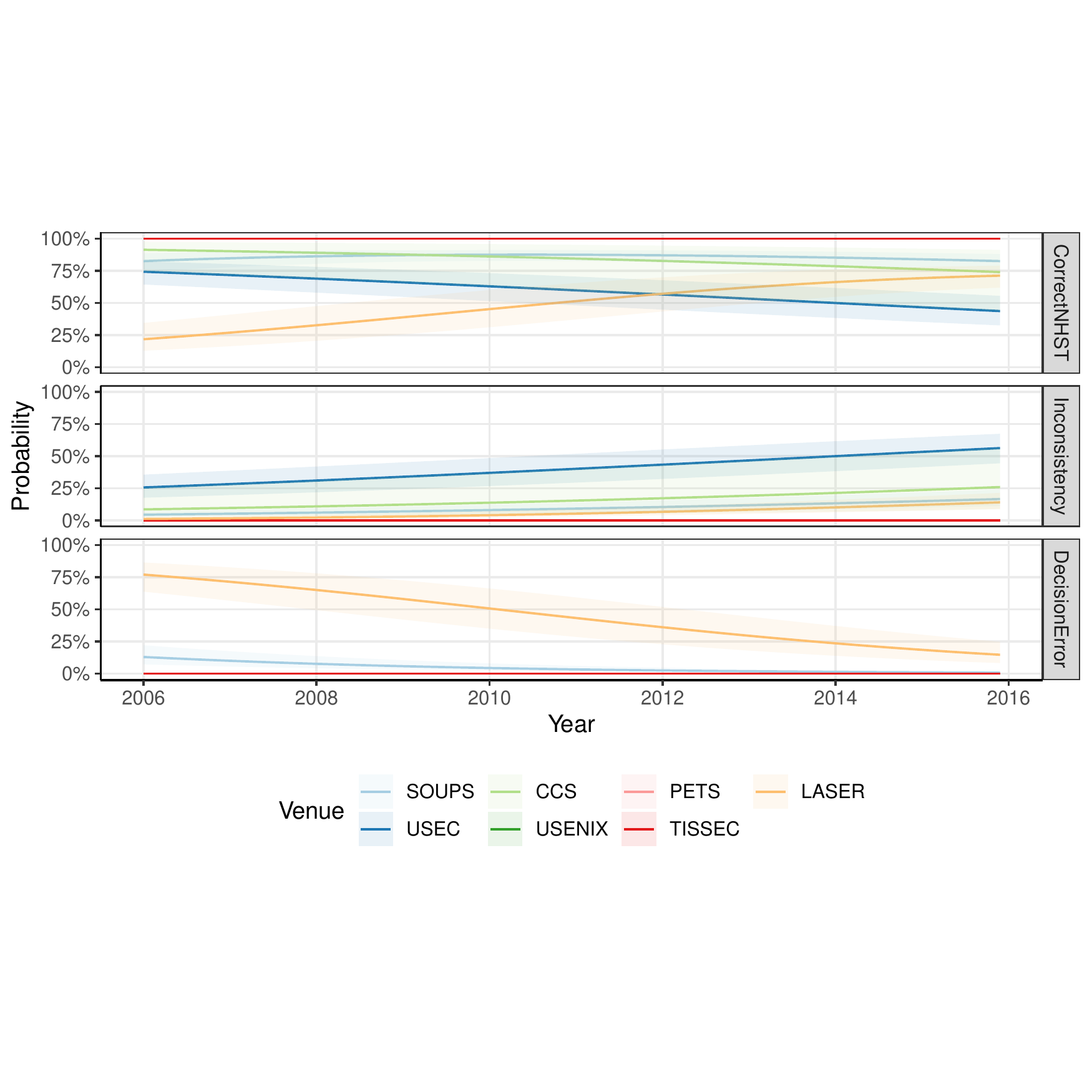} 
\end{minipage}
}
\label{fig:mlrplotVYtestsWoUnp}
\end{minipage}
}

\renewcommand{\thesubfigure}{c}\subfloat[Results aggregated for papers, MLR not significant, $\chi^2(3) = 3.331, p = .343$]{
\begin{minipage}{\linewidth}
\vspace{-0.7cm}
\subfloat{
\begin{minipage}{0.49\linewidth}

\includegraphics[width=\maxwidth]{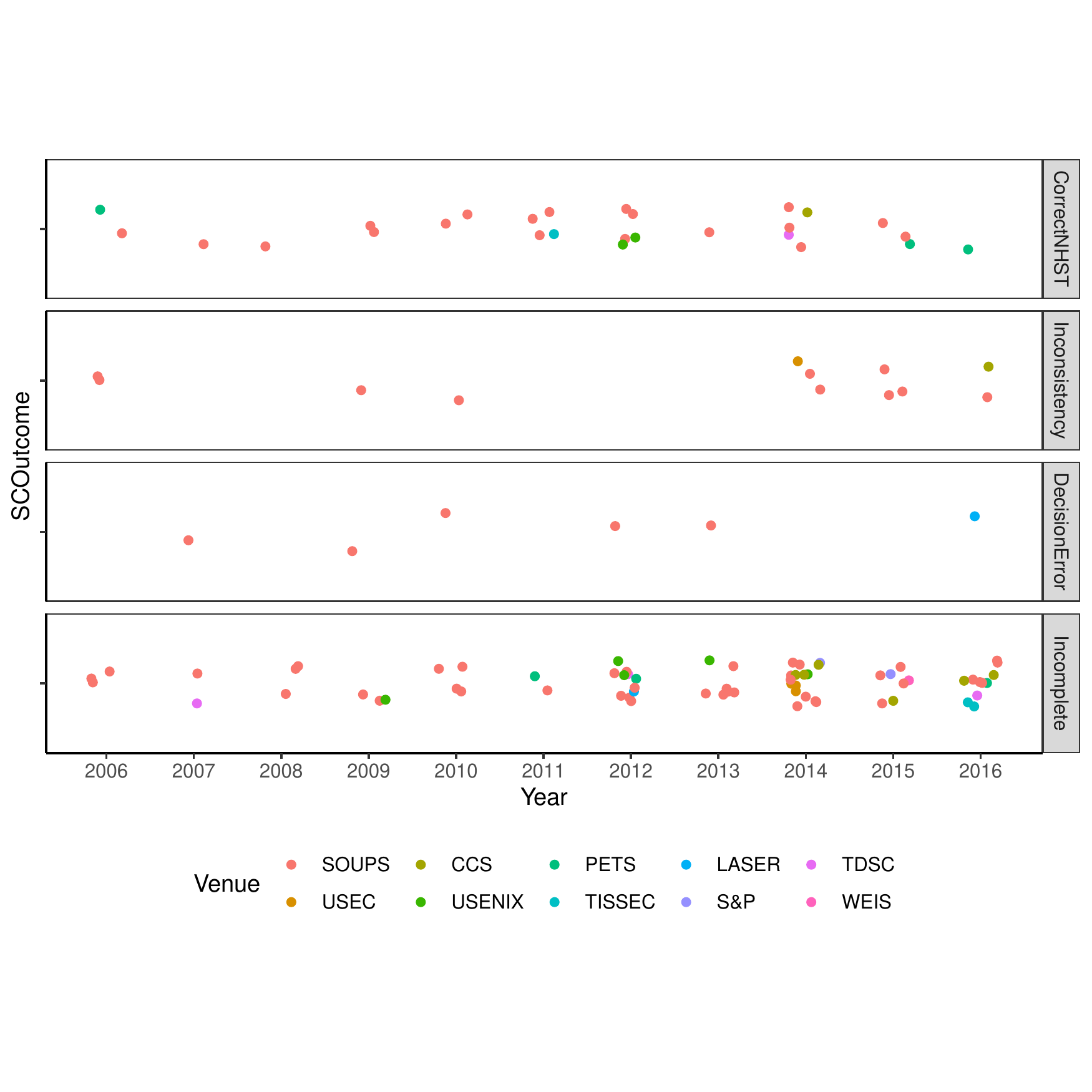} 
\end{minipage}
}~
\subfloat{
\begin{minipage}{0.49\linewidth}
\includegraphics[width=\maxwidth]{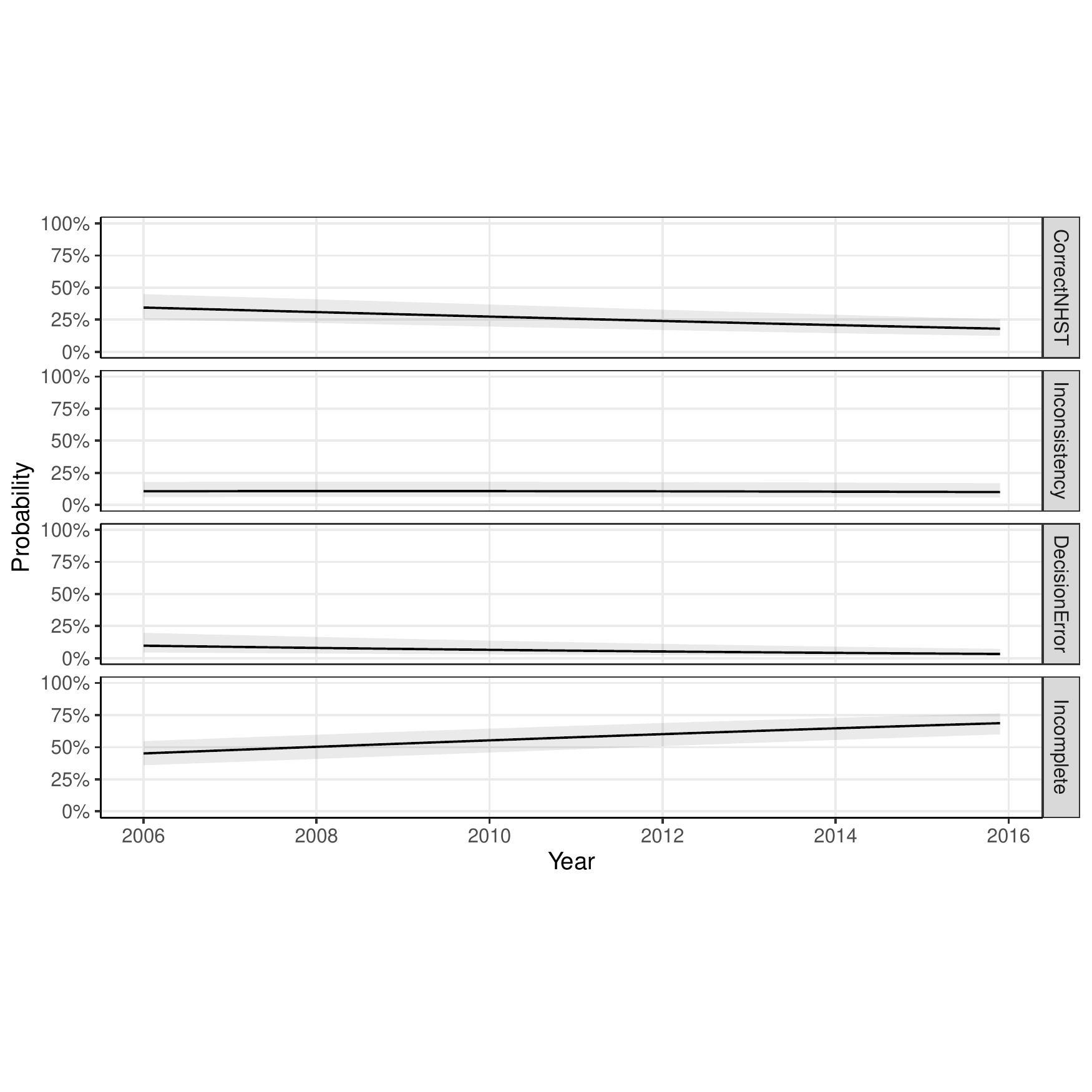} 
\vspace{0.6cm}\end{minipage}
}
\label{fig:mlrplotYear}
\end{minipage}
}
\caption{Comparison of scatter plots and corresponding multinomial logistic regression (MLR, 95\% confidence bands) of \textsf{statcheck} results via \textsf{polytomous\_effects}~\cite{fox2006effect}. We consider (\ref{fig:mlrplotVYtests}) reported tests, (\ref{fig:mlrplotVYtestsWoUnp}) complete test statistics only, and (\ref{fig:mlrplotYear}) aggregates for papers; (\ref{fig:mlrplotVYtests}) and (\ref{fig:mlrplotVYtestsWoUnp}) by \textsf{venue} and \textsf{year}; (\ref{fig:mlrplotYear}) is by \textsf{year} only.}
\label{fig:combinedplotTestsVYall}
\end{figure}
}

\newcommand{\plotObsTypesCombinedVert}{
\begin{figure}[tb]
\centering\subfloat[\textsf{Inconsistency}/\textsf{DecisionError} classes.]{
\label{fig:plotObsTypes:errorClasses}
\begin{minipage}{0.8\textwidth}
\vspace{-2cm}
\includegraphics[width=\maxwidth]{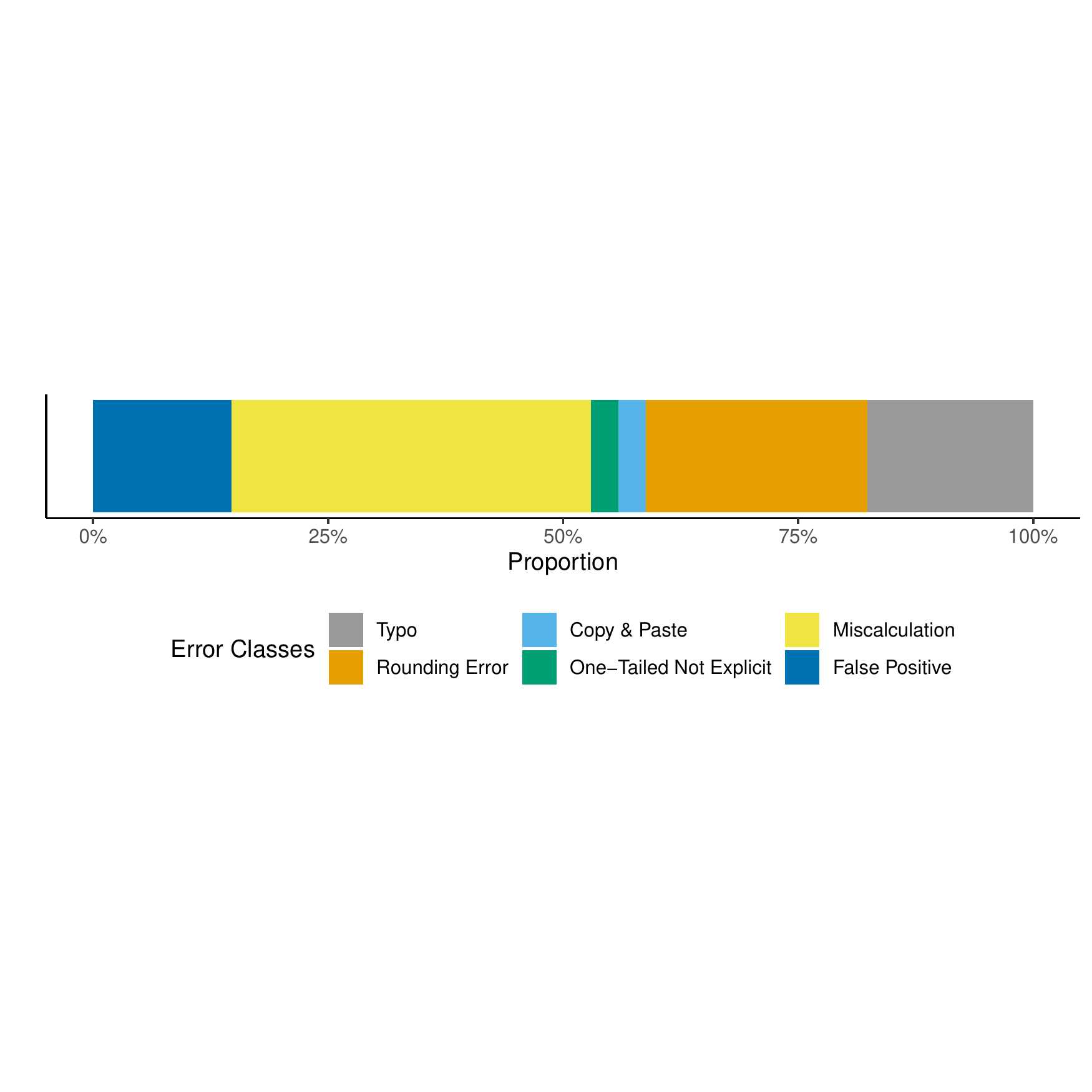} 
\vspace{-4cm}\end{minipage}
}~

\subfloat[\textsf{Incomplete} classes.]{
\label{fig:plotObsTypes:incompleteClasses}
\begin{minipage}{0.8\textwidth}
\vspace{-2cm}
\includegraphics[width=\maxwidth]{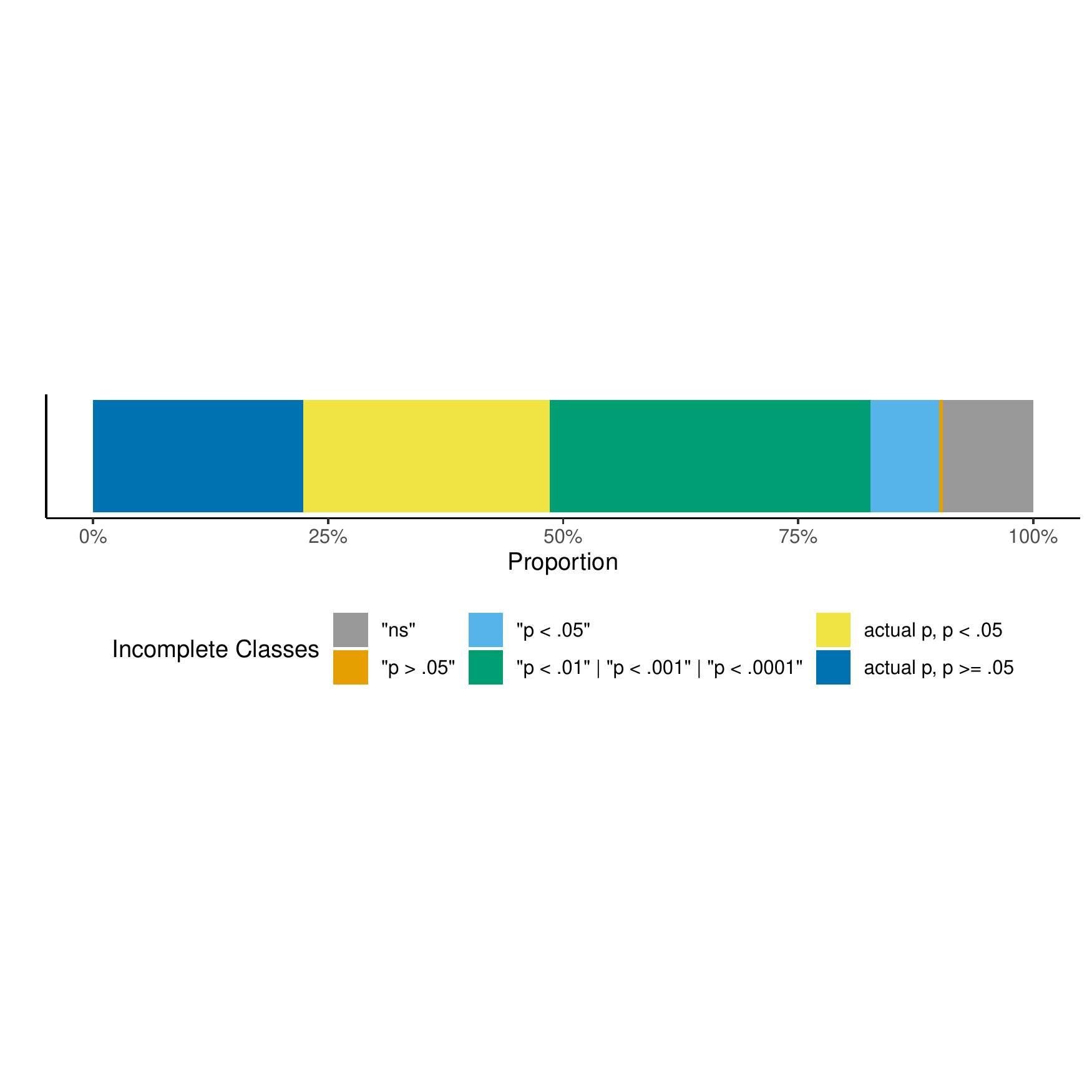} 
\vspace{-4cm}\end{minipage}
}
\caption{Classification of reported \textsf{statcheck} outcomes.}
\label{fig:plotObsTypesCombined}
\end{figure}
}

\newcommand{\plotPValuesCombined}{
\begin{figure}[tb]
\centering\subfloat[Number of reported $p$-Values]{
\label{fig:plotReportedPValues}\centering\begin{minipage}{0.49\linewidth}
\vspace{-1.5cm}

\includegraphics[width=\maxwidth]{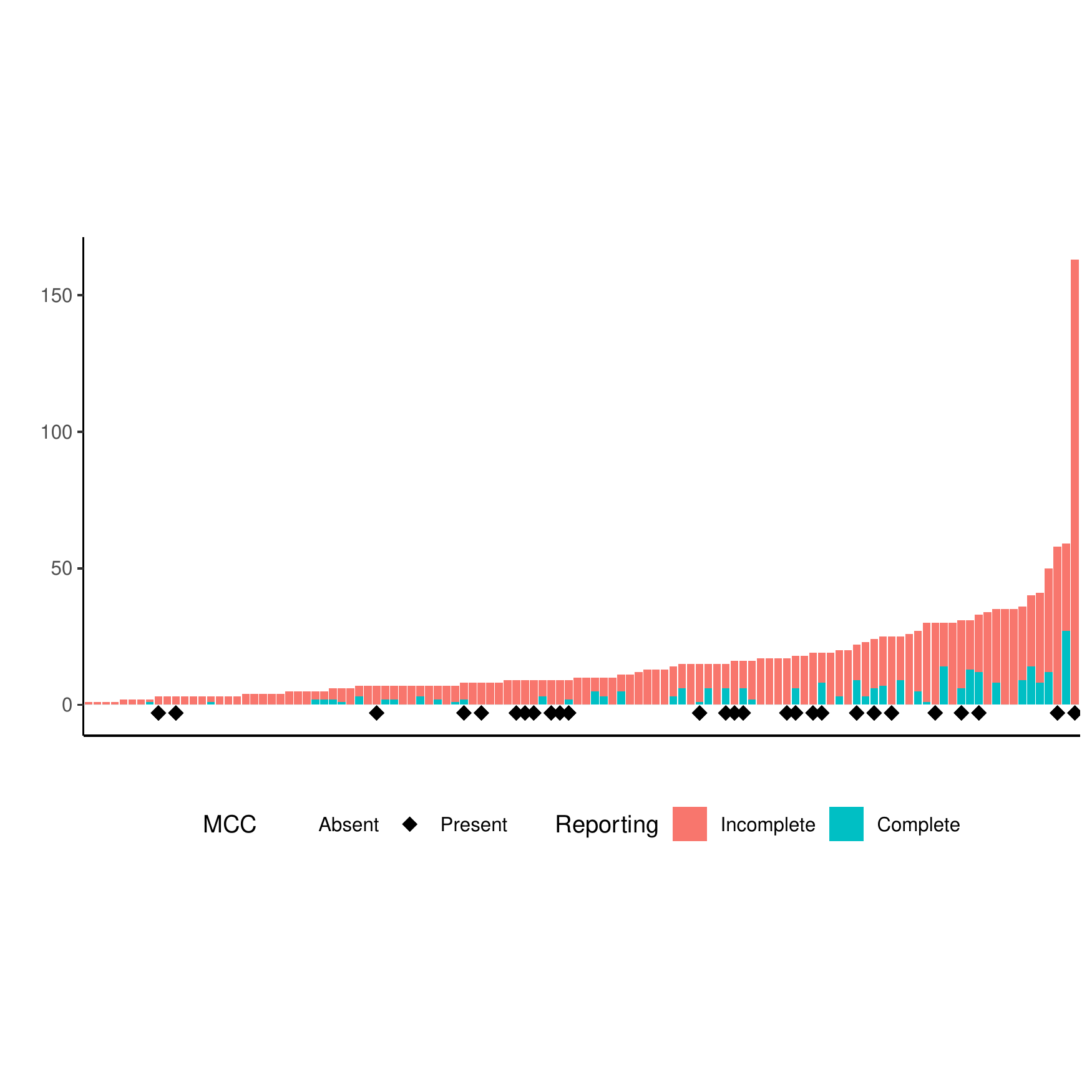} 
\vspace{-1.5cm}
\end{minipage}
}~
\subfloat[Distribution by \textsf{Year}]{
\label{fig:plotPValueByYear}\centering\begin{minipage}{0.49\textwidth}
\vspace{-1.5cm}

\includegraphics[width=\maxwidth]{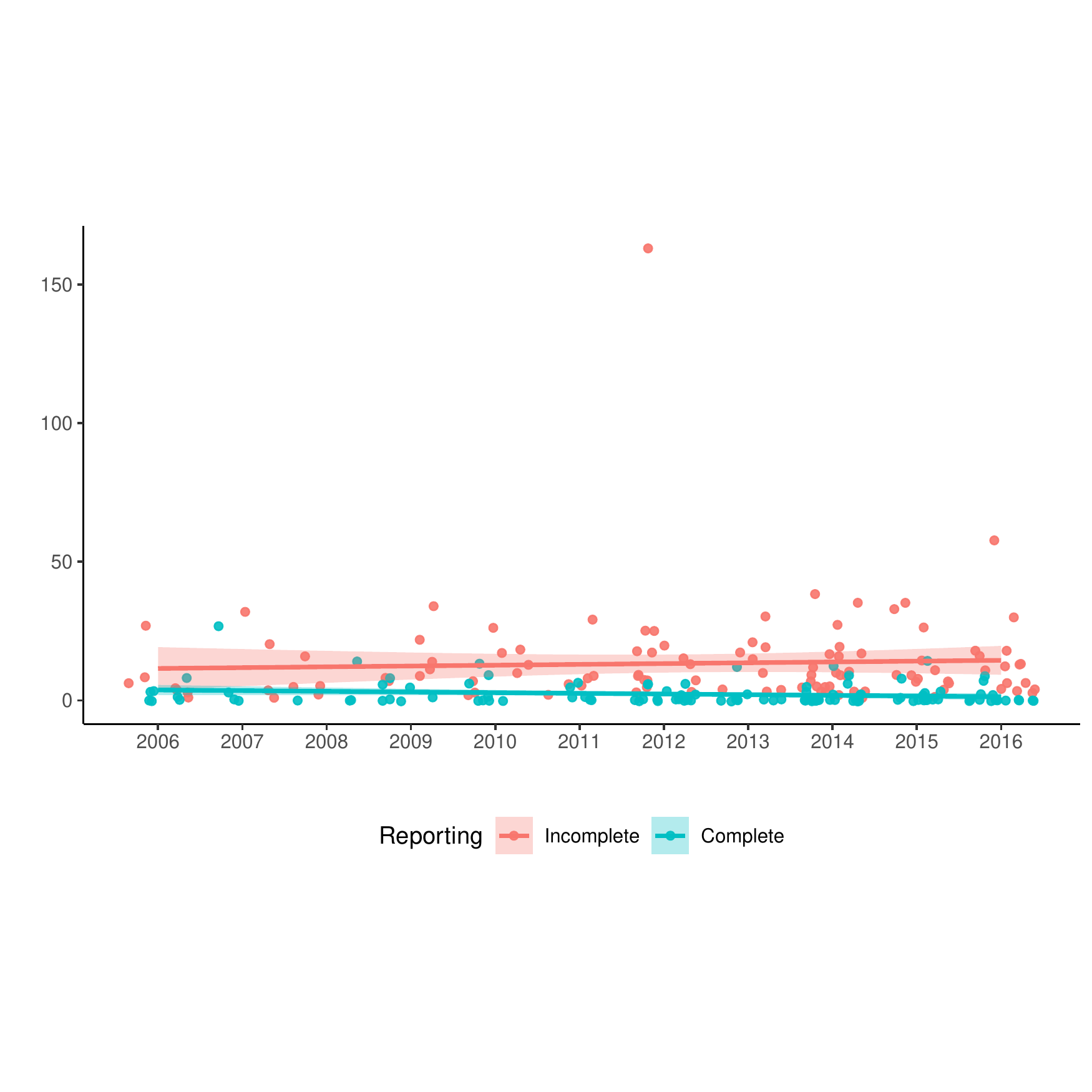} 
\vspace{-1.5cm}
\end{minipage}
}
\caption{Distribution of statistical reporting of papers, that is, how many $p$-values per paper are reported \textsf{Incomplete} or \textsf{Complete}. MCC = Multiple-Comparison Corrections.}
\label{fig:plotPValuesCombined}
\end{figure}
}
\newcommand{\plotPValuesCombinedVert}{
\begin{figure}[p]
\centering\subfloat[Number of reported $p$-Values per paper]{
\label{fig:plotReportedPValues}\centering\begin{minipage}{0.8\linewidth}
\vspace{-1.5cm}

\includegraphics[width=\maxwidth]{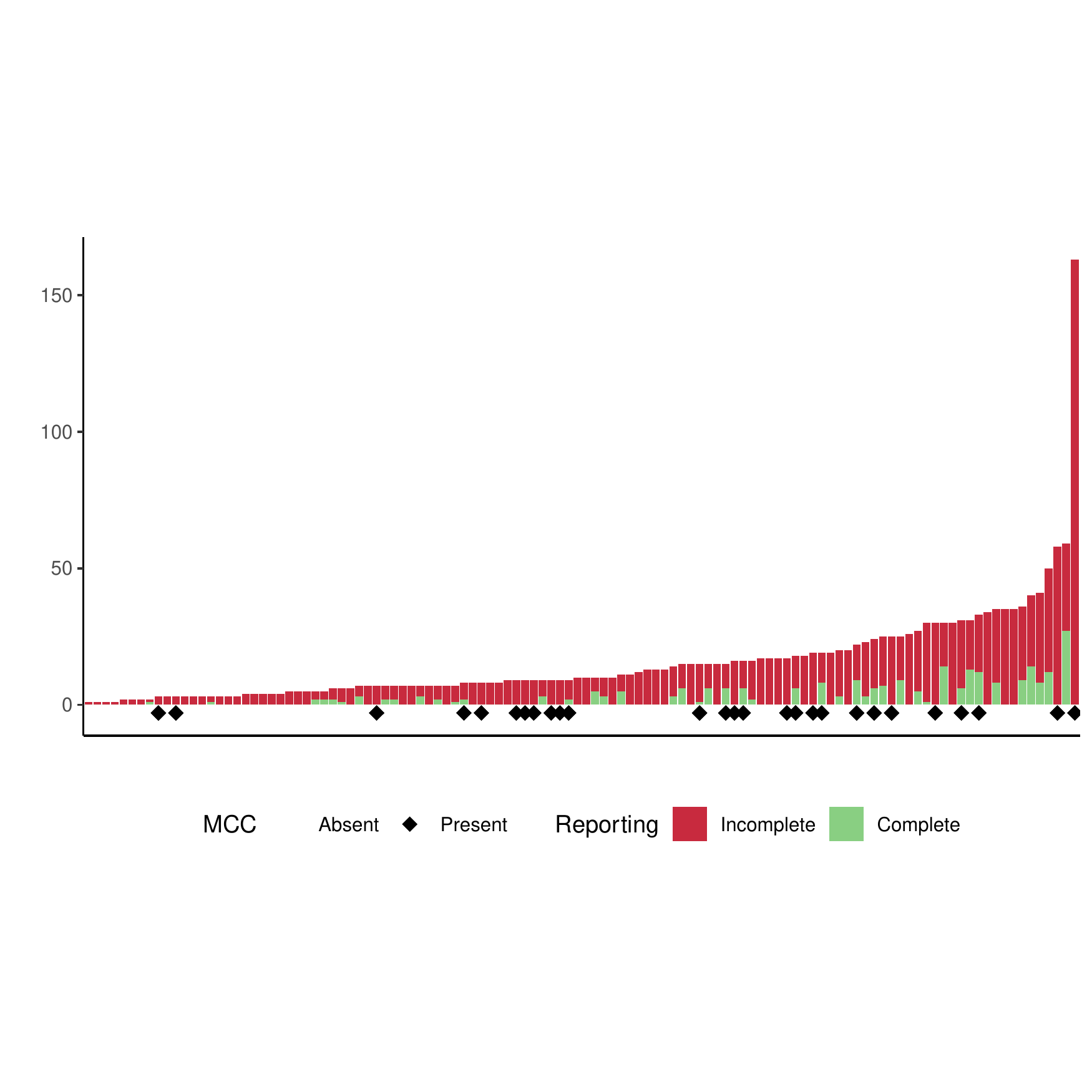} 
\vspace{-2.5cm}
\end{minipage}
}~

\subfloat[Distribution by \textsf{Year}]{
\label{fig:plotPValueByYear}\centering\begin{minipage}{0.8\textwidth}
\vspace{-1.5cm}

\includegraphics[width=\maxwidth]{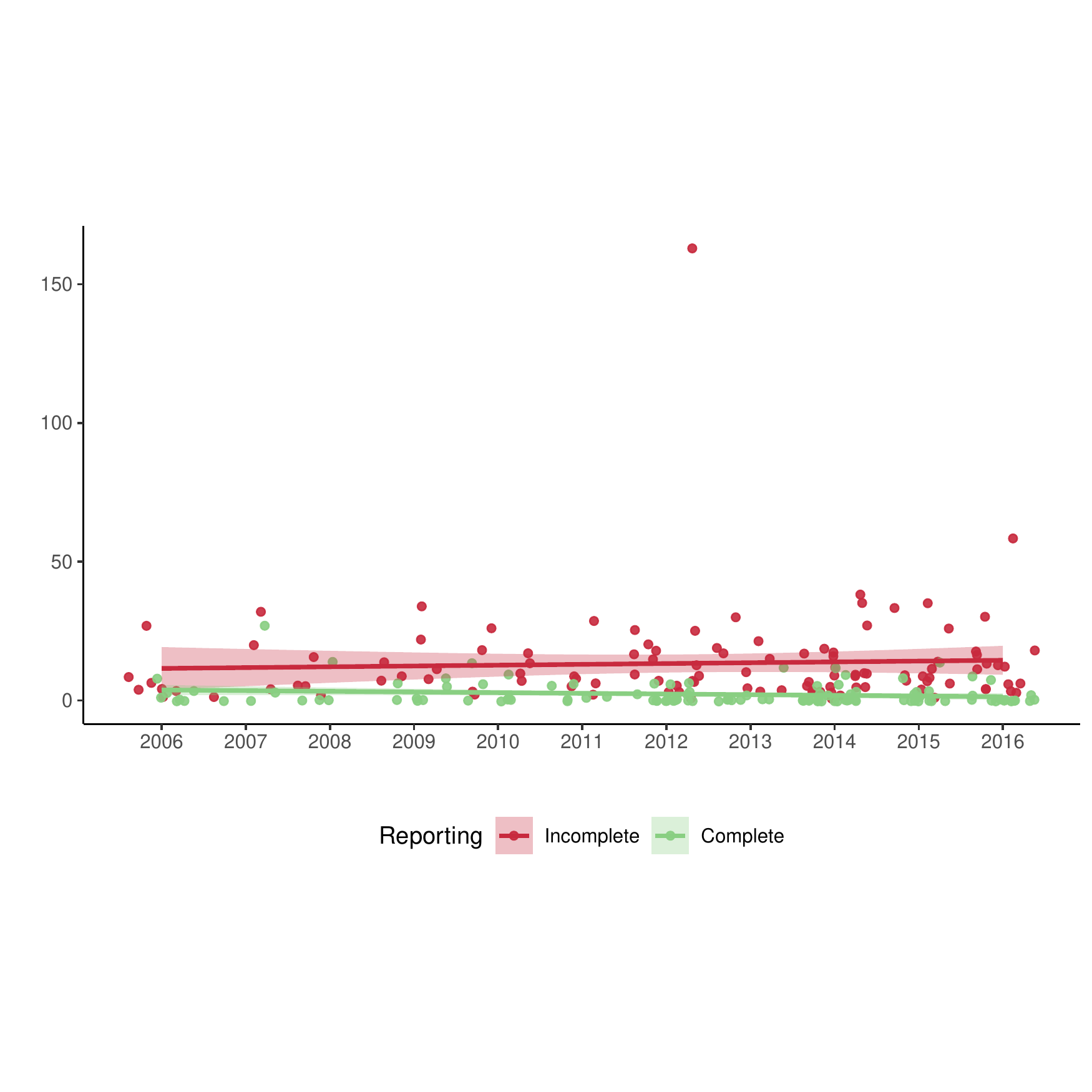} 
\vspace{-2.5cm}
\end{minipage}
}
\caption{Distribution of statistical reporting of papers, that is, how many $p$-values per paper are reported \textsf{Incomplete} or \textsf{Complete}. MCC = Multiple-Comparison Corrections.}
\label{fig:plotPValuesCombined}
\end{figure}
}

\newcommand{\confusionMatrixTests}{
\begin{table}[tbp]
\centering
\caption{Confusion matrix for \textsf{statcheck} evaluating tests.} 
\label{tab:confusionMatrixTests}
\begingroup\footnotesize
\begin{tabular}{rcc}
  \toprule
Predicted & \multicolumn{2}{c}{Reference}\\
 \cmidrule{2-3}
 & Positive & Negative \\ 
 Positive & 29 & 5 \\ 
  Negative & 0 & 218 \\ 
   \bottomrule
 \multicolumn{3}{c}{Accuracy: $.98$, 95\% CI $[.95, .99]$, $\mathit{Acc} > \mathit{NIR}(.88)$, $<.001$***,}\\
 \multicolumn{3}{c}{Sensitivity = 1.00,\quad Specificity = .98,\quad PPV = .85,\quad $F_1 = .92$}\\
\end{tabular}
\endgroup
\end{table}
}
\newcommand{\confusionMatrixScTests}{
\begin{table}[tbp]
\centering
\caption{Confusion matrix for researchers determining significance.} 
\label{tab:confusionMatrixScTests}
\begingroup\footnotesize
\begin{tabular}{rcc}
  \toprule
Predicted & \multicolumn{2}{c}{Reference}\\
 \cmidrule{2-3}
 & Significant & NS \\ 
 Significant & 191 & 12 \\ 
  NS & 1 & 47 \\ 
   \bottomrule
 \multicolumn{3}{c}{Accuracy: $.95$, 95\% CI $[.91, .97]$, $\mathit{Acc} > \mathit{NIR}(.76)$, $<.001$***,}\\
 \multicolumn{3}{c}{Sensitivity = .99,\quad Specificity = .80,\quad PPV = .94,\quad $F_1 = .97$}\\
\end{tabular}
\endgroup
\end{table}
}

\date{}

\begin{DocumentVersionConference}
\title{Fidelity of Statistical Reporting in 10 Years of Cyber Security User Studies\thanks{Preregistered at the Open Science Framework: \protect\url{osf.io/549qn/}}}
\end{DocumentVersionConference}
\begin{DocumentVersionTR}
\title{Fidelity of Statistical Reporting in 10 Years of Cyber Security User Studies (Technical Report)\thanks{Preregistered at the Open Science Framework: \protect\url{osf.io/549qn/}. The short version of this paper is appearing in the Proceedings of the 9\textsuperscript{th} Workshop on Socio-Technical Aspects in Security (STAST 2019), pp. 1--24, LNCS, Springer Verlag (2020).}}
\end{DocumentVersionTR}

\begin{DocumentVersionConference}
\author{Thomas Gro{\ss}}
\institute{Newcastle University, Newcastle upon Tyne, UK}
\end{DocumentVersionConference}
\begin{DocumentVersionTR}
\author{Thomas Gro{\ss}}
\institute{Newcastle University, Newcastle upon Tyne, UK}
\end{DocumentVersionTR}

\maketitle

\begin{abstract}
\processifversion{DocumentVersionTR}{\noindent\textbf{Background.}}
Studies in socio-technical aspects of security often rely on user studies and statistical inferences on investigated relations to make their case.
They, thereby, enable practitioners and scientists alike to judge on the validity and reliability of the research undertaken.

\processifversion{DocumentVersionTR}{\noindent\textbf{Aim.}}
To ascertain this capacity, we investigated the reporting fidelity of security user studies.

\processifversion{DocumentVersionTR}{\noindent\textbf{Method.}}
Based on a systematic literature review of $114$ user studies in cyber security from selected venues in the 10 years 2006--2016, we evaluated fidelity of the reporting of $1775$ statistical inferences using the \textsf{R} package \textsf{statcheck}. We conducted a systematic classification of incomplete reporting, reporting inconsistencies and decision errors, leading to multinomial logistic regression (MLR) on the impact of publication venue/year as well as a comparison to a compatible field of psychology.

\processifversion{DocumentVersionTR}{\noindent\textbf{Results.}}
We found that half the cyber security user studies considered reported incomplete results, in stark difference to comparable results in a field of psychology. Our MLR on analysis outcomes yielded a slight increase of likelihood of incomplete tests over time, while SOUPS yielded a few percent greater likelihood to report statistics correctly than other venues.

\processifversion{DocumentVersionTR}{\noindent\textbf{Conclusions.}}
In this study, we offer the first fully quantitative analysis of the state-of-play of socio-technical studies in security. 
While we highlight the impact and prevalence of incomplete reporting, we also offer fine-grained diagnostics and recommendations on how to respond to the situation.

\end{abstract}

\section{Introduction}
Statistical inference is the predominant method to ascertain that effects observed in socio-technical aspects of security are no mere random flukes, but considered to be ``the real McCoy.''

In general, statistical inference sets out to evaluate a statistical hypothesis stated \emph{a priori}. It employs observations made in studies to establish the likelihood as extreme as or more extreme than the observations made, assuming the statistical hypothesis not to be true. This likelihood is colloquially referred to as a $p$-value.
Alternatively to Null Hypothesis Significance Testing (NHST)---and often used complementarily---studies may estimate the magnitude of effects in reality and confidence intervals thereon~\cite{cumming2013understanding}.
\processifversion{DocumentVersionTR}{Due to the inherent variability of the behavior of human subjects these methods invariably come into focus in quantitative studies of human-factor or social dimensions.}

The onus of proof is generally on the authors of a study.
There are numerous factors influencing whether a study's results can be trusted---%
\begin{inparaenum}[a)]
  \item sound research questions and hypotheses, 
  \item vetted and reliable constructs and instruments,
  \item documentation favoring reproducibility,
  \item sound experiment design, yielding internal and external validity,
  \item randomization and blinding,
  \item systematic structured and standardized reporting%
\end{inparaenum}%
---in the end, it is the outcomes of the statistical inference that often render a final verdict.

These outcomes do not only indicate whether an effect is likely present in reality or not.
They also yield what magnitude the effect is estimated at.
Thereby, they are the raw ingredient for 
\begin{inparaenum}[(i)]
  \item establishing whether an effect is practically relevant, 
  \item evaluating its potential for reuse, and 
  \item including it further quantitative research synthesis.
\end{inparaenum}

\processifversion{DocumentVersionTR}{Ultimately, one would hope that individual studies are prepared to advance the knowledge of a field with a high degree of certainty.
One would hope that a field is prepared to reinforce robust and reliable research; that it engages in self-correction if studies are missing the mark.}

While there have been a number of publications in socio-technical aspects of security offering guidance to the community to that end~\cite{peisert2007design,maxion2011making,schechter2013common,CooGro2016,CooGro2017} as well as proposals in other communities~\cite{APAGuidelines6th2009,moher2010consort,lebel2018unified}, the evidence of the state-of-play of the field has been largely anecdotal~\cite{schechter2013common} or in human-coded analysis~\cite{SLR2017}.
While this field is arguably quite young, we argue that it would benefit greatly from attention to statistical reporting, from attaining fault tolerance through reporting fidelity and from preparing for research synthesis. (cf. Section~\ref{sec:importance_reporting})

In this study, we aim at systematically evaluating the fidelity of statistical reporting in socio-technical aspects of security.
We analyze
\begin{inparaenum}[(i)]
  \item whether statistical inferences are fault-tolerant, in the sense of their internal consistency being publicly verifiable, and
   \item whether the reported $p$-values are correct.
\end{inparaenum}
Through the semi-automated empirical analysis of 114 publications in the field from 2006--2016, we offer a wealth of information including meta-aspects\processifversion{DocumentVersionTR}{such as the use of Amazon Mechanical Turk (AMT), multiple-comparison corrections, and estimation methods}.
We compare statistical reporting fidelity of this field with a related field of psychology as well as analyze the trajectory of the field, that is, the trends found over time.
We substantiate the these results with qualitative coding of errors observed to elucidate what to watch out for.

\paragraph{Contributions.} 
We are the first to subject our own field to a systematic empirical analysis of statistical reporting fidelity.
In that, we offer a well-founded introspection in the field of socio-technical aspects of security that can serve program committees and authors alike to inform their research practice.

\todo[inline]{A! Include in abstract and intro: impact on decision making in security and privacy!}

\section{Background}
\subsection{Importance and Impact of Statistical Reporting}
\label{sec:importance_reporting}

Null Hypothesis Significance Testing (NHST) establishes statistical inference by stating \emph{a priori} statistical hypotheses, which are then tested based on observations made in studies. Such statistical inference results in a $p$-value, which gives the conditional probability of finding data as extreme as or more extreme than the observations made, assuming the null hypothesis being true. Many fields combine NHST with point and interval estimation, that is, establishing an estimate of the magnitude of the effect in the population and the confidence interval thereon.

\subsubsection{Reporting Fidelity and Fault Tolerance.}
Different reporting practices yield different degrees of information and fidelity.
It goes without saying that a simple comparison with the significance level $\alpha$, e.g., by stating that $p < .05$, yields the least information and the least fidelity. Reporting the actual $p$-value observed offers more information as well as a means to quantify the likelihood of the effect.\processifversion{DocumentVersionTR}{\footnote{We note that the $p$-value itself does \emph{not} state the likelihood that a positively reported result is actually true in reality (inverse fallacy) and refer interested readers to literature on the Positive Predicted Value (PPV)~\cite{ioannidis2005most}.}}

To gain further reporting fidelity and fault tolerance, one would not only report the exact $p$-value, but also the chosen test parameters (e.g., independent-samples or one-tailed), the test statistic itself (e.g., the $t$-value) and the degrees of freedom (\vari{df}) of the test.
We, then, obtain a consistent triplet (test statistic, \vari{df}, $p$-value) along with the test parameters.
Table~\ref{tab:fidelity} exemplifies degrees of fidelity.

The upshot of a diligent reporting procedure including full triplets is that it enables cross-checks on their internal consistency and, thereby, a degree of fault tolerance. Vice versa, if only the $p$-value or a comparison with a significance level is reported, the capacity to validate inferences is impaired.

\subsubsection{Impact on Research Synthesis.}
Published studies usually do not stand on their own. To learn what relations are actually true in reality and to what degree, we commonly need to synthesize the results of multiple studies investigating the same relations. 
More mature fields (such as evidence-based medicine or psychology) engage in systematic reviews and meta analyses to that end.

For these down-stream analyses to be viable, the original studies need to contain sufficient data for subsequent meta-analyses. If the original studies omit the actual test statistics and degrees of freedom, the synthesis in meta analyses is hamstringed or rendered impossible altogether.

\begin{table}[tb]
\centering
\footnotesize
\caption{Degrees of fidelity in statistical reporting for the same two-tailed independent-samples $t$-test on a relation with a large effect size (ES).\hspace{\textwidth}
\emph{Note:} \Circle{} = impossible \qquad \RIGHTcircle{} = can be estimated \qquad  \CIRCLE{} = supported}
\label{tab:fidelity}
\footnotesize
\begin{minipage}{\linewidth}
\centering
\begin{tabular}{lcccc}
\toprule
                       & \multicolumn{2}{c}{\textbf{Incomplete Triplet}}                            & \multicolumn{2}{c}{\textbf{Complete Triplet}}\\
                       \cmidrule(lr){2-3} \cmidrule(lr){4-5}
                       & \textbf{Sig.} & \textbf{$p$-Value} & \textbf{ES Inferable} & \textbf{ES Explicit}\\
\midrule
\textbf{Example}       & $p < .05$
                       & $p = .019$
                       & $t(24) = 2.52$ , 
                       & $t(24) = 2.52, p = .019$, \\
                       &&& $p = .019$
                       & Hedges' $g = 0.96$,\\
                       &&&& CI $[0.14, 1.76]$\\
\textbf{$p$ Quantifiable}    & \Circle & \CIRCLE & \CIRCLE & \CIRCLE \\
\textbf{Cross-Checkable}   & \Circle & \Circle & \CIRCLE & \CIRCLE \\
\textbf{ES Quantifiable} & \Circle & \Circle & \RIGHTcircle & \CIRCLE \\
\textbf{Synthesizable} & \Circle & \Circle & \RIGHTcircle & \CIRCLE \\
\bottomrule
\end{tabular}
\end{minipage}
\end{table}

\subsection{Reporting and Methodology Guidelines}
Reporting fidelity is usually one of the goals of reporting standards. 
Given that the field of socio-technical research in cyber security is a young and does not have its own established reporting standards, it is worthwhile to consider ones of other fields. 
Psychology seems a sound candidate to consider as a guiding example in this study. 
Other fields, such as behavioral economics, are equally viable.

The publication guidelines of the American Psychology Association (APA)~\cite{APAGuidelines6th2009} require that inferences are reported with their full test statistics and degrees of freedom. Exact $p$-values are preferred. The APA guidelines require to report appropriate effect sizes and their confidence intervals.
\processifversion{DocumentVersionTR}{Beyond the publication manual itself, we draw attention to the summary of a related working group~\cite{publications2008reporting}. Fidler~\cite{fidler2010american} offers a quick summary of important revisions with an eye on passing them on.}

Of course, there are also methodological guidelines that go far beyond reporting statistical tests. For instance, the CONSORT guidelines~\cite{moher2010consort} cover reporting for randomized trials\processifversion{DocumentVersionTR}{; the PRISMA statement~\cite{moher2009preferred} covers systematic review and meta analyses}. Furthermore, recently LeBel et al.~\cite{lebel2018unified} proposed a unified framework to quantify and support credibility of scientific findings%
\processifversion{DocumentVersionTR}{, drawing on the four areas: 
\begin{inparaenum}[(i)]
  \item method and data transparency,
  \item analytic reproducibility,
  \item analytic robustness, and
  \item effect replicability
\end{inparaenum}}.

Even though socio-technical aspects of security is a young field, there have been initiatives to advance research methodology, considered in chronological order:
\begin{inparaenum}[(i)]
\item In 2007, Peisert and Bishop~\cite{peisert2007design} offered a short position paper scientific design of security experiments.
\item Maxion~\cite{maxion2011making} focused on making experiments dependable, focusing on the hallmarks of good experiments with an eye on validity.
\item In 2013, Schechter~\cite{schechter2013common} considered common pitfalls seen in SOUPS submissions and made recommendation on avoiding them, incl. statistical reporting and multiple-comparison corrections.
\item Coopamootoo and Gro{\ss} proposed an introduction for evidence-based methods~\cite{CooGro2016}, incl. sound statistical inference and structured reporting.
\item The same authors published an experiment design and reporting toolset~\cite{CooGro2017}, considering nine areas with reporting considerations, incl. test statistics and effect sizes.
\end{inparaenum}

\processifversion{DocumentVersionTR}{Notably, the SLR used as sample in this study covers the same years (2006-2016) that have seen these different guidelines being proposed by members of the community.}

\subsection{Analysis of Statistical Reporting}
We analyze statistical reporting of publications with the \textsf{R} package \textsf{statcheck}~\cite{EpsNui2018}.
The \textsf{statcheck} tool extracts Strings of the form $\vari{ts} (\vari{df}) = x, p~\const{op}~y$, where $\vari{ts}$ is the test statistic, \vari{df} the degrees of freedom, and \const{op} a infix relation, such as, $<$. It recognizes $t$, $F$, $r$, $\chi^2$, and $z$ as test statistics and recomputes the corresponding $p$-values from them. 
It, hence, enables a consistency check of reported triplets of test statistic, degrees of freedom and $p$-values.

In this analysis, \textsf{statcheck} recognizes one-tailed tests to some extent from searching keywords and computing if a test were valid if considered one-tailed.
It adheres to the rounding guidelines of the American Psychology Association (APA)~\cite{APAGuidelines6th2009}.
Nuijten et al.~\cite{nuijten2017validity} concede that \textsf{statcheck} does not recognize $p$-values adjusted for multiple-comparison corrections.

While the creators of \textsf{statcheck} have argued for its validity and reliability~\cite{nuijten2017validity,nuijten2016prevalence},
the tool faced scrutiny and controversy~\cite{schmidt2016sources} over its false positive and false negative rates.
Schmidt~\cite{schmidt2016sources}, for example, criticized that \textsf{statcheck}'s inability to recognize corrected $p$-values, such as from Greenhouse-Geisser corrections.
Lakens~\cite{Lakens2015statcheck} found reported errors typically to be minor.
\processifversion{DocumentVersionTR}{However, we find that multiple-comparison, variability or degrees-of-freedom corrections are rarely applied in the field of cyber security. 

While we perceive \textsf{statcheck} as a useful tool to evaluate statistical reporting, we perceive it as crucial to cross-check results manually.}

For this study, we prepare to mitigate possible \textsf{statcheck} mis-classifications by manually checking and coding its outcomes.

\subsection{Related Works}

In 2016/17 Coopamootoo and Gro{\ss}~\cite{SLR2017} conducted a Systematic Literature Review (SLR) on cyber security user studies published in the years between 2006--2016. This research was first presented at a 2017 community meeting of the UK Research Institute in the Science of Cyber Security (RISCS). Their study contained three parts:
\begin{inparaenum}[(i)]
  \item the SLR itself, yielding a sample of 146 cyber security papers,
  \item a qualitative coding of nine ``completeness indicators,'' based on an \emph{a priori} codebook\processifversion{DocumentVersionTR}{~\cite{coopamootoo2017CIcodebook}}.
  \item a quantitative analysis on a sub-sample using parametric tests on differences between means (e.g., $t$-tests).
\end{inparaenum}

While this study uses the same set of papers as a sample to enable a comparison of results, this study takes an entirely different approach to the analysis:
\begin{DocumentVersionConference}
\begin{inparaenum}[(i)]
  \item Instead of manual coding of reporting completeness, we focus on the automated analysis reporting fidelity on extracted $p$-values,
  \item we evaluate quantitative properties on inconsistencies and decision errors of a large part of the sample, and
  \item we obtain a fine-grained understanding of ``things going wrong'' through grounded coding,
\end{inparaenum}
\end{DocumentVersionConference}
\begin{DocumentVersionTR}
Firstly, instead manually coding completeness indicators, which is invariably largely based on human judgment whether statistical reporting is appropriate, we focus on an automated approach extract $p$-values and respective test statistics. To a very large extent equally automated, we ascertain whether the the statistical reporting is internally consistent.

Secondly, while also conduct systematic coding, it is in tandem with the automated analysis and includes a level of detail of individual statistical tests. We, thereby, obtain a more fine-grained understanding of ``things going wrong'' as well as the magnitude of the deviation from a correct result. 
We also introduce a feedback-loop in the coding that enables the error-correction of faults of the tool.

Thirdly, whereas Coopamootoo and Gro{\ss}~\cite{SLR2017} focused on the quantitative analysis of a small sub-sample of tests and their observed (post-hoc) effect sizes, we focus our quantitative lens on inconsistencies and decision errors of all statistical tests conducted in the sample.
\end{DocumentVersionTR}

\section{Aims}

\processifversion{DocumentVersionTR}{\subsection{Outcome Definition.}}
We define the classes of \textsf{statcheck} outcomes for test statistics and papers.
\begin{definition}[SC Outcome Categories]
\label{def:sc_outcome}~\\
\textbf{Individual Tests}: \textsf{SCOutcome} has the following cases for individual tests:
\begin{compactenum}
  \item \textsf{CorrectNHST}: The NHST is reported with its test statistic triplet. The given triplet is correct, where ``correct'' is defined as matching triplet of test statistic, degrees of freedom and corresponding re-computed $p$-value.
  \item \textsf{Inconsistency}: The reported triplet (test statistic, $\vari{df}$, $p$-value) is inconsistent.
  \item \textsf{DecisionError}: The reported triplet (test statistic, $\vari{df}$, $p$-value) is grossly inconsistent, that is, the re-computed $p$-value leads to a different decision on rejecting the null hypothesis.
  \item \textsf{Incomplete}: A $p$-values is reported without sufficient data for an evaluation of the triplet (test statistic, $\vari{df}$, $p$-value).
  \end{compactenum}
\textbf{Entire Papers}: \textsf{SCOutcome} has the following cases for aggregated over papers:
\begin{compactenum}
  \item \textsf{CorrectNHST}: There exist one or more NHSTs reported with correct test statistic triplets. The given complete triplets are correct throughout, where ``correct'' is defined as matching triplet of test statistic, degrees of freedom and corresponding re-computed $p$-value. A paper can be classified as \textsf{CorrectNHST} even if there exist incomplete test statistics.
  \item \textsf{Inconsistency}: There exists an inconsistent triplet (test statistic, $\vari{df}$, $p$-value).
  \item \textsf{DecisionError}: There exists a gross inconsistency in any reported triplet (test statistic, $\vari{df}$, $p$-value), in which a re-computed $p$-value leads to a different decision on rejecting the null hypothesis.
  \item \textsf{Incomplete}: For all $p$-values reported, it holds that there is insufficient data for a correct triplet (test statistic, $\vari{df}$, $p$-value). For a paper classified as \textsf{Incomplete}, there is not a single $p$-value with complete test statistic found.
\end{compactenum}
We call \textsf{Complete} the complement of \textsf{Incomplete}.
\end{definition}
\processifversion{DocumentVersionTR}{This scale is conservative against false positives in that it deems it acceptable if some $p$-values are reported without their test statistics and only classifies an entire paper as \textsf{Incomplete} if \emph{all} $p$-values are reported without test statistics.}

\processifversion{DocumentVersionTR}{\subsection{Descriptives.} We will analyze and visualize the prevalence of statistical misreporting along the following lines.}

\begin{researchquestion}[Prevalence]
\label{rq:prevalence}
How many papers report on Null Hypothesis Significance Testing (NHST) and fall into one of the defined SC outcome categories according to Def.~\ref{def:sc_outcome}
\begin{inparaenum}
  \item \textsf{CorrectNHST},
  \item \textsf{Inconsistency},
  \item \textsf{DecisionError},
  \item \textsf{Incomplete}.
\end{inparaenum}
Which papers use
\begin{inparaenum}
  \processifversion{DocumentVersionTR}{\item MTurk,}
  \item multiple-comparison corrections (MCC),
  \item effect sizes.
\end{inparaenum}
\end{researchquestion}
\processifversion{DocumentVersionConference}{While we originally investigated the use of Amazon Mechanical Turk (AMT) and similar recruiting services, we have declared this aim out of scope for this publication. MCCs and effect sizes are also relevant in relation to power and Positive Predictive Value (PPV) of the studies in question, however, we will consider these inquiries in future work.}

\processifversion{DocumentVersionTR}{\subsection{Comparison to Other Fields.}}
We intend to compare the \textsf{statcheck} results in this field with analyses that have been conducted in other fields that seem related. We are most interested in fields at the intersection of human behavior and technology, such as HCI.
Granted that \textsf{statcheck} surveys have not been that widely conducted yet, we consider the \emph{Journal of Media Psychology (JMP)}~\cite{ElsPry2017} as a primary candidate. This choice is made because of similarities
\begin{compactenum}[(i)]
  \item media psychology is concerned with human subjects and socio-technical aspects,
  \item media psychology includes topics that might also have been published in user studies in cyber security, such as adversarial behavior (e.g., violence) vis-{\`a}-vis of HCI, cyber bullying, behavior on social media, 
  \item media psychology is a relatively young field, JMP having been founded in 1989 and gained its current name 2008.
\end{compactenum}
The distinct difference we are interested in is that JMP is subject to reporting standards (APA).
We note that the selection of JMP as comparison sample may be controversial and that---at the same time---comparisons to further fields are easily done, yet out of the scope for this study.

\begin{researchquestion}[Comparison]
\label{rq:comparison}
  To what extent do the \textsf{statcheck} \textsf{SCOutcomes} differ between our sample in this field and a comparable field in psychology?
  
  \noindent\const{H_{C,0}}: The distribution of the \textsf{SCOutcomes} in cyber security user studies is the same as the distribution in the comparison field. \const{H_{C,1}}: There is a systematic difference of \textsf{SCOutcome} in cyber security user studies to the comparison field.
\end{researchquestion}

\processifversion{DocumentVersionTR}{\subsection{Statistical Model on Venue\&Year.}
We establish a statistical model on in a correlational study on the question:}
\begin{researchquestion}[Influence of Venue and Year]
\label{rq:venue_year}
Considering outcome categories \textsf{SCOutcome} from Def.~\ref{def:sc_outcome} as response variable, what is the influence of predictors publication \textsf{Venue} and \textsf{Year}?

\begin{compactenum}
  \item \const{H_{V,0}}: There is no influence of the publication \textsf{Venue} on the occurrence of the \textsf{statcheck} outcome \textsf{SCOutcome}. \const{H_{V,1}}: There is a systematic influence of the publication \textsf{Venue} on the occurrence of the \textsf{statcheck} outcome \textsf{SCOutcome}.
   \item \const{H_{Y,0}}: There is no influence of the publication \textsf{Year} on the occurrence of the \textsf{statcheck} outcome \textsf{SCOutcome}. \const{H_{Y,1}}: There is a systematic influence of the publication \textsf{Year} on the occurrence of the \textsf{statcheck} outcome \textsf{SCOutcome}.
\end{compactenum}
\end{researchquestion}

As an exploratory inquiry, we employ the \textsf{statcheck} analysis to the submissions of STAST 2019, testing its usefulness in supporting PC members.

\section{Method}
The study has been pre-registered at the Open Science Framework (OSF)\footnote{\url{osf.io/549qn/}}, which also contains Online Supplementary Materials, such as a summary of the SLR specification and the sample itself.
All analyses, graphs and tables are computed directly from the data with the \textsf{R} package \textsf{knitr}, where the \textsf{statcheck} output was cached in \textsf{csv} files.

All statistical tests are computed at a significance level of $\alpha = .05$. The Fisher Exact Tests (FETs) for cases with low expected cell frequency are computed with simulated $p$-values with $10^5$ replicates.

\subsection{Ethics}
This study followed the guidelines of the ethical boards of its institution.
While we make the entire list of analyzed papers available for reproducibility, we decided not to single out individual papers.
We are aware that the the descriptive statistics presented allow making a link to the respective papers; we accept that residual privacy risk.
\emph{Full disclosure:} one of the sample's papers belongs to the author of this study; \textsf{statcheck} flagged it.

\subsection{Sample}
The target population of this study was cyber security user studies.
The sampling frame for this study is derived from a 2016/17 Systematic Literature Review (SLR) conducted by Coopamootoo and Gro{\ss}~\cite{SLR2017} whose results were first published at a 2017 Community Meeting of the Research Institute in the Science of Cyber Security (RISCS). This source SLR's search, inclusion and exclusion criteria are reported in \processifversion{DocumentVersionTR}{Appendix~\ref{sec:SLR}.}\processifversion{DocumentVersionConference}{Online Supplementary Materials.} 

We have chosen this sample to gain comparability to earlier qualitative and quantitative analyses on it~\cite{SLR2017}. This sample restricts the venues considered to retain statistical power for a regression analysis.
We stress that the automated the analysis methodology can be easily applied to other samples.

\subsection{Procedure}
\begin{figure}[tb]
  \centering
  \vspace{-1.5cm}
  \includegraphics[keepaspectratio,width=\textwidth]{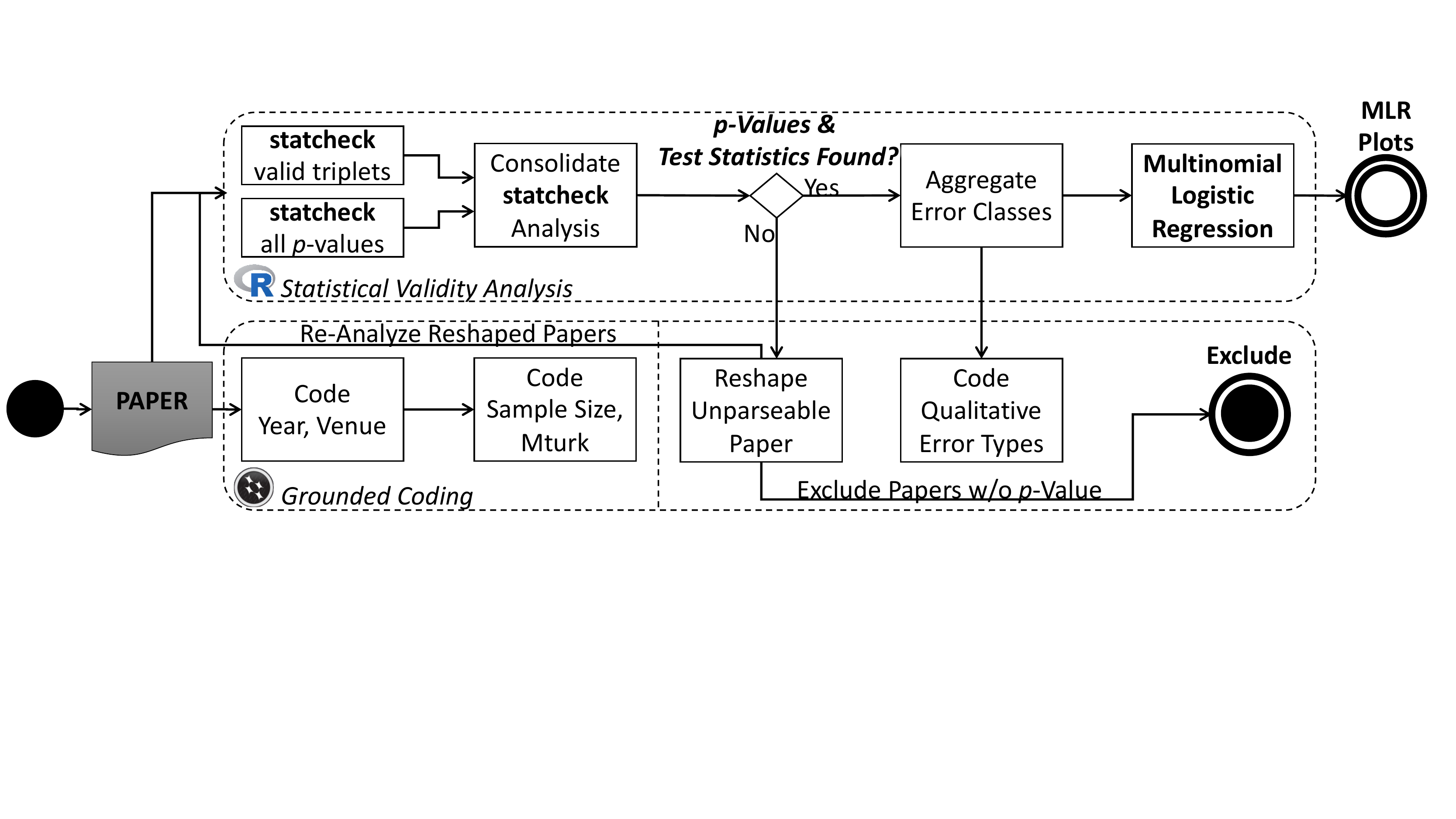}
  \vspace{-2.8cm}
  \caption{Flow chart of the study's procedure with two interlinked analyses.}
  \label{fig:procedure}
\end{figure}

Our procedure, as depicted in Figure~\ref{fig:procedure}, constituted a mixed-methods approach that fusing two interlinked analysis processes: \begin{inparaenum}[(i)] 
\item Statistical Validity Analysis and 
\item Grounded Coding of paper properties and errors detected.
\end{inparaenum}
Our analysis script received as input the PDFs of studies included from the source SLR.

\paragraph{Statistical Validity Analysis.} we computed two iterations of \textsf{statcheck}, one only considering statistical statements in standard format and one including all $p$-values found.
The \textsf{statcheck} results were subjected to a manual cross-check, possibly resulting the the reshaping of papers that \textsf{statcheck} could not parse out of the box.
Subsequently, we merged the results of both analyses and aggregated their events (counting number of correct tests, inconsistencies, decision errors and $p$-values without parseable test statistics). We, thereby, established the dependent variable \textsf{SCOutput} per statistical test and per paper.

\paragraph{Grounded Coding.} We coded paper properties in NVivo.
We evaluated the \textsf{statcheck} results in a second lane of grounded coding, classifying errors of \textsf{statcheck} as well as errors committed by authors of the papers. 

As a part of this analysis, we ``reshape'' papers that could not be parsed by \textsf{statcheck} for reasons outside of the research aims of this study. For instance, if a paper embedded statistical tests as image rather than text, we would transcribe the images to text and re-run \textsf{statcheck} on the ``reshaped'' input.

Once these results are coded, we amend the \textsf{statcheck} outcomes recorded in \textsf{SCOutcome} to ensure that this variable reflects an accurate representation of the sample.

\processifversion{DocumentVersionTR}{Finally, based on the accumulated meta-data on the publications, we then evaluated the \textsf{statcheck} results with respect to publication year and venue.}

\subsection{Grounded Coding}
Grounded coding refers to the the code being grounded in properties found in the data, instead of being based on an \emph{a priori} codebook.
\paragraph{Paper Properties.}
We conducted a systematic coding in \textsf{NVivo} with the purpose to establish overall properties of all papers.
We were extracting especially:
\begin{inparaenum}[(i)]
  \item sample size,
  \processifversion{DocumentVersionTR}{\item use of recruiting platforms (e.g., MTurk),}
  \item use of multiple-comparison corrections, and
  \item use of dependent-samples tests.
\end{inparaenum}

\paragraph{Analysis Outcomes.}
After having run \textsf{statcheck} on the sample, we first conducted a grounded coding of statistical tests marked as 
inconsistency or decision error. 
We re-computed the $p$-values from the test statistics ourselves and interpreted the results in the context of the reporting of the paper. 
We took into account the formulation around the test as well as overall specification of hypotheses, test parameters (e.g., one-tailed) and multiple-comparison corrections.
We include the resulting emergent codebook presented in Table~\ref{tab:codebook}.

Secondly, we analyzed the outcomes \textsf{statcheck} marked as neither inconsistency nor decision error. For those results, we compared the raw text with \textsf{statcheck}'s parsed version as well as recomputed $p$-value. We ignored small rounding differences as \textsf{statcheck} as authors rounding test statistics for reporting will naturally cause small differences. In cases of a mismatch between raw text and interpretation (e.g., in degrees of freedom accounted for), we re-computed the statistics manually.

Finally, we coded whether a mistake by \textsf{statcheck} would be considered a \textsf{FalsePositive} or \textsf{FalseNegative}.
After this evaluation, we adjusted the \textsf{SCOutcome} to ensure that the subsequent analysis is based on a correct representation of the sample.

\begin{table}
\caption{Codebook of the grounded coding of error types.}
\label{tab:codebook}
\footnotesize
\begin{tabular}{@{}lp{4cm} lp{4cm}@{}}
\toprule
\multicolumn{2}{c}{Errors of \textsf{statcheck}} & \multicolumn{2}{c}{Errors of authors}\\
\cmidrule(lr){1-2} \cmidrule(lr){3-4}
Code & Definition & Code & Definition\\
\midrule
\textsf{scParsedOK} & parsed the PDF correctly& 
	\textsf{Typo} & Likely mis-typed\\
\textsf{scCorrect} & \textsf{statcheck} result validated&
	\textsf{RoundingError} & incorrect rounding rules\\
\textsf{scMisclassified} & misclassified test & 
	 \textsf{OneTailedUS} & unspecified one-tailed test\\
\textsf{scMissedMC} & missed multiple-comparison corrections specified paper &
	 \textsf{Miscalculation}& miscalculated the statistics, wrong $p$-value for statistic\\
\bottomrule
\end{tabular}
\end{table}

\todo[inline]{A! - Review incomplete papers manually to find whether \textsf{statcheck} has missed tests.}

\subsection{Evaluation of \textsf{statcheck}}
Appendix~\ref{sec:qualitative_coding} contains the details of the corresponding qualitative coding.

\paragraph{Reshaping of Unparseable Papers.}
There were eight of papers for which \textsf{statcheck} could neither extract $p$-values nor test statistics due to encoding issues (e.g., embedding statistics as images). For all of those, we recorded them as unparseable, yet transformed them into parseable text files for further analysis.

\paragraph{Errors Committed by \textsf{statcheck}.}
Of the total 252 parsed tests, 34 contained an error, 10 of which a decision error. We compared those outcomes against the grounded coding of results and our re-computation of the statistics.

We found that
\begin{inparaenum}[(i)]
  \item \textsf{statcheck} parsed papers that were correctly reported without fail,
  \item it misclassified two tests,
  \item it detected one-tailed tests largely correctly,
  \item it treated dependent-samples tests correctly,
  \item it did not recognize the specified multiple-comparison corrections in three cases.
\end{inparaenum}
This leaves us with 5 false positives and no false negatives, marked in Sub-Figure~\ref{fig:plotObsTypes:errorClasses}.

\paragraph{Detection Performance of \textsf{statcheck}.}
\label{sec:statcheck_detection_perf}
For the analysis of complete test triplets, we analyzed the confusion matrix of \textsf{statcheck} results vs. our coding (Table~\ref{tab:confusionMatrixTests}).
The Positive Predictive Value (PPV) of $85\%$ indicates a decent likelihood of a positive \textsf{statcheck} report being true.

\confusionMatrixTests

\subsection{Multinomial Logistic Regression}
We conducted multinomial logistic regressions with the \textsf{R} package \textsf{nnet}\processifversion{DocumentVersionConference}{~\cite{nnet2016}}, relying on Fox's work~\cite{fox2006effect} for visualization. The models were null, year-only, venue-only and year and venue combined. The dependent variables was \textsf{SCOutput}. The independent variables were \textsf{Year} (interval) and \textsf{Venue} (factor).

\section{Results}
\subsection{Sample}

We have refined the inputed sample of $146$ publications by excluding publications that do neither contain empirical data nor  significance tests ($p$-value), retaining $114$ publications for further analysis. We illustrate the sample refinement in Table~\ref{tab:sample}. We include the final sample in \processifversion{DocumentVersionTR}{Appendix~\ref{sec:SLRsample}}\processifversion{DocumentVersionConference}{the Online Supplementary Materials} and outline its distribution by publication venue and year in Table~\ref{tab:contingencyVenueYear}. We note that the sample is skewed towards SOUPS and more recent publications. We note that the sample was drawn only from 10 specific venues in an effort to retain power in a logistic regression with \textsf{venue} as a categorical factor.

\begin{table}[tb]
\centering
\caption{Sample Refinement and Final Composition}
\label{tab:sample}
\begin{tabular}{lr rl}
\toprule
\textbf{Phase}  & Excluded & Retained Sample & \\ 
\midrule
\textit{Source SLR}~\cite{SLR2017} (Google Scholar) &    ---         & 1157\\
\quad Inclusion/Exclusion             & 1011 & 146\\
\midrule
\textit{This study}\\
\quad Studies with Empirical Data           &       24       & 122\\
\quad Studies with NHST/$p$-Value           &       8            & 114 & $\rightarrow$ \textbf{Final Sample}\\
\bottomrule
\end{tabular}
\end{table}

\contingencyVenueYear

\subsection{Exploration of the Distribution}
\processifversion{DocumentVersionTR}{We offer explorative descriptive analyses of the SLR cohort, before we move on to the pre-registered analyses.}

\paragraph{Distribution of Qualitative Properties.}
We visualize the presence of qualitative properties of papers over time in Figure~\ref{fig:proportionsplotPropertiesYear}. 
We observe 
\begin{inparaenum}[(i)]
  \processifversion{DocumentVersionTR}{\item MTurk being used from 2010 (\ref{fig:plotMTurkUse}),}
  \item Multiple-Comparison Corrections seeing adoption from 2009 (\ref{fig:plotMCCUse}),
  \item Effect sizes being on and off over the years (\ref{fig:ESUse}).
\end{inparaenum}
\processifversion{DocumentVersionTR}{\proportionsplotPropertiesYear}
\processifversion{DocumentVersionConference}{\proportionsplotPropertiesYearwoMTurk}

\paragraph{Distribution of $p$-Values.}
We analyze the distribution of $p$-values per paper. Therein we distinguish incomplete and complete triplets including test statistic and degrees of freedom. In Figure~\ref{fig:plotPValuesCombined}, we depict this $p$-value distribution; \ref{fig:plotReportedPValues} is ordered by number of the tests reported on, distinguishing between complete/incomplete triplets while annotating the presence of multiple comparison corrections (MCC);
\ref{fig:plotPValueByYear} is organized by publication year. 
The included linear regression lines indicate little to no change over time.
\plotPValuesCombinedVert

\subsection{Prevalence of Statistical Misreporting}
For RQ\ref{rq:prevalence}, we compare statistical misreporting by \textsf{venue} and \textsf{year}, considering individual tests as well as entire papers (cf. contingency tables in \processifversion{DocumentVersionTR}{Appendix~\ref{sec:contingencyTables}}\processifversion{DocumentVersionConference}{the Online Supplementary Materials}). 

\paragraph{Misreported Tests.}
For individual tests, there is a statistically significant association between the \textsf{statcheck} outcomes
and the publication venue, FET $p = .033$, as well as the publication year, FET $p < .001$. This offers first evidence to reject the null hypotheses \const{H_{V,0}} and \const{H_{Y,0}}.

Table~\ref{tab:contingencyTestsSCOutcomeVenue} contains the corresponding contingency table.

\contingencyTestsSCOutcomeVenue

\paragraph{Papers with Misreporting.}
Sub-Figure~\ref{fig:waffleplotSLR} on p.~\pageref{fig:waffleplotSLR} shows a hierarchical waffle plot of the \textsf{statcheck} outcomes. For aggregated outcomes per paper displayed in Figure~\ref{fig:proportionsplotCombinedSLR}, the associations per venue and year are not statistically significant, FET $p = .964$ and FET $p = .458$ respectively. A likely reason for this result is visible in the histograms of Figure~\ref{fig:errorHistogramPlot}: errors are at times clustered, in that, some papers contain multiple errors.

\proportionsplotCombinedSLRArea

\errorDoubleHistogram

\subsection{Comparison with JMP}
With respect to RQ\ref{rq:comparison}, the \textsf{statcheck} outcomes of the included SLR and Journal of Media Psychology (JMP) are statistically significantly different, $\chi^2(3) = 88.803, p < .001$. Hence, we reject the null hypothesis $\const{H_{C,0}}$ and conclude that there is a systematic difference between fields. We find an effect of Cram{\'e}r's $V = 0.646$, 95\% CI $[0.503, 0.773]$.

If we restrict the analysis to the papers containing \textsf{Complete} tests and, thereby, exclude papers marked \textsf{Incomplete}, we find that the difference between fields is not statistically significant any longer, $\chi^2(2) = 0.197, p = .906$, Cram{\'e}r's $V = 0.037$, 95\% CI $[0, 0.139]$.

\waffleplotCombined

\subsection{Reporting Test Outcomes by Venue and Year}

\combinedplotTestsVYso

While we analyzed tests and aggregated paper \textsf{SCOutcome} by \textsf{venue} and \textsf{year}, we found that these multinomial logistic regressions were not stable. Even if the models were statistically significant, this missing stability was evidenced in extreme odds-ratios, which was likely rooted in the sparsity of the dataset. (We report all MLR conducted in \processifversion{DocumentVersionTR}{Appendix~\ref{sec:all_mlr}}\processifversion{DocumentVersionConference}{the Online Supplementary Materials} for reference). To overcome the sparsity, we chose to collapse the \textsf{venue} factor into \textsf{SOUPS} and \textsf{OTHER} levels, called \textsf{venue'} (and the corresponding null hypothesis \const{H_{V',0}}).

A multinomial logistic regression on individual tests with \textsf{SCOutcome} $\sim$ \textsf{venue'}+\textsf{year} with \textsf{Incomplete} as reference level is statistically significant, LR, $\chi^2(6) = 15.417, p = .017$. Because the model explains McFadden $R^2$= .01 of the variance, we expect little predictive power.

The corresponding predictors are statistically significant as well. Hence, we reject the null hypotheses \const{H_{V',0}} and \const{H_{Y,0}}. Figure~\ref{fig:combinedplotTestsVYso} contains an overview of the scatter plot vs. the predicted probabilities from the MLR.

While we find that there is an effect of \textsf{year} in increasing likelihood of \textsf{Incomplete} outcomes, this only accounts for an increase of $0.2\%$ per year, barely perceptible in the graph. Everything else being equal, a transition from \textsf{venue} \textsf{SOUPS} to \textsf{OTHER} yields an increase of likelihood of the \textsf{Incomplete} outcomes, by a factor of roughly 2. However, these changes are dwarfed by the overall intercept of tests being correct (in comparison to \textsf{Incomplete}).

In absolute terms, the expected likelihood of tests being \textsf{Incomplete} is $80\%$, with \textsf{OTHER} venues having a few percent greater \textsf{Incomplete} likelihood. \textsf{SOUPS} exhibits an expected likelihood of $13\%$ of being \textsf{CorrectNHST}, while \textsf{OTHER} venues yield a few percent lower likelihood.

\todo[inline]{Promised model validation: residuals etc. fit, 10-fold validation. df. Field.}

\subsection{Qualitative Analysis}
We offer a summary of the analysis here, a detailed account is included in Appendix~\ref{sec:qualitative_coding}.

\plotObsTypesCombinedVert

\paragraph{Composition of Incomplete $p$-Values.}
Sub-Figure~\ref{fig:plotObsTypes:incompleteClasses} contains an overview of the classes of incompletely reported $p$-values.
Less than half the cases of incomplete triplets contain an actual $p$-values (half of them, in turn, significant or not significant).
$31\%$ of the incomplete cases compared to lower significance bound than $\alpha = .05$.
$9\%$ of the tests are simply declared non-significant, another $7\%$ reported as significant wrt. $p < .05$.

\paragraph{Distribution of $p$-values.}
Figure~\ref{fig:histogramPValueDifference} shows the difference between reported and computed $p$-values.
When comparing reported and re-computed $p$-values, we found that in 22 out of 34 cases, the reported $p$-value was more significant than the computed one ($65\%$).

\histogramPDiff

\subsection{Significance Detection Performance}
\label{sec:detection_perf}

We analyzed the decision making of authors on statistical significance of reported results vis-{\`a}-vis of recomputed $p$-values (Table~\ref{tab:confusionMatrixScTests}). We observe a somewhat low specificity of $80\%$. Note that this analysis only refers to a reported significance decision is valid with respect to a corresponding correct $p$-value, and \emph{not} whether a positive reported result is true.
\confusionMatrixScTests

\subsection{Supporting the STAST 2019 PC in Checking Statistics}
\label{sec:support_stast_2019}

Aligned with Recommendation~2 in Section~\ref{sec:support_pcs}, we offered a \textsf{statcheck} analysis to the STAST PC members to support the workshop's discussion phase. Of 28 submitted papers, 9 papers ($32\%$) included a statistical inference.

Let us consider these 9 papers in detail as an exploratory analysis.
One paper contained a major error in terms of statistics being invalid, two papers used the wrong statistical method for the experiment design at hand (e.g., independent-samples statistics in a dependent-samples design).
Two of those three papers were also flagged by \textsf{statcheck}.
These errors themselves, however, were detected by program committee members, not by the \textsf{statcheck} analysis.

On third of the papers reported statistics in an APA compliant format.
6 papers ($66\%$) reported exact $p$-values, 4 papers ($44\%$) reported effect sizes as required by the STAST submission guidelines.
Of the 9 papers, 7 needed multiple-comparison corrections, which only two provided in their initial submission.

In terms of \textsf{statcheck} evaluation with the methodology of this study, we found 5 papers ($56\%$) to be \textsf{Incomplete}, one paper \textsf{Inconsistent}, three papers ($33\%$) \textsf{CorrectNHST}.
This distribution is not significantly different from the SLR sample shown in Figure~\ref{fig:waffleplotSLR}, $\chi^2(3) = 0.829, p = .843$, Cram{\'e}r's $V = 0.082$, 95\% CI $[0, 0.188]$.

\section{Discussion}
\subsubsection{Incomplete reporting holds back the field.}
Nearly two thirds of the papers with $p$-values did not report a single complete test triplet (cf. Fig.~\ref{fig:waffle.SLR}).
This impairs the ability to cross-check internal consistency of tests and, thereby, undermines fault-tolerance. 
Hence, such papers have limited credibility and fidelity of statistical information.

The incomplete reporting observed in this study is in stark contrast to the analysis of the Journal of Media Psychology (JMP), in which not a single paper was \textsf{Incomplete}.
Hence, we conclude that mandated reporting standards are an effective tool.

It is further troubling that the likelihood of incomplete reporting did not seem to decrease over time (cf. Fig.~\ref{fig:mlrplotVYso}).

In terms of research reuse and synthesis, the situation is aggravated, because effect sizes are vastly under-reported in this field.
Only a small minority reports them explicitly; one third of the papers allows to infer them (cf. Fig~\ref{fig:ESUse}).

There are three consequences to this phenomenon: 
\begin{inparaenum}[(i)]
  \item It is exceedingly difficult for practitioners to ascertain the magnitude of effects and, thereby, their practical significance.
  \item It is near-impossible to compare research results in meta-analyses and to synthesize well-founded summary effects.
  \item Hence, disputes over differences between original studies and replications are hard to settle satisfactorily.
\end{inparaenum}.

\subsubsection{While some errors are minor, we caution against clustered errors and miscalculations.}
Of the 44 papers with complete test statistic triplets analyzed, 60\% were deemed correct; more than one quarter had at least one inconsistency; 14\% had at least one decision error.
Of all tests with complete triplets analyzed 14\% were erroneous. 
Here, the socio-technical security sample showed similar error rates as the psychology sample.

Especially the 26 papers with complete test triplets and correct reporting---one quarter of the sample---stand testament to efforts of authors and program committees ``get it right.''

The errors observed by \textsf{statcheck} were often minor typos and rounding errors that could have been easily avoided, however nearly 40\% seemed to be serious miscalculations.
We found that these errors were at times clustered: there are a few papers with a number of errors.

\processifversion{DocumentVersionTR}{Of course, we would need to assume that the 68 papers without complete test triplets have at least the same error rates as the ones with complete triplets, yielding another dark figure.}

\subsubsection{There is a dark figure of decision errors lurking in the underuse of multiple-comparison corrections.}
This study leaves the detailed analysis of power and multiple-comparison corrections (MCCs) to future work.
Still, we do not want to withhold insights already apparent from Fig.~\ref{fig:plotReportedPValues}: 
There is a Damocles sword hanging over many papers: Multiple-Comparison Corrections (MCCs).

We have seen in Fig.~\ref{fig:plotMCCUse} that even though MCCs came in use from year 2009, only about one third of the papers employed them.
From Fig.~\ref{fig:plotReportedPValues}, we observe that there are papers with a considerable number of reported $p$-values without MCCs.
Hence, there may well be a sizable dark figure of papers with decision errors in store once adequate MCCs are employed.

These observations inform Recommendation 3 in that observing studies with many comparisons but without corrections can be an indication of the number of comparisons, multiple-comparison corrections as well as the power needed to sustain them only being considered as an afterthought.

\subsubsection{Automated checking of statistical reporting is viable.}
The \textsf{statcheck} detection rates were very good and comparable to the rates reported by Nuijten et al.~\cite{nuijten2017validity}.
We note, however, that \textsf{statcheck} did not operate completely autonomously, but was complemented with human coding to overcome parsing issues.
We find the approach viable for the use in socio-technical aspects of security.

\processifversion{DocumentVersionTR}{We believe that we encountered problems reported by Schmidt~\cite{schmidt2016sources,schmidt2017statcheck} to a lesser extent 
as this field is largely operating with simple statistical tests and few corrections of $p$-values.}

\subsection{Limitations}

\subsubsection{Generalizability.} The study is based on an existing SLR sample that largely consists of SOUPS publications and only contains few cases for other venues. Dealing with a sparse matrix, the likelihoods computed for non-SOUPS venues as well as overall logistic regressions suffer from more uncertainty.

\processifversion{DocumentVersionTR}{Also, the use of the SLR sample instead of a statistical sampling method with a complete sampling frame limits generalizability.}

\subsubsection{Syntactic Validity Checks.} While we have made good experiences with \textsf{statcheck} and only found few false positives and negatives, we observe that \textsf{statcheck} results can suffer from hidden errors.
While we complemented the automated analysis with a human review and coding of reported errors, we observe that \textsf{statcheck} could have missed or misinterpreted individual tests. However, based our inspection of the 114 analyzed papers, we expect that the number of \textsf{statcheck} errors is small compared to the 1775 tests analyzed.
In the end, an automated tool cannot replace the trained eye of a knowledgable reviewer. However, this study is about the overall distribution of errors, which will be hardly skewed by rare false positives or negatives.

\subsubsection{Deviations from the Pre-Registration.}
\begin{DocumentVersionConference}
We deviated from the OSF pre-registration by
\begin{inparaenum}
  \item not attempting the exploratory analysis of the impact of authors,
  \item not attempting an exploratory logistic regression on completeness indicators,
  \item abandoning the planned ordinal logistic regression in favor of the MLR, because \textsf{SCOutcome} did not yield an ordinal scale,
  \item merging non-SOUPS \textsf{venue} levels to overcome the sparsity of the dataset,
  \item not attempting further cross-validation due to low variance explained.
\end{inparaenum}
\end{DocumentVersionConference}
\begin{DocumentVersionTR}
\begin{compactenum}
  \item We did not attempt the exploratory of author\&institution as the sample seemed too small and dimensionality reduction may introduce artifacts.
  \item We pre-registered an ordinal logistic regression as primary analysis, however found that \textsf{SCOutcome} is not a valid ordinal variable and retained the also pre-registered multinomial logistic regression as tool of choice.
  \item We merged the non-SOUPS \textsf{venue} levels to overcome the sparsity of the dataset,
  \item We did not pursue a logistic regression on Coopamootoo and Gro{\ss}'s completeness indicators~\cite{SLR2017}, as the nine additional regressions would yield a higher Type-I error rate.
\end{compactenum}
\end{DocumentVersionTR}

\section{Recommendations}
The recommendations made here need to be seen as part of a greater paradigm shift.
Instead of focusing on single publications, one may consider that a study does not stand on its own.
Truly advancing the knowledge of a field calls for creating robust studies that prepare the ground for systematic replications, reuse and research synthesis.

\subsubsection{1. Establish sound reporting standards.}
Sound and generally accepted reporting standards could greatly improve the credibility of the field.
This could either mean developing systematic reporting standards for socio-technical aspects of security or adopting existing standards.

Developing systematic reporting standards would involve a stable coalition of program committee chairs and members as well as journal editors forming a working group to that effect. Such a working group would likely take into account requirements for this field as well as examples of mature reporting standards from other fields.

Given that considerable thought has gone into APA standards~\cite{APAGuidelines6th2009}\processifversion{DocumentVersionTR}{ and Psychology Journal standards~\cite{publications2008reporting}} and that these standards apply to human dimensions, they are a viable and sufficiently mature candidate, at least when it comes to statistical reporting. Our analysis showed that the majority of papers reporting complete test statistics triplets were actually compliant to APA requirements.

While not perfect, their recommendations on statistical reporting could have considerable benefits for reporting fidelity, research reusability and synthesis. One option in this context would be to only adopt a subset of recommendations directly benefiting reporting fidelity.

In any case, one would consider sound reporting for test statistics themselves, effect sizes and their confidence intervals, as well as essential information on the sample, design and procedure.
Again, this field can well take into account more comprehensive initiatives from other fields~\cite{lebel2018unified}.

\subsubsection{2. Support PCs in checking statistics.}
\label{sec:support_pcs}
From our experience researching this study, we can attest that checking statistics can be a tedious affair.
Even with all their failings, tools like \textsf{statcheck} can support program committee members in detecting incorrect results.
Such an approach certainly requires human mediation to avoid false positives, yet can offer insights at low cost.

As reported in Section~\ref{sec:support_stast_2019}, we tested this recommendation on the STAST 2019 program committee. While \textsf{statcheck} correctly identified reporting issues and did not produce a false positive, major errors were discovered by program committee members in the analysis of experiment designs vis-{\`a}-vis their statistical inferences. This yields an indication that an automated tool, such as \textsf{statcheck}, will only support but never replace the expert judgment of the reviewers.

There are organizational methods, such as pre-registrations or registered reports, that can support a PC further in ascertaining the integrity of results.

\subsubsection{3. Embrace \emph{a priori} power and multiple-comparison corrections.}
We make this recommendation with a grain of salt, as we have not reported on a dedicated study on power, yet.
However, even this study on reporting fidelity shows that this consideration would benefit the community.

Low power and missing adequate MCCs can well undermine the results of a good study and increase the likelihood of a positive result being a false positive.
We encourage researchers to plan in advance for the power required, accounting for the MCCs necessary for the planned tests.

\section{Conclusion}
This study is the first systematic analysis of a large sample of security user studies with respect to their statistical reporting fidelity.
For the first time, we offer a comprehensive, quantitative, and empirical analysis of the state-of-play of the field of socio-technical aspects of security.
We offer a wealth of different perspectives on the sample, enabling us to obtain a fine-grained analysis as well as broad recommendations for authors and program committees alike.

We stress that the research and reviewing process for security user studies constitutes a socio-technical system in itself that impacts the decision making in security and privacy.
Because scientists and practitioners alike seek to re-use research results, the fidelity or uncertainty of those results---especially their statistical inferences---plays a major role in the credibility of the field and the confidence of its audience.
Hence, self-reflection of the field will ultimately impact the decision making by users in security and privacy, as well.

As future work, we consider expanding the sample, including further venues, such as CHI, as well as offering a dedicated analysis of statistical power and Positive Predictive Value (PPV) present in the field.

\section*{Acknowledgment}

We would like to thank Malte Elson for the discussions on \textsf{statcheck}, on the corresponding analyses in psychology, and on general research methodology.
We thank the anonymous reviewers of STAST 2019 for their discussion and insightful comments, as well as the volume co-editor Theo Tryfonas for offering additional pages to include the requested changes. 

This study was in parts funded by the UK Research Institute in the Science of Cyber Security (RISCS) under a National Cyber Security Centre (NCSC) grant on ``Pathways to Enhancing Evidence-Based Research Methods for Cyber Security'' (Pathway I led by Thomas Gro{\ss}).
The author was in parts funded by the \CASCAde.

\bibliographystyle{splncs04}

\newpage

\begin{appendix}
\section{Details on Qualitative Analysis}
\label{sec:qualitative_coding}

\subsection{Errors Committed by \textsf{statcheck}.}
\paragraph{Parsing Accuracy.} 
In all 34 error cases, \textsf{statcheck} parsed the PDF file correctly, and its raw test representation corresponded to the PDF.
In all but two tests, \textsf{statcheck} recognized the test correctly. In said two cases, it mistook a non-standard-reported Shapiro-Wilk test as $\chi^2$ test, creating two false positives. There was one case in which the \textsf{statcheck} computed $p$-value for an independent-samples $t$-test differed slightly from our own calculation, yet only marginally so, presumably because of a unreported Welch correction.

\paragraph{One-Tailed Tests.}
In seven cases, \textsf{statcheck} recognized one-tailed tests correctly. For three of those tests, the authors framed the hypotheses as one-tailed. In three other tests, the authors used one-tailed test results without declaring their use. There was one additional case in which the authors seemed to have used a one-tailed test, yet the rounding was so far off the one-tailed result that \textsf{statcheck} did not accept it as ``valid if one-tailed'' any longer.
There was one test marked as ``one-tail'' which \textsf{statcheck} did not recognize as one-tailed, yet that test also suffered from rounding errors.

\paragraph{Dependent-Samples Tests.}
There were 7 papers using dependent-samples methods (such as matched-pair tests or mixed-methods regressions). We found that \textsf{statcheck} treated the corresponding dependent-samples statistics correctly.

\paragraph{Multiple Comparison Corrections.} 
In three cases, \textsf{statcheck} did not recognize $p$-values that were correctly Bonferroni-corrected, counting as three false positives. %
It is an open point, however, how many paper should have employed multiple-comparison corrections, but have not done so, an analysis \textsf{statcheck} does not perform.

\subsection{Errors Committed by Authors}
\paragraph{Typos.}
We considered 6 to be typos or transcription errors ($18\%$). Another 1 error seemed to be a copy-paste error ($3\%$)

\paragraph{Rounding Errors.} Of all 34 reported errors, we found 8 to be rounding errors ($24\%$).

\paragraph{Miscalculations.}
We found 13 cases to be erronious calculations ($38\%$).

\subsection{Composition of Incomplete $p$-Values}
Of 1523 incomplete cases, 134 were declared ``non-significant'' without giving the actual $p$-value ($9\%$). Further, 6 were shown as $p > .05$. ($0\%$). 

Of the incomplete cases, 102 were reported statistically significant at a $.05$ significance level ($7\%$).

Of the incomplete cases, 477 were reported statistically significant at a lower significance level of $.01$, $.001$, or $.0001$ ($31\%$).

Of 1523 incomplete $p$-values, 680 gave an exact $p$-value ($45\%$). Of those exatly reported $p$-values, half (367) were claimed statistically significant at a significance level of $\alpha = .05$ ($54\%$). Of those exatly reported $p$-values, 19 claimed an impossible $p$-value of $p = 0$ ($3\%$).

\newpage
\begin{DocumentVersionTR}
\section*{\LARGE Online Supplementary Materials}
We made the materials of the study (specification of the inputted SLR, included sample, contingency tables) also publicly available at its Open Science Framework Repository\footnote{\url{osf.io/549qn/}}.

\section{Underlying Systematic Literature Review}
\label{sec:SLR}
This meta-analytic study is based on a Systematic Literature Review (SLR), which was conducted in 2016/17 for the UK Research Institute in the Science of Cyber Security (RISCS). We adapt this description of the SLR's search from its technical report~\cite{SLR2017}.

\subsection{Search Strategy of the SLR Sample}
\label{sec:SLR_search}
The SLR included security and privacy papers published between 2006 and 2016 (inclusive).

The search was restricted to the following security and privacy venues:
\begin{compactitem}
  \item journals: IEEE Transactions on Dependable \& Secure Computing (TDSC), ACM Transactions on Information and System Security (TISSEC), 
  \item flagship security conferences: IEEE S\&P, ACM CCS, ESORICS, and PETS or 
  \item specialized conferences and workshops: LASER, SOUPS, USEC and WEIS.
\end{compactitem}  

The search was conducted on Google Scholar. 
Each query extracts articles mentioning ``\emph{user study}'' and at least one of the words ``\emph{experiment},'' 
``\emph{evidence}'' or ``\emph{evidence based}.'' The described query was executed for each of the $10$ publication venues.
In the advanced search option of Google Scholar, each of the following fields were set:
\begin{compactitem}
\item with all words = \emph{user study}
\item at least one of the words = \emph{experiment evidence ``evidence based''}
\item where my words occur = \emph{anywhere in the article}
\item return articles published in = [publication venue]
\item return articles dated between = \emph{2006--2016}
\end{compactitem}

The search yielded $1157$ publications.

\subsection{SLR Inclusion/Exclusion Criteria}
\label{sec:SLR_incl-excl}
We adapt the inclusion/exclusion criteria of the 2017 SLR~\cite{SLR2017} for this pre-registration.
The SLR focused on human factors studies including a human sample.
The following \emph{Inclusion Criteria} were applied to its overall pool of $1157$ publications:
\begin{compactitem}
 \item Studies including a user study with human participants.
  \item Studies concerned with evidence-based methods or eligible for hypothesis testing and statistical inference.
  \item Studies that lend themselves to quantitative evaluation, quoting statements of statistical significance, $p$-values or effect sizes.
  \item Studies with true experiments, quasi-experiments or observational analysis.
\end{compactitem}
Of the papers included, the ones fulfilling the following \emph{Exclusion Criteria} were excluded:
\begin{compactitem}
  \item Papers that were not subject to research peer-review, key note statements, posters and workshop proposals.
  \item Position papers or informal arguments.
  \item Papers not including a study with human participants,
  \item Theoretical papers.
  \item Studies with qualitative methodology.
\end{compactitem}
This inclusion/exclusion process yielded a final sample of $146$ publications.

\section{SLR Sample}
\label{sec:SLRsample}
\begin{small}
\begin{longtable}{lp{6cm}ccc}
\caption{Sample of inputted SLR~\cite{SLR2017} and this study with marked exclusions (Ex.).}\\
\toprule
\textbf{Tag} & \textbf{Tilte} & \textbf{Venue} & \textbf{Year} & \textbf{Ex.}\\
\midrule
\endfirsthead
\multicolumn{5}{c}%
{\tablename\ \thetable\ -- \textit{Sample. Continued from previous page}} \\
\toprule
\textbf{Tag} & \textbf{Tilte} & \textbf{Venue} & \textbf{Year} & \textbf{Ex.} \\
\midrule
\endhead
\bottomrule 
\multicolumn{5}{r}{\textit{Continued on next page}} \\
\endfoot
\bottomrule
\endlastfoot
AcqGro2006 & Imagined Communities Awareness Information Sharing and Privacy on Facebook & PETS & 2006\\
AdAcBr2013 & Sleights of Privacy Framing disclosures and the limits of transparency & SOUPS & 2013\\
AfBrGr2012 & Detecting Hoaxes Frauds and Deception in Writing Style Online & S\&P & 2012\\
AfCaSt2014 & Doppelg{"a}nger Finder Taking Stylometry to the Underground & S\&P & 2014\\
AgShJa2013 & Do not embarass Re-examining user concerns for online tracking and advertising & SOUPS & 2013\\
AhmIss2007 & A New Biometric Technology Based on Mouse Dynamics & TDSC & 2007\\
AkhPor2013 & Alice in warningland a large-scale field study of browser security warning effectiveness & USENIX & 2013\\
AlbMai2015 & Evaluating the Effectiveness of Using Hints for Autobiographical Authentication A field Study & SOUPS & 2015\\
AlFaWr2015 & The Impact of Cues and User Interaction on the Memorability of System Assigned Recognition-Based Graphical Passwords & SOUPS & 2015\\
AlPoRe2014 & Your Reputation Precedes You History Reputation and the Chrome Malware Warning & SOUPS & 2014\\
AngOrt2015 & WTH Experiences Reactions and Expectations Related to Online Privacy Panic Situations & SOUPS & 2015\\
AtBoHe2015 & Leading Johnny to Water Designing for Usability and Trust & SOUPS & 2015\\
BaMaLi2014 & The Privacy and Security Behaviors of Smartphone App Developers & USEC & 2014\\
BeGiKr2015 & User Acceptance Factors for Anonymous Credentials & WEIS & 2015\\
BeLoSi2007 & Establishing Darknet Connections An evaluation of Usability and Security & SOUPS & 2007\\
BelShe2016 & Crowdsourcing for Context Regarding Privacy in Beacon Encounters via Contextual Integrity & PETS & 2016\\
BenRei2013 & Should users be informed On risk-perception between Android and iPhone users & SOUPS & 2013\\
BeWaLi2010 & The Impact of Social Navigation on Privacy Policy Configuration & SOUPS & 2010\\
BiCoIn2015 & What the App is That Deception and Countermeasures in the Android User Interface & S\&P & 2015\\
BonSch2014 & Towards reliable storage of 56-bit secrets in human memory & USENIX & 2014\\
BoSaRe2012 & Neuroscience Meets Cryptography Designing Crypto Primitives Secure Against Rubber Hose Attacks & USENIX & 2012\\
BrCrDo2013 & Your Attention Please - Designing security-decision UIs to make genuine risks harder to ignore & SOUPS & 2013\\
BrCrKo2014 & Harder to Ignore - Revisiting Pop-up Fatigue and Approaches to Prevent it & SOUPS & 2014\\
BrGrSt2011 & Indirect content privacy surveys - measuring privacy without asking about it & SOUPS & 2011 & \\
BruVil2007 & Improving Security Decisions with Polymorphic and Audited Dialogs & SOUPS & 2007\\
BrViDj2008 & Evaluating the Usability of Usage Controls in Electronic Collaboration & SOUPS & 2008\\
BuBeFa2010 & How good are Humans at Solving CAPTCHAs - A Large Scale Evaluation & S\&P & 2010 & \EXCL \\
BuBePa2011 & The failure of Noise-Based Non-Continuous Audio Captchas & S\&P & 2011\\
BuWoVo2014 & Introducing Precautionary Behavior by Temporal Diversion of Voter Attention from Casting to Verifying their Vote & USEC & 2014\\

CaMiVa2016 & Hidden Voice Commands & USENIX & 2016 & \EXCL \\
CaoIve2006 & Intentional Access Management - Making Access Control Usage for End-Users & SOUPS & 2006\\
ChBiOr2007 & A second look at the usability of click-based graphical passwords & SOUPS & 2007\\
ChBoKa2014 & On the Effectiveness of Obfuscation Techniques in Online Social Networks & PETS & 2014\\
ChChBa2015 & You shouldnt collect my secrets - Thwarting sensitive keystroke leakage in mobile IME apps & USENIX & 2015\\
ChMuAs2015 & On the impact of touch id on iphone passcodes & SOUPS & 2015\\
ChObSt2009 & Sanitizations slippery slope- the design and study of a text revision assistant & SOUPS & 2009 & \EXCL \\
ChPoSe2012 & Measuring user confidence in smartphone security and privacy & SOUPS & 2012\\
ChStFo2012 & Persuasive cued click-points - Design implementation and evaluation of a knowledge-based authentication mechanism & TDSC & 2012\\
CzDeYa2010 & Parenting from the pocket - Value tensions and technical directions for secure and private parent-teen mobile safety & SOUPS & 2010\\
DaKrDa2014 & Increasing security sensitivity with social proof - A large-scale experimental confirmation & CCS & 2014\\
DaPuRa2012 & Impact of spam exposure on user engagement & USENIX & 2012\\
DewKul2006 & Aligning usability and security - a usability study of Polaris & SOUPS & 2006\\
DuHeAs2010 & A closer look at recognition-based graphical passwords on mobile devices & SOUPS & 2010 & \\
DuNiOl2008 & Securing passfaces for description & SOUPS & 2008\\
EgJaPo2014 & Are you ready to lock & CCS & 2014\\
FaFeSh2015 & Anatomization and Protection of Mobile Apps Location Privacy Threats & USENIX & 2015\\
FaHaAc2013 & On the ecological validity of a password study & SOUPS & 2013\\
FaHaMu2012 & Helping Johnny 2.0 to encrypt his Facebook conversations & SOUPS & 2012\\
FoChOo2008 & Improving text passwords through persuasion & SOUPS & 2008\\
GaCaCo2012 & Risk communication design - video vs. text & PETS & 2012\\
GaCaMa2011 &Designing risk communication for older adults & SOUPS & 2011\\
GaChLi2014 & Effective risk communication for android apps & TDSC & 2014\\
GawFel2006 & Password management strategies for online accounts & SOUPS & 2006\\
GiEgCr2006 & Power Streip Prophylactics and Privacy Oh My & SOUPS & 2006\\
GrCoAl2016 & Effect of cognitive depletion on password choice & LASER & 2016\\
GroBar2014 & Social status and the demand for security and privacy & PETS & 2014\\
HaChDh2008 & Use your illusion- secure authentication usable anywhere & SOUPS & 2008 & \EXCL \\
HaChHa2009 & New directions in multisensory authentication & SOUPS & 2009\\
HaCrKl2014 & Targeted threat index - Characterizing and quantifying politically-motivated targeted malware & USENIX & 2014\\
HaDeSm2015 & Where Have You Been - Using Location-Based Security Questions for Fallback Authentication & SOUPS & 2015\\
HaRiSt2012 & Goldilocks and the two mobile devices - going beyond all-or-nothing access to a devices applications & SOUPS & 2012\\
HaScWr2014 & Applying psychometrics to measure user comfort when constructing a strong password & SOUPS & 2014\\
HaZeFi2014 & Its a hard lock life - A field study of smartphone un-locking behavior and risk perception & SOUPS & 2014\\
HuMoWa2012 & Clickjacking - attacks and defenses & USENIX & 2012\\
HuOhKi2015 & Surpass - System-initiated user-replaceable passwords & CCS & 2015\\
JaRaBe2014 & To authorize or not authorize - helping users review access policies in organizations & SOUPS & 2014\\
JeSaJe2007 & Tracking website data-collection and privacy practices with the iWatch web crawler & SOUPS & 2007\\
JoEgBe2012 & Facebook and privacy - its complicated & SOUPS & 2012\\
JusAsp2009 & Personal choice and challenge questions - a security and usability assessment & SOUPS & 2009\\
KaBrDa2014 & Privacy Attitudes of Mechanical Turk Workers and the US Public & SOUPS & 2014\\
KaFlRo2010 & Two heads are better than one - security and usability of device associations in group scenarios & SOUPS & 2010\\
KaMaSo2015 & Sound-proof - Usable two-factor authentication based on ambient sound & USENIX & 2015\\
KaTyWa2009 & Conditioned-Safe Ceremonies and a User Study of an Application to Web Authentication & SOUPS & 2009\\
KayTer2010 & Textured agreements - re-envisioning electronic consent & SOUPS & 2010\\
KeBrCr2009 & A nutrition label for privacy & SOUPS & 2009\\
KeCaLi2012 & Self-identified experts lost on the interwebs - The importance of treating all results as learning experiences & LASER & 2012\\
KhHeVo2015 & Usability and security perceptions of implicit authentication - Convenient secure sometimes annoying & SOUPS & 2015\\
KilMax2012 & Free vs. transcribed text for keystroke-dynamics evaluations & LASER & 2012\\
KluZan2009 & Balancing usability and security in a video CAPTCHA & SOUPS & 2009 & \EXCL \\
KorBoh2014 & Too Much Choice - End-User Privacy Decisions in the Context of Choice Proliferation & SOUPS & 2014\\
KoShCr2014 & Telepathwords - Preventing weak passwords by reading users minds & SOUPS & 2014\\
KoSoTs2009 & Serial hook-ups - a comparative usability study of secure device pairing methods & SOUPS & 2009\\
KrHuHo2016 & Use the Force- Evaluating Force-Sensitive Authentication for Mobile Devices & SOUPS & 2016\\
KuCrAc2009 & School of phish - a real-world evaluation of anti-phishing training & SOUPS & 2009 & \EXCL \\
KuRoCr2006 & Human selection of mnemonic phrase-based passwords & SOUPS & 2006\\
LeMoPe2016 & Privacy Challenges in the Quantified Self Movement - An EU Perspective & PETS & 2016\\
LiAnSc2016 & Follow my recommendations - A personalized privacy assistant for mobile app permissions & SOUPS & 2016\\
LiAsCa2008 & Risk communication in security using mental models & USEC & 2008\\
LiBrYe2011 & Demographic Profiling from MMOG Gameplay & PETS & 2011\\
LiLiSa2014 & Modeling users' mobile app privacy preferences - Restoring usability in a sea of permission settings & SOUPS & 2014\\
LiXiPe2011 & Smartening the crowds- computational techniques for improving human verification to fight phishing scams & SOUPS & 2011\\
LlPoAt2015 & Face-off - Preventing Privacy Leakage From Photos in Social Networks & CCS & 2015\\
MaDeKe2011 & Using data type based security alert dialogs to raise online security awareness & SOUPS & 2011\\
MaLeAd2012 & The PViz comprehension tool for social network privacy settings & SOUPS & 2012\\
MalPre2013 & Sign-up or give-up- Exploring user drop-out in web service registration & SOUPS & 2013\\
MoGaSa2014 & Dynamic cognitive game captcha usability and detection of streaming-based farming & USEC & 2014\\
MohaBe2010 & Do windows users follow the principle of least privilege - investigating user account control practices & SOUPS & 2010\\
MoLiVi2014 & Understanding and specifying social access control lists & SOUPS & 2014\\
NoBlCa2014 & Why Johnny Cant Blow the Whistle - Identifying and Reducing Usability Issues in Anonymity Systems & USEC & 2014\\
PanCut2010 & Usably secure low-cost authentication for mobile banking & SOUPS & 2010\\
PaNoKa2012 & Reasons rewards regrets - privacy considerations in location sharing as an interactive practice & SOUPS & 2012\\
PanPra2014 & Crowdsourcing attacks on biometric systems & SOUPS & 2014 & \EXCL \\
PeKoBu2014 & Cloak and swagger - Understanding data sensitivity through the lens of user anonymity & S\&P & 2014\\
PoHaEg2012 & Android permissions - User attention comprehension and behavior & SOUPS & 2012\\
PoIlMa2014 & Faces in the distorting mirror- Revisiting photo-based social authentication & CCS & 2014\\
PuGros2015 & Towards a Model on the Factors Influencing Social App Users Valuation of Interdependent Privacy & PETS & 2015\\
RaBoJa2014 & To befriend or not - a model of friend request acceptance on facebook & SOUPS & 2014\\
RaDeGr2016 & Privacy Wedges- Area-Based Audience Selection for Social Network Posts & SOUPS & 2016\\
Rader2014 & Awareness of Behavioral Tracking and Information Privacy Concern in Facebook and Google & SOUPS & 2014\\
RaHaBe2009 & Revealing hidden context- improving mental models of personal firewall users & SOUPS & 2009\\
RajCam2016 & Influence of Privacy Attitude and Privacy Cue Framing on Android App Choices & SOUPS & 2016\\
RaWaBr2012 & Stories as informal lessons about security & SOUPS & 2012\\
ReKrMa2016 & How I Learned to be Secure- a Census-Representative Survey of Security Advice Sources and Behavior & CCS2016\\
RiBoMo2016 & Measuring the influence of perceived cybercrime risk on online service avoidance & TDSC & 2016\\
RiQiSt2012 & Progressive authentication- deciding when to authenticate on mobile phones & USENIX & 2012 & \EXCL \\
RoCuJo2014 & Behavioral Experiments Exploring Victims Response to Cyber-based Financial Fraud and Identity Theft Scenario Simulations & SOUPS & 2014\\
RuKiBu2013 & Confused Johnny- when automatic encryption leads to confusion and mistakes & SOUPS & 2013\\
RuOnYo2016 & User Attitudes Toward the Inspection of Encrypted Traffic & SOUPS & 2016\\
SchBon2015 & Learning assigned secrets for unlocking mobile devices & SOUPS & 2015\\
SchRee2009 & 1 plus 1 equal you- measuring the comprehensibility of metaphors for configuring backup authentication & SOUPS & 2009\\
ScMcPa2011 & Empowering end users to confine their own applications - The results of a usability study comparing SELinux AppArmor and FBAC-LSM & TISSEC & 2011\\
ScWaKo2013 & Exploring the design space of graphical passwords on smartphones & SOUPS & 2013\\
ShBeRo2016 & Behavioral Study of Users When Interacting with Active Honeytokens & TISSEC & 2016\\
ShKeKo2012 & Correct horse battery staple- Exploring the usability of system-assigned passphrases & SOUPS & 2012\\
ShKoDu2016 & Designing Password Policies for Strength and Usability & TISSEC & 2016\\
ShKoKe2010 & Encountering stronger password requirements- user attitudes and behaviors & SOUPS & 2010\\
ShKrVi2015 & Portrait of a Privacy Invasion & PETS & 2015 & \EXCL \\
ShKuSe2014 & Beware your hands reveal your secrets & CCS & 2014 & \EXCL \\
ShMaKo2007 & Anti-phishing phil- the design and evaluation of a game that teaches people not to fall for phish & SOUPS & 2007\\
SmeGoo2009 & How users use access control & SOUPS & 2009 & \EXCL \\
StHuBr2012 & Are privacy concerns a turn-off- engagement and privacy in social networks & SOUPS & 2012\\
StoBid2013 & Memory retrieval and graphical passwords & SOUPS & 2013\\
SuEgAl2009 & Crying Wolf - An Empirical Study of SSL Warning Effectiveness & USENIX & 2009\\
TaOzHo2006 & A comparison of perceived and real shoulder-surfing risks between alphanumeric and graphical passwords & SOUPS & 2006 & \\
ThLiCh2016 & What Questions Remain - An Examination of How Developers Understand an Interactive Static Analysis Tool & SOUPS & 2016 & \EXCL\\
UrKeKo2012 & How does your password measure up - the effect of strength meters on password creation & USENIX & 2012\\
VItak2015 & Balancing privacy concerns and impression management strategies on Facebook & SOUPS & 2015\\
WaGeCh2016 & On the Security and Usability of Segment-based Visual Cryptographic Authentication Protocols & CCS & 2016\\
WaRaBe2016 & Understanding Password Choices - How Frequently Entered Passwords are Re-used Across Websites & SOUPS & 2016\\
WrPaBi2012 & Do you see your password- applying recognition to textual passwords & SOUPS & 2012\\
WuMiLi2006 & Web wallet- preventing phishing attacks by revealing user intentions & SOUPS & 2006\\
XuReCh2012 & Security and usability challenges of moving-object CAPTCHAs- decoding codewords in motion & USENIX & 2012\\
YaLiCh2016 & An Empirical Study of Mnemonic Sentence-based Password Generation Strategies & CCS & 2016\\
YeHeOp2014 & An epidemiological study of malware encounters in a large enterprise & CCS & 2014\\
ZhPaWa2016 & An Efficient User Verification System Using Angle-Based Mouse Movement Biometrics & TISSEC & 2016 & \EXCL\\
ZhWaJi2014 & Privacy Concerns in Online Recommender Systems- Influences of Control and User Data Input & SOUPS & 2014\\
\end{longtable}
\end{small}

\section{Contingency Tables}
\label{sec:contingencyTables}
We include a number of contingency tables on the distribution of papers and test results per venue and year. Table~\ref{tab:contingencyVenueYear} shows the distribution of the sample, that is, included papers by venue and year.

\contingencyVenueYear

Tables~\ref{tab:contingencySCOutcomeVenue} and~\ref{tab:contingencySCOutcomeYear} contain the \textsf{statcheck} outcomes aggegated per paper, by venue and year, respectively.

\contingencyTestsSCOutcomeVenue
\contingencyTestsSCOutcomeYear

Tables~\ref{tab:contingencyTestsSCOutcomeVenue} and~\ref{tab:contingencyTestsSCOutcomeYear} show the corresponding \textsf{statcheck} results for individual tests, by venue and year, respectively.
\contingencySCOutcomeVenue
\contingencySCOutcomeYear

\begin{DocumentVersionTR}
\section{Statistics Tools}
We used \textsf{R} (version 3.4.1), with \textsf{statcheck}~\cite{EpsNui2018} (version 1.3.0).

We calculated $p$-values for $t$, $F$, $\chi^2$, and $Z$ test statistics with the \textsf{R} functions
\textsf{pt()}, \textsf{pf()}, \textsf{pchisq()}, and \textsf{pz()}, respectively\footnote{We calculate \textsf{pz()} as
$p_z = 2*\mathsf{pnorm}(-\mathsf{abs}(\vari{z}))$.}.

We computed the waffle plots with the \textsf{R} package \textsf{waffle}~\cite{Rudis2006waffle} (version 0.7.0).

We computed the multinomial logistic regression with the \textsf{R} package \textsf{nnet} (version 7.3-12)~ using \textsf{lmtest} (version 0.9-35) for the likelihood-ratio tests.
We used John Fox's package \textsf{car}~\cite{FoxWei2017car,fox2018r} for regression diagnostics (version 2.1-5).
To display the multinomial logistic regressions over time we used the \textsf{R} function \textsf{polytomous\_effects}, which was originally developed by John Fox for the effect display of multinomial odds~\cite{fox2006effect}.
\end{DocumentVersionTR}

\section{Root Causes for Unparseable Papers}
One paper could not be parsed by \textsf{statcheck} because its $p$-values were inconsistently reported (partially as capital P, partially as capital ``Pr'', ``Pr $<$ t'' etc.).

One paper gave a $p$-value as the Greek letter rho ($\rho$), which was embedded in the PDF as image.

One paper only reported statistics, but unparseable as the test statistics were not transcribed to text. The tables of $\chi^2$-results as well as in-text statistics were embedded as bitmaps. We resorted to consulting the publisher's page for an HTML that could be translated to text.

Three papers wrote out equations as text, e.g., ``Chi-Sq = \dots'' or ``p-value = \dots''

One paper  only reported regression tables with significance codes, but no $p$-values.

One paper had a single statistical statement, which could not be parsed.

\section{All Multinomial Logistic Regressions Conducted}
\label{sec:all_mlr}

The primary regression tables are available in Figures~\ref{tab:coefTabMLRanasoCorrectNHST}, \ref{tab:coefTabMLRanasoInconsistency}, and \ref{tab:coefTabMLRanasoDecisionError}.
\coefTabMLRanasoCorrectNHST
\coefTabMLRanasoInconsistency
\coefTabMLRanasoDecisionError

\paragraph{Reporting of Test Statistics.}
We analyzed \textsf{statcheck} outcome for all statistical tests found ($N=1775$) by \textsf{venue} and \textsf{year}. We conducted a multinomial logistic regression on \textsf{SCOutcome} per test.

Having conducted a likelihood-ratio test between the \textsf{venue}+\textsf{year} model and the null model, the overall model is statistically significant, $\chi^2(30) = 90.713, p < .001$. The model explains McFadden $R^2$= .05 of the variance.

The corresponding predictors are statistically significant as well. Hence, we reject the null hypotheses \const{H_{V,0}} and \const{H_{Y,0}}.

Figure~\ref{fig:combinedplotTestsVYall} on p.~\pageref{fig:combinedplotTestsVYall} contains an overview of the scatter plot vs. the predicted probabilities from the MLR in the top pane (\ref{fig:mlrplotVYtests}).

\paragraph{Reporting Excluding Incomplete Test Statistics.}
We note that the cases with incomplete test statistics dominate the analysis of the MLR on tests.
We, therefore, conduct a second MLR solely on tests with complete test statistics triplets.
A likelihood-ratio test \textsf{venue}+\textsf{year} vs. null shows that the model is marginally statistically significant, $\chi^2(16) = 24.491, p = .079$, McFadden $R^2$= .10.

The predicted probabilities are shown in the middle pane (\ref{fig:mlrplotVYtestsWoUnp}) of Figure~\ref{fig:combinedplotTestsVYall} on p.~\pageref{fig:combinedplotTestsVYall} along with the corresponding scatter plot, for information.

\paragraph{Reporting per Paper.}
We conducted an analysis of the \textsf{statcheck} outcome per paper by \textsf{year}. For that, we have aggregated the \textsf{statcheck} results for each paper and then conducted a multinomial logistic regression on the aggregate $N=114$.

Testing for the overall significance by a likelihood-ratio test between the designated model and the null model, we find the overall model non-significant, $\chi^2(3) = 3.331, p = .343$, McFadden $R^2$= .01.

The scatterplot for this analysis and the predicted probabilities are shown in the bottom pane (\ref{fig:mlrplotYear}) of Figure~\ref{fig:combinedplotTestsVYall} on p.~\pageref{fig:combinedplotTestsVYall}, for information.

\combinedplotTestsVYall

\afterpage{\clearpage}

\end{DocumentVersionTR}

\end{appendix}

\end{document}